\documentstyle[12pt,epsf,epsfig]{report}
\def\theequation{\arabic{section}\arabic{equation}}

\renewcommand\thesection{\arabic{chapter}.\arabic{section}}

\renewcommand\theequation{\arabic{chapter}.\arabic{equation}}

\newcommand{\av}{\vec{a}}

\newcommand{\apm}{{\hat{A}}^{+}_{1}{\hat{A}_2}}
\newcommand{\aone}{\disc{\hat{A}_1}}
\newcommand{\atwo}{\disc{\hat{A}_2}}
\newcommand{\adotmu}{ \dot{a_\mu} }
\newcommand{\dotamu}{ \dot{a_\mu} }
\newcommand{\bra}[1]{\langle{#1}\vert}
\newcommand{\ket}[1]{\vert{#1}\rangle}
\newcommand{\matel}[3]{\bra{#1}{#2}\ket{#3}}
\newcommand {\bzero}{\disc{B_{0}}(\vecr)}

\newcommand{\cg}{{C}_{\mbox{g}}}

\newcommand{\cl}{\cal {L}}

\newcommand{\delxsq}{ \delta_{x}^{2} }
\newcommand{\delzsq}{ \delta_{z}^{2} }
\newcommand{\dd}{\partial}
\newcommand{\dmuup}{\dd^{\mu}}
\newcommand{\dmud}{\dd_{\mu}}
\newcommand{\dnud}{\dd_{\nu}}

\newcommand{\deltax}{\disc\delta_{x}}
\newcommand{\deltaz}{\disc\delta_{z}}
\newcommand{\delb}{ \Delta^B  }
\newcommand{\deff}{D_{\mbox{eff}}}

\newcommand{\dsf}{\displaystyle\frac}

\newcommand{\expon}[1]{{e}^{\disc{#1}}}
\newcommand{\eps}{\varepsilon}

\newcommand{\fpi}{f_{\pi}}
\newcommand{\fomegann}{ F_{ {\omega}NN}}

\newcommand{\grhonn}{ G_{ {\rho}NN}}
\newcommand{\frhonn}{ F_{ {\rho}NN}}

\newcommand{\ga}{\disc{g_{A}}}
\newcommand{\gomegann}{ G_{ {\omega}NN}}

\newcommand{\gzero}{g^{\prime}_o}
\newcommand{\gsnn}{g_{\sigma NN}}
\newcommand{\gpinn}{\disc{g_{{\pi}NN}}}
\newcommand{\gastar}{\disc{g_A^{*}}}
\newcommand{\gsnnstar}{g_{\sigma NN}^*}
\newcommand{\gpinnstar}{\disc{g_{\pi NN}^{*}}}
\newcommand{\half}{\displaystyle \frac{1}{2}}
\newcommand{\intdt}{\disc{ \int_{0}^{1}dt}}
\newcommand{\integ}{\int_{0}^{\infty}dx}
\newcommand{\il}{\int\limits}

\newcommand{\lhisb}{{\cl}_{{\disc\chi}\mbox{sb}}}

\newcommand{\ltwo}{{\cl}_{\mbox{2}}}

\newcommand{\lfora}{{\cl}_{\mbox{4a}}}
\newcommand{\lfors}{{\cl}_{\mbox{4s}}}

\newcommand{\lsk}{ {\cl}_{{\disc\mbox{sk}}} }

\newcommand{\lkin}{ {\cl}_{{\disc\mbox{kin}}} }

\newcommand{\lamm}{{\disc\lambda_{M}}}

\newcommand{\lbr}{\lbrack}
\newcommand{\rbr}{\rbrack}
\newcommand{\ra}{\rightarrow}

\newcommand{\msig}{m_{\sigma}}
\newcommand{\mod}[1]{\vert{#1}\vert}
\newcommand{\mpi}{{\disc{m}_{\pi}}}

\newcommand{\mn}{M_{N}}

\newcommand{\md}{M_{\Delta}}
\newcommand{\mh}{M_{H}}
\newcommand{\mpiq}{{\disc{m}_{\pi}^{2}}}

\newcommand{\nbn}[1]{{#1}\bar{#1}}
\newcommand{\nwl}{\\[2mm]}
\newcommand{\ofo}[1]{ \bra{0} {#1} \ket{0} }

\newcommand{\pal}{\partial}

\newcommand{\pirow}{\pi\rho\omega}
\newcommand{\pirows}{\pi\rho\omega\sigma}
\newcommand{\piros}{\pi\rho\omega\sigma}
\newcommand{\pr}[1]{{#1}^\prime}
\newcommand{\piv}{\vec{\pi}}

\newcommand{\qzero}{ Q_{I=0} }
\newcommand{\qone} { Q_{I=1} }

\newcommand{\qkw}{q^2}

\newcommand{\rhat}{\hat{r}}

\newcommand{\rv}{\vec{r}}
\newcommand{\rhoxdar}[1]{\disc{\rho^{#1}(x)}}

\newcommand{\rhoprim}{{\rho}'}
\newcommand{\rhoprimprim}{{\rho}''}
\newcommand{\rhozero}{{\disc\rho_{0}}}
\newcommand{\rhox}{\rho(x)}
\newcommand{\rhoone}{{\disc\rho_{1}}}
\newcommand{\rhotwo}{{\disc\rho_{2}}}

\newcommand{\riso}[1]{ {\langle {#1}^2 \rangle} }
\newcommand{\rmsscal} { \disc\riso{r} _{I=0} }

\newcommand{\rmsmag} { \disc\riso{r} _{M,I=0} }
\newcommand{\sigmat}{\tilde{\sigma}}

\newcommand{\sigmav}{\vec{\sigma}}
\newcommand{\sigmavo}{{\sigmav}_{1}}
\newcommand{\sigmavt}{{\sigmav}_{2}}
\newcommand{\sinkwt}{\sin^2 \theta}
\newcommand{\stwo}{{\disc}s_{2}}
\newcommand{\summa}{\disc\sum}
\newcommand{\tauv}{\vec{\tau}}
\newcommand{\tauone}{ \tauv_{1}   }
\newcommand{\tautwo}{ \tauv_{2}   }

\newcommand{\thprim}{{\theta}'}
\newcommand{\thprimprim}{{\theta}''}
\newcommand{\tetone}{{\tilde{\theta}}_{1}}

\newcommand{\Tr}{\mbox{Tr\,}}
\newcommand{\uzero}{\disc{U_0}}
\newcommand{\uone}{{\disc}U_{1}}
\newcommand{\utwo}{\disc{U_2}}

\newcommand{\vecr}{\vec{r}}
\newcommand{\vectau}{\vec{\tau}}

\newcommand{\vnn}{V_{NN}}

\newcommand{\vcr}{V_{C}(r)}
\newcommand{\vtwo}{\disc{V}_{2}(x,\xone)}
\newcommand{\vfora}{\disc {V}_{4a}(x,\xone)}
\newcommand{\vw}{\disc {V}_{W}(x,\xone)}

\newcommand{\vhisb}{ V_{{\disc\chi}\mbox{sb}}(x,\xone)}

\newcommand{\vecx}{\vec{x}}
\newcommand{\vecrone}{\vec{r_{1}}}
\newcommand{\vecrtwo}{\vec{r_{2}}}
\newcommand{\vecri}{\vec{r_{i}}}
\newcommand{\vvtwo}{V_{\mbox{2}}}
\newcommand{\vvfora}{V_{4a}}
\newcommand{\vvhisb}{V_{{\chi}sb}}
\newcommand{\vvw}{V_{W}}

\newcommand{\veps}{\varepsilon}

\newcommand{\xpr}{\pr{x}    }
\newcommand{\xonetwo}{{x_{1,2}}}
\newcommand{\xone}{x_{1}}
\newcommand{\zone}{\disc{z_1}}

\newcommand{\zonetwo}{{z_{1,2}}}
\newcommand{\be}{\begin{equation}}
\newcommand{\ee}{\end{equation}}
\newcommand{\beq}{\begin{equation}}
\newcommand{\eeq}{\end{equation}}
\newcommand{\ba}{\begin{array}{l}}
\newcommand{\ea}{\end{array}}
\newcommand{\bea}{\begin{eqnarray}}
\newcommand{\eea}{\end{eqnarray}}
\newcommand{\banonum}{\begin{eqnarray*}}
\newcommand{\eanonum}{\end{eqnarray*}}
\newcommand{\banum}{\begin{eqnarray}}
\newcommand{\eanum}{\end{eqnarray}}
\newcommand{\lab}[1]{\label{#1}}
\newcommand{\bfr}{\begin{flushright}}
\newcommand{\bfl}{\begin{flushleft}}
\newcommand{\efl}{\end{flushleft}}
\newcommand{\efr}{\end{flushright}}
\newcommand{\bi}[1]{\bibitem{#1}}
\newcommand{\bb}{\begin{thebibliography}}
\newcommand{\eb}{\end{thebibliography}}
\newcommand{\bc}{\begin{center}}
\newcommand{\ec}{\end{center}}
\newcommand{\ci}[1]{\cite{#1}}
\newcommand{\disc}{\displaystyle}
\newcommand{\ds}{\displaystyle}
\newcommand {\bdc}{\begin{document}}
\newcommand{\edc}{\end{document}}
\newcommand{\re}[1]{(\ref{#1})}
\newcommand{\vecnab}{\mbox{\boldmath $\nabla$}}
\newcommand{\veck}{{\bf k}}
\newcommand{\vecV}{{\bf V}}
\newcommand{\vecB}{{\bf B}}
\newcommand{\tochka}{\, .}
\newcommand{{\vergul}}{\, ,}
\begin{document}
\newcounter{fignum}
\newcounter{tabnum}
\thispagestyle{empty}
\normalsize
\renewcommand{\baselinestretch}{1.1}
\bc {\bf PROPERTIES OF MESONS AND  NUCLEONS IN CHIRAL TOPOLOGICAL
MODELS OF QCD.} \ec \bc (Thesis of the doctoral dissertation -
second degree) \ec
\smallskip
\bc {\bf Abdulla Rakhimov} \ec
%\smallskip
\bc Institute of Nuclear Physics, Tashkent - 700132,
Uzbekistan\footnote{ E-mail: rakhimov@rakhimov.ccc.uz;
rakhimovabd@yandex.ru} \ec
\medskip
\bc \bf Abstract \ec

    The problem under consideration has, actually,
two aspects. The first one concerns the case when a
 nucleon is in free space that is in vacuum,
 while the second one studies a nucleon embedded into a nuclear environment.
 Both of these two aspects have been considered in the framework
 of chiral topological models of QCD.  In this sense the whole content
 of the thesis may be divided into two main parts.

   In the first part the original Skyrme model
(with $m_\pi\neq 0$) has been extended by inclusion of the
 light scalar - isoscalar
 $\sigma$ - meson. The lagrangian has following advantages:\\
 It satisfies the scale invariance and trace anomaly constraint
 of QCD;\\
 It gives a realistic nucleon - nucleon (NN) interaction potential,
which is quite close to the phenomenological one.\\
 The lagrangian has been further extended by explicit inclusion of $\sigma$,
$\rho$
 and $\omega$ - mesons as well.
 In order to get a more complete picture of NN
 potential the appropriate meson - nucleon vertex form - factors
were obtained.

   The second part of the thesis considers a nucleon immersed into a
nuclear medium. For this purpose a medium modified Skyrme
lagrangian
 has been proposed. The lagrangian describes well such well known medium
effects as decreasing of nucleon mass and increasing of its size
( swelling). We studied also a system with finite temperature
($T\neq 0$) also.
 The temperature effects were taken into account by using the method of
 termofield dynamics (TFD).
 The explicit calculations show that the temperature dependence
 of meson - nucleon vertex form - factors is not simple.
 When the temperature increases they remain almost unchanged until
 a certain temperature $T_C$ then start to decrease dramatically.
 The corresponding critical temperatures for each meson - nucleon
system were calculated and the appropriate conclusion about their
origin was brought.

    The methods and results of this thesis may be used in
 the studies of heavy -  ion collisions as well as  of cosmology and
 of astrophysics.
\newpage

%\large
%\renewcommand{\baselinestretch}{1.3}
\setcounter{page}{2}
\tableofcontents
%\listoffigures
%\listoftables
%  Title delaem v WORD\include{title}
%\include{intro}
\section*{}
\addcontentsline{toc}{section}{{\bf Introduction}}

\bc {\Large\bf Introduction} \ec \indent

The main goal of this thesis  is the investigation of hadron
properties in free space as well as in nuclear medium.
%%%%%    VIN MAu-----------------
It has been generally believed that, properties of hadrons and
their interaction should be described in framework the of Quantum
Chromodynamics (QCD) which is the underlining fundamental theory
of hadron physics. However, in this theory one is failed with the
difficulty related to the running coupling constant
 $\alpha(\qkw)$. In fact, for large momentum transfer $q$, as in the deep
inelastic scattering of leptons by hadrons, $\alpha(\qkw)$ is
small, and hence one can use the perturbation theory to make
theoretical predictions which turned out to be in good agreement
with the experiment. On the other hand, for the  processes of
interest in nuclear physics or low energy hadron physics, the
length scale is typically $\sim 1 fm$,
 that corresponds
to small momentum transfer $q$. The running coupling constant
$\alpha(\qkw)$ is then large and perturbation theory is useless.
In this strong coupling regime, non perturbative methods are
indispensable,
 but so far, not much  successes
has been achieved in this respect. In this regime  one is led to
modelling non - perturbative QCD. So, the challenge to nuclear
physicist is thus to find models which can  bridge the gap
between the fundamental theory and our wealth of  knowledge about
low energy phenomenology.

 Our
choice of models should be inspired by QCD, perhaps one day will
be derived from it, and embody its important features, such as
confinement and chiral symmetry, while avoiding the full
complexity of a non - abelian guage theory. The parameters of
these models must, for now at least, be determined
phenomenologically, although  the hope is that eventually they
should be calculated from QCD.
%%%%%%%%  MEissner and Zahed, p. 144

At present there is no systematic method for treating continuum
QCD in the perturbative regime. With the exception of lattice
Monte Carlo methods which still encounter both conceptual and
technical problems mainly related to chiral fermions, the
understanding of low - energy  physics at the hadronic scale is
mostly qualitative and relies heavily on the use of such
phenomenological models as potential models \ci{isgurkarl}, bag
models \ci{thomas},  hybrid- bag models \ci{chodosthorn} and
effective chiral models \ci{leebook}.

In the non relativistic potential models of Isgur and Karl
\ci{isgurkarl} the quarks are treated as  constituent of protons
 much like in the conventional parton model.
The confinement character of QCD is put in by hand using a
linearly rising potential between the constituent quarks at large
distances. The remaining gluonic effects are treated
perturbatively, leading to Breit - Wigner hyperfine interactions.
While these models are justified  for the heavy flavor hadrons
such as charmonium and bottonium, they are questionable for the
light hadrons such as  $N$ and $\Delta$, where we know that,
the     constituent quarks are highly relativistic.

In the usual bag models, for instance, in the
 MIT bag model, the quarks are confined  into a region
of hadronic size $ V\approx \Lambda^3 $,
 inside of which the quarks carry their
current masses and are treated fully relativistically. The bag is
assumed to be a bubble of perturbative vacuum immersed into a non
perturbative environment  characterized by a bulk parameter $B$ (
the bag pressure) \ci{chodosjaffe}. Because of asymptotic freedom
the remaining gluonic effects are treated perturbatively inside
the bag. Although  rather  successful in describing the
spectroscopy of the lightest hadrons, the MIT bag model suffers
from a serious drawback, namely, axial - vector conservation. The
confining  bag wall breaks explicitly chiral symmetry, allowing
left  and right- handed quarks to mix at the boundary. We know
that, in the chiral limit
 ($m_f=0$ )
QCD is chirally symmetric.
%They lack, however, flexibility.
            The standard treatment of
the MIT and cloudy bag model (CBM) lack translational invariance
and they do not provide a dynamical description of hadron. This
makes center of mass (c.m.)
 and recoil corrections difficult to calculate
\ci{goldflam, ourcbm1, ourcbm2, ourcbm3}. Besides, these models
are not convenient to describe nucleon - nucleon dynamics as they
make explicit use of the quark degrees of freedom.

%%%%%%%%%%%%%%%%%%%%%%%%%%%%%%%%%%%%%%%%%%%%%%%%%%
%%%%%%%  VIn MAu ____________
On the other hand, since the fundamental constituents in QCD,
quarks and gluons are confined, it might be more suitable to
eliminate their degrees  of freedom which are not directly
observable and incorporate their effects into an effective theory
framed in terms of observable hadronic degrees of freedom
including as many relevant features of QCD as possible.

 This alternative theoretical route can be traced back to the works of
't Hooft  \ci{thooft} and Witten \ci{witten}.
 't Hooft proposed to look for the
parameter in QCD, other than the coupling "constant"
-$\alpha(\qkw)$ in terms of which one can perform expansions and
arrived at the $1/N_c$ expansion, $N_c$ being the number of
colors. Here one can note that, there is nothing fundamental in
the value $N_c=3$; this is required by phenomenological
considerations as the fit to the $\pi_0\rightarrow 2\gamma$ decay
rate or to the ratio of $e^+e^-$ cross sections,
$\sigma_{\mbox{tot}}(e^+e^-)/\sigma(e^+e^-\ra \mu^+\mu^-) $.

In a detailed analysis of QCD diagrams in the $1/N_c$ expansion
Witten \ci{witten} was led to the conjecture that in the large $
N_c$ limit, QCD is equivalent to an effective
 field theory involving only {\bf mesons}
(and glueballs). Soon after this result, Witten \ci{anw} also
recognized that {\bf baryons} can be regarded as {\bf solitons}
of this meson theory in the manner first proposed by Skyrme
\ci{skyrme}.
%*********T.H.Skyrme  Proc. Roy. Soc. A260, 127

All this results give support to the idea that low energy hadron
physics can be described by effective theories constructed from
meson fields alone. Therefore nonlinear chiral models including
only meson degree of freedom will be used to study hadron
properties in free space and in nuclear medium.
%%%%%%%%%%%  Mathiot************
Before entering into more details, we would like first to clearly
identify what we mean by nuclear medium. There are obviously two
different regimes, leading to two different physical
representations of the medium.
 At very high density one expects
a phase transition in which quarks and gluons are deconfined and
more or less at the same time chiral symmetry is restored. Above
the critical density
 $\rho_c$
- which is expected to be of order of
 several normal nuclear matter density $\rhozero$
- one  thus expects a uniform medium of weakly interacting
 (almost) massless
quarks and gluons.

Below this critical density, quarks and gluons are confined and
one is
 dealing
with a medium composed of more or less densely packed hadrons.
Near the critical density it may be difficult
 to clearly identify nucleons in the
medium dominated by all types of hadrons  ( mainly mesons and
baryon resonances). At normal nuclear density or less the
situation is however much simpler. We know that, the relevant
degrees of freedom are nucleons interacting by the exchange of
many correlated mesons \ci{mathiot}.

We shall concentrate in this thesis  on the low energy regime.
%%-----------------------------------------------------------
One of the most exiting topics in this regime is the study of
variation of hadron properties as the nuclear environment
changes. Recent experiments from HELIOS-3 \ci{helios} and the
CERES \ci{ceres} at SPS/CERN energies have shown that there
exists a large excess of $e^+e^-$ pairs in central $S+Au$
collisions. Those experimental results may give a hint of some
change of a hadron properties in nuclei. Ultrarelativistic heavy
- ion experiments ( e.g. at RHIC) \ci{last} are also expected to
give significant information on the strong interaction (QCD)
through the detection of changes in hadronic properties.

Theoretically lattice QCD simulations may eventually give the most
reliable information on the density  and or temperature dependence
of hadron properties in matter. However current simulations have
been performed only for finite temperature $(T\neq0)$. Therefore
many authors have studied hadron masses in matter using effective
theories and have reported that the mass decreases in nuclear
medium. In the present thesis we shall make an attempt to modify
Skyrme model to study the changes in mass of hadrons and in their
interaction.

The present thesis is organized as follows. In Chapter 1 the
basic ideas and techniques of the Skyrme model "for beginners" is
reviewed. In Chapters 2 - 4 we shall consider nucleon properties
and nucleon - nucleon (NN) interaction in free space. Here the
original  Skyrme model is extended by inclusion of
 scalar - isoscalar meson which plays a crucial role in  central
the nucleon - nucleon interaction.

It is believed that the nucleon properties and NN interaction may
change in the nuclear medium. This modification is considered in
Chapters 5-7. The main conclusion of the present thesis are
summarized in the last chapter.
%%%%%%%%   end OF INTR0
%%%%%%%%%%%%%%%%%%%%%%%%%%%%%%%%%%%
%%%%%%%%%%%%%%%%%%%%%%%%%
%\include{ch1rev}
%%%%%%%%  BEGIN CHAP1
\chapter
[
 The physics   of Skyrme solitons.\\
(A brief review) ]{} \bc { \Large\bf
 The physics   of Skyrme solitons.\\
(A brief review) } \ec
%
%}%
%\markboth{\thechapter\hspace{1em}%
%Chapter 2}{}
%\renewcommand{\chaptermark}[1]{\markboth
%{\uppercase{\thechapter\hspace{1em}#1}} %
%{\uppercase{\thechapter\hspace{1em}#1}} %
%}% kones
\indent

An important property of QCD is its chiral symmetry. This
invariance must be preserved in the corresponding effective field
theories \footnote { A reader is referred to  more extended
review articles \ci{zb,holzrev,mz,nik} }.
    \section { The $\sigma$ model}
\indent

The simplest chirally invariant effective field theory is the
$\sigma$ model proposed by Gell Mann and Levy \ci{gellmanlevy}.
Its version {\bf without fermions} falls precisely in the class
of effective field theories contemplated  by Witten. This version
of the $\sigma$  model can be formulated in the following way.
Consider a $2\times 2$ matrix field \be U(r)= \frac{1}{\fpi}
[\sigma(r)+i\vectau\piv(r)] \ee where $\sigma(r)$  is a scalar
and isoscalar field, $\piv(r)$ is a pion field and
 $\fpi$ is the pion decay
constant. The Lagrangian density can than be written as \be \ba
\ltwo(r)=\half {(\dmud\sigma)}^2+\half
{(\dmud\pi)}^2=\frac{\fpi^2} {4}\Tr [\dmud U(r)\dmuup U^{+}(r)]
\ea \lab{lskyrm2} \ee

In the non-linear realization of the $\sigma$ model, the $\sigma$
field is eliminated from the model via the chirally invariant
constraint, \be \sigma^2+{\piv}^2=\fpi^2 \ee implying that $U$ is
unitary. Defining
 $\sigma=\fpi cos\theta(r) $  and $\piv=\fpi \hat\pi sin\theta(r) $,
we can write $U=\expon{i\vectau\hat\pi\theta(r)} $ with
$\hat\pi=\vec\pi/\mod{\pi}$ .

The Lagrangian \re{lskyrm2} possesses chiral $SU(2)_V \otimes
SU(2)_{A} $ symmetry. Left and right transformations are
associated with left and right multiplication by elements of
$SU(2)$ group. The chiral $SU(2)_{R}\otimes SU(2)_{L} $
 group of transformations of the field $ U$
will correspond to the direct product of left and right
transformations of the elements of $SU(2)  $ group:
  $ U\rightarrow AUB^{+}$ with arbitrary constant
$SU(2) $ matrices $A$ and $B$. The chirally invariant action is
usually written by means of left (or right) invariant Cartan
forms: \be \ba \ltwo=-\dsf{\fpi^2}{4}\Tr(L_\mu L^\mu), \nwl
L_\mu=U^{+}\dmud U ,\;\;\;R_\mu=U \dmud U^{+}. \ea \ee
%%%%%%  ISPRAVLEN ZNAK 02.02.2005
In terms of basic $\pi$ and $\sigma$  fields the corresponding
vector $\vec{V}_{\mu}=[\piv \dmud \piv]  $ and axial currents
$\vec{A}_{\mu}=\sigma\dmud \piv-\piv\dmud\sigma $
 may be obtained by infinitesimal rotations in
isospin space
 $ \sigma  \ra\sigma$, $\piv\ra\pi-[\vec\eps\piv]$
and $\sigma\ra\sigma-(\vec\eps\piv)$,
 $\piv\ra\piv+\vec\eps\sigma$ respectively. Clearly
 $\dmud V^{\mu}=0 $ and
 $\dmud A^{\mu}=0  $ in accordance  with chiral invariance.

The trouble with the non-linear $\sigma$ model is that it leads
to energetically unstable solutions; by dimensional arguments,
one can see that the energy scales like $ R $, where $ R$ is the
size of the system, and it vanishes when the system shrinks to
zero size.

\section {  The Skyrme Model}%
\markboth{\thesection\hspace{1em}%
model }{} \indent

To remedy this shortcoming, Skyrme \ci{skyrme}  proposed  the
addition of higher-order terms in the derivative $\dmud U $ to the
form $\ltwo(r) $. He suggested to introduce additional term
 of the form
\be \ba \lfora(r)=\dsf{1}{32 e^2}\Tr[(\dmud U)U^{+},(\dnud
U)U^{+}]^2= \dsf{1}{32{e}^{2}}{\Tr}{[L_{\mu},L^{\nu}]}^{2}, \ea
\ee with the four derivative term.
 The resulting Lagrangian density
\be \lsk(r) =\ltwo(r)+\lfora(r) \lab{lsk} \ee
  yields a
complicate non-linear Euler-Lagrange equation which is difficult
to solve. However, if one makes the ansatz that the pion field is
directed radially in configuration space, $\hat\pi=\rhat $,
 $\rhat$ being a unit vector in coordinate space
  $ \hat{r}=\vec{r}/\mod{r}$,
the U field is of the "hedgehog" form \ci{chodosthorn,skyrme} \be
U=U_0=\exp{ (i\vec{\tau}\vec{r}\theta(r))}. \lab{ejik} \ee
 In this case, the Lagrangian is
\be \ba L=\ds\int d^3 r \lsk(r)=\dsf{\pi \fpi}{e} \integ dx
[x^2(\theta'{^2}+ \dsf {2\sinkwt} {x^2})+ \nwl +4 \sinkwt (
\dsf{\sinkwt} {x^2} +2{\theta'}^2)] \ea \lab{ch1mass} \ee
 with the choice of length scale $x=2e\fpi r  $ and
 $\theta'=d\theta/dx$. Minimizing L
with respect to $\theta$, i.e. using $\frac{d}{dx} (\frac{ \dd
L}{\dd \theta'}) =\frac{\dd L}{\dd \theta} $
 one gets the following Euler-Lagrange
equation \be \ba \theta''(\dsf{x^2}{4}+2\sinkwt)+\dsf{x\theta'}{2}
-\dsf{\sin2\theta}{4} + \nwl + {\theta'}^2
\sin2\theta-\dsf{\sinkwt \sin2\theta}{x^2}=0. \ea \ee Soliton
solutions can be found  with the boundary conditions
 $\theta(\infty)=0  $ and  $\theta(0)=n\pi  $, $n - $ integer.
 This solution
is now energetically stable since the  contribution of the term
$\lfora $ of \re{lsk} to the energy is proportional to $1/R$ so
that the total energy can be minimized with respect to R to give
a finite value. It can also be shown \ci{skyrme} that the
solutions characterized by this boundary conditions are
topological solitons whose winding number is $ n$.

\subsection { The Baryonic Current.}
\indent

 Besides the usual Noether currents, such as $V_\mu$ and $A_\mu$,
associated with various symmetries fulfilled by the Lagrangian
$\lsk(r)$, Skyrme showed that one can construct an "anomalous"
current \be B_\mu=\dsf{1}{24\pi^2}\eps_{\mu\alpha\beta\gamma}\Tr
L^\alpha L^\beta
 L^\gamma ,
\ee
 which is automatically conserved.
A new quantum number can be therefore associated with
 $B_\mu$ since
this current  does not correspond to any initial symmetry of the
Lagrangian. With the hedgehog solution, the time component of
$B_\mu  $  is \be \quad B_0(r)=\frac{(\cos2\theta-1) }{4\pi^2r^2
}\frac{d\theta}{dr}. \quad \ee Integrating this density over all
space $\vecr$ and taking into account the boundary conditions one
gets $B=\ds\int d^3 r B_0(r)  =n $ where $n$ is winding number
characterizing  the solution $U$. At this point Skyrme identified
$n$ with the byron number, or equivalently $B_\mu$ with the
byronic current. {\bf Baryons} therefore emerge from the model as
{\bf topological solitons}.

   \subsection{ Quantization.}
\indent

 So far, the Lagrangian as well as the Euler-Lagrange
equation are classical. $U$  is a classical field and therefore
carries no definite spin and isospin values. A simple
quantization method starting from the classical solution
$U=U_0=\exp{ (i\vec{\tau}\vec{r}\theta(r))}$
 was proposed by Adkins, Nappi and Witten
 \ci{anw}. These authors introduce collective coordinates
through the unitary transformation $A=a_4+i\vectau\av  $
 where the
$a_\mu  $'s are independent of $\vecr$ but can depend on time $t$.
The  $a_\mu  $'s also satisfy the constraint $a_{4}^{2}+\av^2=1  $
 since $A$ is unitary.
Under this transformation, $U_0$ becomes $U=A(t)U_0A^{+}(t)  $.
Substituting U into the Lagrangian yields \be \ba L=\ds\int d^3 r
\lsk\lbr U=A(t)U_0A^{+}(t) \rbr = -M+\lambda\Tr(\dot A{\dot{
A}}^+)= \nwl =-M+2\lambda\disc\sum_{\mu=1}^{4}({\dotamu})^2 \ea
\ee
 where e.g. $\dot a=da/dt$, M is given by
eq. \re{ch1mass}  and \be \lambda=\dsf{\pi}{3\fpi e^3} \integ x^2
\sinkwt[1+4(\theta'{^2}+\sinkwt/ {x^2})] \ee From the Lagrangian
one gets the Hamiltonian \be H=\Pi_\mu\adotmu-L , \ee where
$\Pi_\mu$ 's
 are the conjugate momenta
$\Pi_\mu=\dd L/\dd\adotmu=4\lambda\adotmu  $ Substitution of this
equation into the previous one yields \be
H=M+\frac{1}{8\lambda}\summa_{\mu=1}^{4} \Pi_{\mu}^{2}\;\;\; . \ee
  With the usual canonical quantization procedure
$\Pi_{\mu}=-i\dsf{\dd}{\dd a_\mu}  $, the Hamiltonian takes the
form \be H=M+\frac{1}{8\lambda}\summa_{\mu=1}^{4}
(-\frac{\dd^2}{\dd a_{\mu}^{2} }) \lab{H} \ee
 with the
constraint $a_{4}^{2}+\av^2=1   $. Because of this constraint the
term $\summa_{\mu=1}^{4}\frac{\dd^2}{\dd a_{\mu}^{2} }  $ is to
be interpreted as the Laplacian on a 3-dimensional sphere. The
wave functions of $H$ are polynomials in  $a_\mu$.

\subsection{The Spin and Isospin Operators}
\indent

 Under a rotation about the z-axis in configuration
space by a small angle $\delta$, it is easy to show that \be
\expon{i\vectau\vecr}\ra\expon{i\delta\tau_z/2}
\expon{i\vectau\vecr} \expon{-i\delta\tau_z/2} \ee
   Hence, under this rotation the field configuration
 $AU_0A^+  $ transforms into a new one with
\be \ba A\ra A \expon{i\delta\tau_z/2},  \nwl (a_4+i\vectau\vec
a)\ra (a_4+i\vectau\vec a)(1+i\delta\tau_z/2) \ea \ee
 This gives
after some algebra \be \ba a_4\ra a_4 -\delta a_z /2,\quad a_z\ra
a_z +\delta a_4  /2,\nwl a_x\ra a_x -\delta a_y /2,     \quad
a_y\ra a_y +\delta a_x /2 . \ea \ee
 Identifying this transformation
with the initial rotation written in terms of the spin operators,
i.e. $a_\mu\ra (1+i\delta J_z)a_\mu$
 one finds
\be
 J_z={i\over{2}} [-a_4\frac{\dd}{\dd a_z}+a_z\frac{\dd}{\dd a_4}
-a_x\frac{\dd}{\dd a_y}+a_y\frac{\dd}{\dd a_x}] \ee
 Under a rotation
about the z-axis of isospin space by a small angle $\delta $
$A\ra  \expon{-i\delta\tau_z/2}A$
 and the same calculation as before leads
to \be
  I_z={i\over{2}}\lbr a_4\frac{\dd}{\dd a_z}-a_z\frac{\dd}{\dd a_4}
-a_x\frac{\dd}{\dd a_y}+a_y\frac{\dd}{\dd a_x}\rbr \ee It is easy
to verify that the spin-isospin wave functions for various
baryons are: \be \ba \ket{p\uparrow}=(a_x+ia_y)/\pi\; , \quad
 \ket{p\downarrow}=-i(a_4+ia_z)/\pi \;,
\nwl \ket{n\uparrow}=i(a_4+ia_z)/\pi \;,\quad
   \ket{n\downarrow}=-(a_x-ia_y)/\pi\;.
\ea \ee
 The
calculation  of observables is then reduced to the evaluation of
appropriate matrix element of operators (constructed from the
field configuration) with these spin-isospin wave functions. For
example, the proton mass is \be \ba \bra{p\uparrow}      H
\ket{p\uparrow} = \dsf{1}{ {\pi}^2} \disc\int d^4 a \delta
(\disc\sum_{\mu}a_{\mu}^{2}-1) (a_x-ia_y)H(a_x+ia_y)=\nwl
=M+\dsf{3}{8\lambda} \ea \ee
 where the expression
\re{H}  for  $ H (U=AU_0A^+)$       has been used.

      \subsection {Results for the Static Properties of Baryons}
\indent

The previous procedure has been applied  by Adkins, Nappi and
Witten  \ci{anw}  to the calculation of baryon masses (nucleon
and delta resonance), charge radii, magnetic moments and coupling
constants. The results are listed in Table \ref{tabch11}. These
authors adjust $\fpi$  and $e$  to fit the nucleon and $\Delta$
masses obtaining $\fpi =64.5 $  MeV and $e= 5.45$. The other
quantities are predicted. They are
 in qualitative agreement with experiment (within
$30 \%$ )  except for $\ga$.

\begin{table}[hbt]
\caption{\large \it
 Baryon properties in the original Skyrme model \ci{anw}. }
%\vspace {1cm}
\begin{center}
\begin{tabular}{|l|c|c|c|}\hline
Quantity &       Prediction &            Experiment \\
\hline
$\mn$  &             input     &              939 MeV   \\
$M_\Delta$     &             input      &            1232 MeV    \\
$\fpi$      &       64.5 MeV        &        93  MeV        \\
$\rmsscal^{1/2}$     &       0.59 fm          &       0.72fm \nwl
$\rmsmag^{1/2}$      &       0.92 fm       &         0.81 fm    \\
$\mu_p$       &         1.87        &           2.79      \\
$\mu_n $      &         -1.31      &            -1.91      \\
$\ga$        &        0.61       &             1.23        \\
$\gpinn$      &         8.9       &             13.5         \\
\hline
\end{tabular}
\end{center}
\label{tabch11}
\end{table}
%%%%%%%  OKA %%%%%%%%%%%%%%%%%
\section{ Two baryon system}
\indent

The principal goal of studying the $B=2$ sector of the Skyrme
model is to apply it to the nucleon  nucleon (NN) interaction.
Characteristic features of the nuclear force shared by
 various semiphenomenological potentials are as follows:
\begin{description}
\item[a)]
a long range ($\ge 1.4$ fm ) interaction dominated by one - pion
exchange potential (OPEP),
\item[b)]
 a medium - range ($\sim 1fm$)
 attraction in the central force, and
\item
[c)]
 a strong short - range
 repulsion ($\le0.7 $fm).
\end{description}
The OPEP is well established, while the microscopic origins of
the medium - and short range interactions are less well
understood.

A simple study has already been made by Skyrme himself,
 which surprisingly contains
many essential features, at least qualitatively. He proved that a
matrix product of two $B=1$  configurations, $U=U_1  U_2$,
belongs to the $B=2$  sector. Then he proposed the "product
ansatz" for two Skyrmions separated by $\vec R$ individually iso
- rotated by $A$ and $B$ : \be U_{12}=AU_0(\rv+\vec
R/2)A^+BU_0(\rv-\vec R/2)B^+ \ee The classical energies
calculated using this ansatz giving  \be E_{12}=2E_0+V(\vec R,C),
\ee
 where the static potential V depends of the relative distance
$\vec R$ and the relative orientation $C=A^{+}B  $. A global
rotation of $U$ does not change the total energy. Skyrme pointed
out the following two facts:
\begin{description}
\item[i.]
 For large $R=\mod{\vec R}$, $\;\;V(\vec R,C)$
 approaches the one -  pion -  exchange potential
 (OPEP):
\be V(\vec R,C) \approx T_{ab}\nabla_a\nabla_b\{\exp(-\mpi R)/R\}
\lab{okaV} \ee with  $T_{ab}=\Tr[\tau_aC\tau_bC^{+}]/2  $. The
Yukawa function in  eq.  \re{okaV} is replaced by $1/R$ for the
massless pion. This is natural since the  Skyrmion contains a one
pion cloud in the tail region.
\item
[ii.]
 At $R=0$ and $ A=B=1$,
the configuration $U_{12}$ takes the form of hedgehog with $B=2$
and its energy is about $1$ GeV higher than the two Skyrmion
threshold. This excess of energy at $R=0$ suggests a short range
repulsion between two nucleons.
\end{description}

More quantitative analyses using the product ansatz
 have been performed by several
authors and the precise form of the potential $V(\vec R,C)  $ was
worked out. The NN interaction is then obtained by the
semiclassical method from $V(\vec R,C) $. Each Skyrmion is
rotated by the collective coordinate $A$ or $B$, and the relevant
spin - isospin component is extracted. One finds that the NN
potential consists of three terms, given by \be V_{NN}(\vec
R)=V_0(R)+(\tauone\tautwo)
(\sigmavo\sigmavt)V_{\sigma}(R)+(\tauone\tautwo)S_{12}V_{T}(R)
\lab{okav22} \ee where $S_{12}$ is the standard tensor operator
for nucleons: \be S_{12}=3(\sigmavo\hat R)(\sigmavt\hat
R)-(\sigmavo\sigmavt) \ee The last two terms in \re{okav22} tend
asymptotically to the OPEP, as expected, while the first term has
a shorter range. Unfortunately, it is found that the spin -
isospin - independent part of the central force $V_0  $ is
strongly repulsive at medium distances
 so that $V_{NN} $ in Eq. \re{okav22}
does not account for nuclear binding.
%%%%%%%%%%%%%%%%%%%%%%%%%%%%%%%%%%%%%%%%%%%%%%%%%%%%%%%
%%%%%%%%%       ERICson wiese

\section{Scaling behavior of the Skyrme model}
\indent

The second   shortcoming of the Skyrme model was revealed by
Shehter \ci{shehter}. He proved that the chiral effective
Lagrangian such as the Skyrme model have different scaling
behavior than QCD.
%%%%%%  WAINZDOCH******************
Classically the QCD Lagrangian (with massless quarks) is invariant
under scale transformations $x_{\mu} \ra \expon{-\lambda}x_{\mu}
$
 with ${\lambda}$ real. The corresponding
conserved current is the    dilaton current,
 $S_{\mu}=T_{\mu \nu}x^{\nu} $ where $T_{\mu \nu } $ is the
energy-momentum tensor. At the quantum level,
 however $S_{\mu}$ has nonvanishing
divergence \ci{collins} \be \ba \dmuup S_\mu=(\dsf {\beta(g)}{2g})
F^{a}_{\mu \nu }F^{\mu \nu }_{a}, \lab{tranom} \ea \ee where $g$
is the coupling constant, ${\beta}(g)$ the QCD beta function and
$F^{a}_{\mu \nu } $ the gluonic field strength tensor. Since
$\dmuup S_\mu =T^{\mu}_{\mu} $
 the trace of $T_{\mu \nu } $ in Eq. \re{tranom} is called
 the "trace anomaly".
To restore the anomalous scaling behavior of QCD at the effective
Lagrangian level one introduces a scalar field ${\chi}(x)$
\ci{shehter} such that $\dmuup S_\mu =-b\chi^4$. Fluctuations in
$\chi $ around $\chi_0 $ may be interpreted as  a scalar glueball
or a quarkonium.

In the next chapter we shall take into account scalar dilaton
field to restore the anomalous scaling behavior and consider NN
interaction. Meanwhile, to make this point clearer and for further
 references we briefly outline
the meaning of "scale dimension" below. Let's consider the
kinetic terms of a fermion field (e.g. quark field in QCD) \be
\lkin^{\psi}(x)=i\bar\psi (x)\dmuup \gamma_\mu \psi (x) \ee
 and a boson (pion) field
\be \lkin^\pi(r)=\half \dmud\pi \dmuup\pi . \ee
 The scale invariance demands that the action
 $S=\int d^4 x \cal L $ should be invariant
under scale transformations $x_{\mu}=\expon{-\lambda}\xpr_\mu  : $
 \be
S=\int d^4 x \lkin(x)=\int d^4 \xpr\lkin^\prime(\xpr) \lab{sef}
\ee \lab{wain5} Let the fermion and  boson fields scale as
 $\psi(x)=\exp{(n\lambda)}       \psi^\prime(\xpr) $  and
 $\pi(x)=\exp{(m\lambda)}       \pi^{\prime}(\xpr) $ respectively.
For fermions we get \be \ba
 S=i\ds\int d^4 x \bar\psi (x)\dmuup \gamma_\mu \psi (x)=
   i\ds\int d^4 \xpr       \bar\psi^{\prime}
 (\xpr)\dmuup \gamma_\mu\psi^\prime (\xpr)
\expon{\lambda}\expon{2n\lambda}\expon{-4\lambda}= \nwl =\ds\int
d^4 \xpr \lkin^\prime(\xpr) \ea \ee and hence
$-4\lambda+\lambda+2n\lambda=\lambda(2n-3)=0 $.
 Since $\lambda$ is arbitrary then $n=3/2$.
 Similarly, for boson fields one can show that $m=1$.
 Therefore we may  conclude
that \be x_{\mu}\ra\expon{-\lambda}x_\mu, \quad
\psi(x)\ra\expon{3\lambda/2} \psi(x) ,\quad
 \pi(x)\ra\expon{\lambda} \pi(x).
\ee i.e. the scale dimension $d$ equals to $d_f=3/2 $ and $d_b
=1$ for fermions and bosons respectively. Obviously, the scale
dimension of chiral nonlinear field
 $U(x)$ is zero. For this reason the four derivative term
of the Skyrme Lagrangian $\lfora $ is scale invariant by itself:
\be \int d^4 x\lfora(x)\ra\int d^4\xpr\expon{-4\lambda}
\lfora^\prime (\xpr) \expon{4\lambda}= \int d^4 \xpr
\lfora^\prime  (\xpr) \ee whereas the kinetic term \be
\ltwo(x)=\dsf{\fpi^2}{4}\Tr [\dmud U(x)\dmuup U^{+}(x)] \ee is
not: \be \ba \ds\int d^4 x\ltwo(x)\ra \dsf{\fpi^2}{4} \ds\int
d^4\xpr\expon{-4\lambda}\expon{2\lambda} \Tr [\dmud
U^\prime (\xpr)\dmuup U^{'+}(\xpr)]= \nwl
=\expon{-2\lambda}\ds\int d^4 \xpr      \ltwo^\prime (\xpr) \ea
\ee How can the invariance  be restored? According to Shehter
\ci{shehter} an additional boson field $\chi(x) $ should be
included into the Skyrme Lagrangian. As a result the minimal
version of scale invariant Skyrme Lagrangian will be  given by \be
\ba {\cl} (U, \chi)=\dsf{1}{2}\dmud\chi \dmuup\chi +
\dsf{\fpi^2}{4} (\dsf{\chi}{\chi_0})^2\Tr
 [\dmud U\dmuup U^{+}]+
\nwl + \dsf{1}{32 e^2}\Tr[(\dmud U)U^{+},(\dnud U)U^{+}]^2
%\frac{\fpi^2\mpi^2}{2} (\dsf{\chi}{\chi_0})^3\Tr [U-1]
\ea \ee where $\chi_0 $ is the vacuum expectation value of a
dilaton
 field $\chi(x) $.

But unfortunately this is not the whole story. The basic QCD
Lagrangian whose main features should be reflected in an
effective one has  the "trace anomaly" mentioned above. To
restore the anomalous scaling behavior of QCD at the effective
Lagrangian level one has to introduce an effective potential
 $V(\chi) $, describing
the self interaction of dilaton fields. It was shown that, the
effective potential could be uniquely determined by the vacuum
energy density $b\chi_{0}^{4}/4 $
 and the condition that the minimum should occur at
$\chi=\chi_0  $ which yield \ci{collins} \be \ba
V(\chi)=\dsf{b\chi_0 }{16}\{1+ (\dsf{\chi}{\chi_0})^4
[4\ln(\chi/\chi_0)-1]  \}. \ea \ee
 The parameters $b $ and $\chi_0 $ are obtained from
the value of the gluon condensate $\cg $ \be \ba \ofo{\dmuup
S_\mu} =\ofo {  (\dsf {\beta(g)} {2g}) F^{a}_{\mu \nu }F^{\mu \nu
}_{a} } {\equiv}\cg=b\chi_{0}^{4} \ea \ee
 and the dilaton mass
\be m_{\chi}^{2}=\dsf{\dd^2
V}{\dd\chi^2}|_{\chi=\chi_0}=4b\chi_{0}^{2} \ee
 which together with $\fpi $ and $e$ specify the parameters
of the Lagrangian. It has been shown that, the suitable choice of
parameters $\chi_0\approx 120MeV $, $\cg\approx (300MeV)^4 $,
 $e\approx 4.5$ and $\fpi=93 MeV$
leads to a large dilaton mass $m_\chi\approx1500MeV $,
 which may be identified
with a glueball. Although this scalar field has nothing to do with
a scalar $\sigma$ meson $(\msig\approx 600 MeV)$, its inclusion
into the Skyrme Lagrangian proved to solve a
 long standing problem of missing
of intermediate range attraction in the central $NN$ interaction
\ci{yabu,waindzoch}.

Nevertheless, it would be quite desirable to include into the
Skyrme Lagrangian a dilaton field with lower mass e.g.
$m_\chi=m_\sigma\approx 600 MeV $. This goal has been achieved by
using the method of bosonization      in ref.s \ci{nik,anov}. As
a result,   a modified Lagrangian which differs from Shehters one
\ci{shehter}  mainly due to the self interaction potential: \be
V(\chi)=\dsf{b\chi_0}{24}\{1-(\dsf{\chi}{\chi_0})^4-
\dsf{4}{\eps}[1- { (\dsf{\chi} {\chi_0})} ^\eps]\} \lab{vsig} \ee
has been  proposed. We shall use this model to study $NN$
interaction   in the next chapter.

\section{PCAC and Goldberger - Treiman relations}
\indent

The chiral invariance of Skyrme model leads to  well known
relations. Here we show how these relations can be derived in a
rather model independent way. Considering pion - nucleon system
we start with the interaction Hamiltonian: \be H_{int}=-ig_{\pi
NN}\bar\psi(x) \gamma_5\vec\tau\psi(x)\vec\varphi_{\pi}(x)
\lab{hint} \ee where $g_{\pi NN}$ is the pion - nucleon coupling
constant, $\psi(x)$ and $\vec\varphi_{\pi}(x)$ are nucleon and
pion fields respectively. The latter satisfies Klein - Gordon
equation: \be
(\partial_\mu\partial^\mu+\mpiq)\vec\varphi_{\pi}(x)= -ig_{\pi
NN}\bar\psi(x) \gamma_5\vec\tau\psi(x) \lab{kg} \ee Axial current
of this system is the sum of the nucleonic, $\vec {A}_{\mu}^{N}$,
\be \vec {A}_{\mu}^{N}=g_A \bar\psi(x)\gamma_\mu
\gamma_5\vec\tau\psi(x)/2 \lab{axnuc} \ee and pionic, $\vec
{A}_{\mu}^{\pi}$, \be \vec {A}_{\mu}^{\pi}=\fpi\partial_\mu
\vec\varphi_{\pi}(x) \lab{axpi} \ee
 axial currents:
 \be
 \vec {A}_{\mu}=\vec {A}_{\mu}^{N}+\vec {A}_{\mu}^{\pi}
 \lab{axtot}
 \ee
 Using Dirac  $(i\gamma_\mu \partial^\mu-M    )\psi(x)=0$ and
  Klein - Gordon equations one may obtain
  \be
  \ba
   \partial^\mu\vec {A}_{\mu}^{N}=
   M g_A \bar\psi(x)i \gamma_5\vec\tau\psi(x),
   \nwl
   \partial^\mu\vec {A}_{\mu}^{\pi}=- \fpi\mpiq \vec\varphi_{\pi}(x)
\lab{pcc} \ea \ee Note that, the last second equation in \re{pcc}
is called as PCAC (partial conservation of axial current)
relation. Now, the requirement of conservation of the total axial
current $ \partial^\mu\vec {A}_{\mu}=0$ gives following equation
for the divergent of $\vec {A}_{\mu}$: \be
(\partial_\mu\partial^\mu+\mpiq)\vec\varphi_{\pi}(x)=
-g_{A}M\bar\psi(x)i \gamma_5\vec\tau\psi(x) \lab{div} \ee
Finally, comparing right hand sides of the Eq.s \re{kg} and
\re{div} one easily gets Goldberger - Treiman relation: \be
Mg_A=\fpi g_{\pi NN} \lab{gt} \ee This utmost important relation
between axial coupling constant $g_A$ and pion - nucleon coupling
constant $g_{\pi NN}$ is in good agreement with the experiment:
$g_{\pi NN}=13.4\pm0.1$, $Mg_A/\fpi=12.78\pm0.1$. The small
deviation reflects the accuracy of the relation. We shall come
back to this point considering in - medium modification of this
relation in Chapter 5.
%
%
 %%%%%%%%%%   END OF CHAP 2 *************
%
%%%%%%%%%%%%%%%%%%%
%\include{ch2sgp}
%%%%%%  ReMOVE  these 2 lines if you want your fig.s (tabs) to be numbered
%%as chapter.number like V.3
\setcounter{fignum}{\value{figure}}
\setcounter{tabnum}{\value{table}}
%%%%%%%%%%%%%%%%%%%
\chapter[Scalar dilaton - quarkonium meson in nucleon - nucleon
           interaction.]{}
\setcounter{figure}{\value{fignum}}
\setcounter{table}{\value{tabnum}} \bc { \Large\bf Scalar Dilaton
- Quarkonium Meson in Nucleon - Nucleon
           Interaction.}\footnote
{The  present chapter is based on following articles by the
author and his collaborators: \ci{ouryafnn,ouruz,oursingap}} \ec
\section{Introduction}
\indent

Skyrme - like models allow for the study of the properties of
individual baryons and offer a framework for investigating
 interactions between classical baryons.
The simplicity of calculations and the existence
 of a transparent connection
between NN and $\nbn{N}$ interactions within the Skyrme model are
distinct advantage over the models such as nonrelativistic quark
models or the bag models. The most of calculations based on the
product approximation  within Skyrme - like models lead to
projected potentials that are  in qualitative agreement with
empirically motivated NN potentials at short and long distances.
However many of them does not take into account the scale anomaly
of the QCD omitting scalar particles and fail to reproduce the
intermediate range attraction in the scalar channel.

At the present time there are two different approaches to include
scalar meson - dilaton into the Skyrme model. In both approaches
the interaction of dilaton field with the chiral field is
dictated by the scale invariance. The main difference is in the
origin of dilaton. On the one hand, there is the approach in
which dilaton is associated with the glueball \ci{gomm}. In this
approach the glueball field saturates the scale anomaly
completely. In the other approach the dilaton is treated as a
quarkonium arising due to fluctuations of the quark condensate
\ci{anov}, which is an order parameter for the chiral symmetry
breaking. In this approach the dilaton - quarkonium saturates
 a part of the scale anomaly, namely by the part which is produced by
the quark loops. This choice of the quarkonium as a dilaton is
based on two principal observations:
\begin {description}
\item [a)]
The experimental studies of scalar resonances \ci{deikman} show
that the real lightest candidate for the glueball state is
$f_0(1590)$ which does not appear in $\pi\pi$ and $K\bar{K}$
productions.
\item[b)]
Consideration of the chiral anomaly \ci{anov} shows that the only
guage - invariant combination of the gluon field
$G_{\mu\nu}G_{\lambda\rho} $ can contact only with a total
antisymmetric tensor $\varepsilon^{\mu\nu\lambda\rho} $. Thus
this combination has $J^{PC}=0^{-+} $ quantum number and
contributes to Wess - Zumino - Witten action producing $U(1)$
anomaly. It suggests that the $0^{++}$ glueball field as a
fluctuation of $ G^{2}_{\mu\nu} $ cannot interact directly with
chiral field, but only through the mixing with the dilaton --
quarkonium. The estimate gives small values for the mixing angle
of glueball and quarkonium states.
\end{description}

In the present chapter we calculate the central $NN$ potential
within a Skyrme model suggested by Andrianov and Novozhilov in
ref. \ci{anov}. The model is based on effective low energy action
for pseudoscalar and scalar (dilaton - quarkonium) meson which
has been derived starting directly from the QCD generating
functional by the joint chiral and conformal bozonization method.

\section{The Lagrangian}
\indent

 The Lagrangian suggested by Andrianov and Novozhilov \ci{anov} is
\be \ba
{\cl}_{AN}(U,\sigma)=\ltwo+\lfora+ W_{\sigma} , \\
\\
{\cl}_{2}=-{\disc\frac{\fpi^{2}}{4}}[{\Tr}L_{\mu}L^{\mu}-2{(\dd_{\mu}
{\sigma})}^{2}]{e}^{-2{\disc\sigma}} , \\
\\
{\lfora}={\disc\frac{1}{32{e}^{2}}}{\Tr}{[L_{\mu} , L^{\nu}]}^{2} , \\
\\
W_{\sigma}=-{\disc\frac{\cg}{24}}[\expon{-4\sigma}-1
+{\disc\frac{4}{\varepsilon}} (1-\expon{-{\varepsilon}\sigma})]   \\
\ea \lab{s1} \ee where $f_{\pi}=93MeV $ is  pion decay constant,
$ e-$ dimensionless parameter,
 $\cg$ - gluon
%\edc
condensate parameter : $\cg={((300-400)MeV)}^{4}$,
  $\sigma$ - scalar meson field ,
$L_{\mu}=U^{+}\dd_{\mu}U  $  ,   $U$- is an $SU(2)$ matrix chiral
field. It is a generalization of the well - known original Skyrme
model \ci{anw,skyrme} (OSM) and takes into account the conformal
anomaly of the QCD. The term $  \ltwo$ includes the kinetic term
of the chiral field and scalar fields. The $ \lfora$ is  a well
known Skyrme term. The effective potential for the scalar field
is a result of the extrapolation of the low energy potential into
the high energy region in one - loop approximation to the Gell -
Mann - Low QCD $\beta$ - function. The parameter $\varepsilon$
depends on the number of flavors $N_f$:
 $ \varepsilon=8N_{f}/(33-2N_f)$.
Note that in the limit of heavy $\sigma$-- meson the potential
becomes equal to the symmetric quartic term
  $\:\:\:{\lfors}={\disc\frac{\gamma}{8e^{2}}}
{[{\Tr}L_{\mu}L^{\mu}]}^{2}\:$, which is necessary to reproduce
the $\pi\pi$ scattering data \ci{donoghue}. We   extend the
Lagrangian \re{s1} by adding chiral symmetry breaking term: \be
{\lhisb}={\disc\frac{{e}^{-3{\disc\sigma}}{\mpi}^{2}{\fpi}^{2}}{2}}\!\Tr(U-1)\\
\lab{s2} \ee
 to provide the correct
asymptotic behavior of the soliton field. Thus our model is given
by: \be \cl=\cl_{\mbox{AN}}+\lhisb \lab{s3} \ee In the static
approximation when some parts of the Lagrangian have simple form
\be \ba
{\ltwo}={\disc\frac{\fpi^{2}}{4}}[{\rho}^{2}{\Tr}{\disc{L}^{2}_{i}
-
2{\rhoprim}^{2}]}, \\
\\
{\lfora}={\disc\frac{1}{32{e}^{2}}}{\Tr}{\disc{[L_{i}, L_{j}]}^{2}}, \\
\ea \lab{s4} \ee where
 $\:\rho\equiv\exp {(-{\sigma}(r))}\:\:$
we may  use the most popular hedgehog ansatz: $U=U_0=\exp{
(i\vec{\tau}\vec{r}\theta(r))} $,
  $ \hat{r}=\vec{r}/\mod{r}$
and propose  spherical symmetry for the scalar field
$\sigma(r)$.  Then the mass functional for classical soliton
is     given by: \be \ba \mh=
{\disc\frac{8{\pi}{\fpi}}{e}}{\ds\integ}(\tilde{M}_{2}+
\tilde{M}_{\mbox{4a}}+ \tilde{M}_{\mbox{w}}+\tilde{M}_{\disc\pi}),
\nwl
\tilde{M}_{2}={\rho}^{2}(\thprim^{2}x^{2}/2+s^{2})/4+\rhoprim^{2}x^{2}/8,
\nwl \tilde{M}_{\mbox{4a}}=s^{2}(d/2+\thprim^{2}), \nwl
\tilde{M}_{\mbox{w}}={\deff}\;x^{2}[{\rho}^{4}-1+{\disc\frac{4}{\varepsilon}}
(1-{\rho}^{\disc\varepsilon})] \nwl
\tilde{M}_{\disc\pi}={\rho}^{3}(1-c)x^{2}{\beta}^{2}/4 \ea
\lab{s5} \ee
%\edc
where $x=2e{\fpi}r,   c\equiv\cos(\theta), $
$s\equiv\sin(\theta),   $ $d={(s/x)}^{2},   $
${\beta}={\mpi}/(2{\fpi}e),   $
$\deff={\cg}/384{e}^{2}{\fpi}^{4}.   $ The Euler -- Lagrange
equations  for the chiral angle $\theta (x)$  and scalar meson
shape function $\rho(x)$ follow to be \be \ba
{\thprimprim}[{\disc\frac{1}{4}}x^{2}{\rho}^{2}+2s^{2}]+
{\disc\frac{1}{4}}[2{x}^{2}\thprim\rho\rhoprim+
2x\thprim{\rho}^{2}-{\rho}^{2}s_{2}]+ \nwl
+s_{2}(\thprim^{2}-d)-{\rho}^{3}x^{2}{\beta}s/4=0 , \ea \lab{s6}
\ee \be \ba
x^{2}{\rhoprimprim}/4-{\rho}{x}^{2}({\thprim}^{2}/2+d)/2+
x\rhoprim/2-4{\deff}{x}^{2}({\rho}^{3}- \nwl
-{\rho}^{\disc{\varepsilon}-1})
-3{\rho}^{2}(1-c)x^{2}{\beta}^{2}/4=0\;, \ea \lab{s7} \ee where
$s_{2}{\equiv}\sin(2\theta)  $ and a prime corresponds to the
derivative with respect to $x$.  For the asymptotic behavior of
$\theta(x)$ at      large distances when $
{\rho}\:{\rightarrow}\: 1 $ we get from \re{s6} and \re{s7}
 a familiar formula: \ci{bhaduri}
\be \theta(x)\sim
{\disc\frac{3{\gpinn}{\fpi}e^{2}}{2{\pi}\mn{x^2}}}
(1+{\beta}x)\exp{(-{\beta}x)}{\equiv}A\frac{(1+{\beta}x)}{x^2}\exp{(-{\beta}x)}
\lab{s8} \ee where $\mn$ is the nucleon mass, $\gpinn=13.5$ --
the pion-nucleon coupling constant.  This formula     gives
connection between $\gpinn$ and Skyrme parameter $e$ and reveals
that $ e$ is not a free parameter of Skyrme model in some sense.
The behavior  of $\rho(x)$ at large $x$ exhibits a rapid decrease
from the unity: \be {\rho}(x)\sim 1-\gamma
\frac{\expon{-\tilde{D}x}}{x} \:,\:\:
\tilde{D}=\sqrt{16{\deff}(4-{\varepsilon})} \lab{s9} \ee
 At
small distances the behavior of $\theta(x)$ and $\rho(x)$ is
given by: \be \ba
\theta(x){\sim}{\pi}-{\alpha}x\\
{\rho}\sim{\rhozero}-a{x}^{2} \ea \lab{s10} \ee with
$a=[3{\alpha}^{2}{\rhozero}+16{\deff}({\rhozero}^{3}-
{\rhozero}^{{\disc{\varepsilon}-1}})
+6{{\rhozero}}^{2}{\beta}^{2}]/6$.

Thus  Eq.s \re{s6} and  \re{s7} may be solved numerically using
well known methods  \ci{jrho,llv89} and the parameters $ A $, $
{\alpha}$,  $\gamma$ and $ {\rhozero} $ in
 Eqs. \re{s6}, \re{s10} may be
adjusted to yield continuity in $\theta, \rho,  \thprim,
\rhoprim $  at a merging point. In addition,  the consequence of
the  virial theorem: \be
\integ[\tilde{M}_{2}-\tilde{M}_{4a}+3(\tilde{M}_{\mbox{w}}+\tilde{M}_{\disc\pi})]=0
\lab{s11} \ee may be used to control the accuracy of stable
solutions of Eq.s \re{s6}, \re{s7}. The resulting solution,
$\theta(r)$, is
 similar to that one in the original Skyrme model
(OSM)  \ci{anw} while
 $\rho(r)$ rapidly goes to the unity starting from  $\rhozero$ at $r=0$ .

Masses of nucleon $M_{N}$  and $\Delta$  - isobar - $M_{\Delta}$
may be calculated from the following expressions: \be \ba
\mn=\mh+3/8{\lamm}\\
\md=\mh+15/8{\lamm} \ea \lab{s12} \ee
 where $\mh$
- soliton mass (5), $\lamm$ is momentum inertia of
 the rotating skyrmion:
\be
\lamm={\disc\frac{4{\pi}}{3{e}^{3}{\fpi}}}{\integ}x^{2}s^{2}[{\rho}^{2}/4+
\thprim^{2}+d ] \lab{s13} \ee

\section{The nucleon -- nucleon interaction}
\indent

 We investigate the skyrmion-skyrmion interaction
  by using the product approximation:
\be \ba U (\vecx;\vecrone , \aone;\vecrtwo , \atwo)=
\aone{\uzero}(\vecx-\vecrone)\apm{\uzero}(\vecx-\vecrtwo){\atwo^{+}}\equiv
\uone\utwo \\
{\rho}(\vecx , \vecrone ,
\vecrtwo)=\rho(\vecx-\vecrone)\rho(\vecx-\vecrtwo)
\equiv\rhoone\rhotwo \\
\ea \lab{s14} \ee where $\uzero(\vecx-\vecri)$ for $i=1, 2$ is
the hedgehog solution $( U_{0}(\vecr)=\exp{
(i\vec{\tau}\vec{r}\theta(r))}, $
  $ \hat{r}=\vec{r}/\mod{r})$
 located at $\vecri$,  and $\disc{A_i}$
 the collective coordinate to describe the rotation.
The skyrmion-skyrmion potential is defined by: \be
V(\rv)=-{\int}d\vecx[{\cl}(\uone\utwo ,
\rhoone\rhotwo)-{\cl}(\uone , \rhoone) -{\cl}(\utwo , \rhotwo)]
\lab{s15} \ee where $\vecr $ is the relative coordinate between
two skyrmions $(\vecr=\vecrone-\vecrtwo)$.
      The product approximation has reasonable properties:
\begin{description}
\item[a)]
 the baryon number is additive,
\item[b)]
it gives a correct behavior at the asymptotic region,  i. e., the
one-pion exchange potential,
\item[c)]
at $\vecr=0 $ the product form is very close to the hedgehog
solution of the baryon number $B=2$.
\end{description}

    The static NN potential may be obtained by using a standard technique
\ci{otofnn,riskann,depacenn,yabuandonn} which gives the following
expressions for the scalar-isoscalar part of central NN
interaction: \be \ba
\vcr={\disc\frac{8\pi\fpi}{e}}\intdt\integ{x}^{2}[\vtwo+\vfora+
     \vw+\\
     \ +\vhisb+1\:\rightarrow{\:}2]
\ea \lab{s16} \ee Here the contributions from $\ltwo$ , $\lfora$ ,
$\lhisb$ and  $W_{\sigma}$ are denoted by $\vvtwo$,  $\vvfora$,
$\vvhisb$ and $\vvw$ respectively:
%\normalsize
\be \ba \vtwo=\{\rhox\phi(x)\rhoone\phi_{1}
\zone+(\rhoone^{2}-1)\times \nwl
\times\lbr\rhoxdar{2}(\tilde{\theta}
(x)+3d(x))+\phi^{2}(x)\rbr\}/8 \lab{s17} \nwl
\vw=\deff\{[\rhoxdar{4}(\rhoone^{4}-2)+1]-4[\rhoxdar{\eps}
(\rhoone^{\eps}-2)+1]/\eps\}/2 \lab{s18} \nwl
\vfora=\{d_{1}[3d(x)+2\tilde{\theta}(x)]+\tilde
{\theta}(x){\tetone}(1-\zone^{2})/2\}/3 \lab{s19} \nwl
\vhisb=-\rhoxdar{3}d^{2}(x)\{\rhoone^{3}[c(x)c_{1}-1]/2-c(x)+1\}/4
\lab{s1720}
%\eanum
\ea \ee
%\large
where the following notations are introduced:
$\:\tilde{\theta}(x)=\thprim^{2}(x)-d(x)\: $ ,
  $\:\phi(x)=\rhoprim(x)\;$,
$\:{(\xonetwo)}^{2}=x^2+\tilde{r}^{2}{\pm}2xt\tilde{r}\;$,
$\:\zonetwo=(x\pm\tilde{r}t)/\xonetwo\:$ ,
$\:\tilde{r}=2re\fpi\:$, $r$ is a relative
 distance between the two nucleons, $\rho_i\equiv\rho(x_i)$,
$d_i{\equiv}d(x_i) $, $(i=1,2)$ e.t.c.

\section{Results and discussions.}
\indent

\begin{figure}[!hbt]
%\vspace{0.5cm}
\bc
\epsfysize=14cm \epsfbox{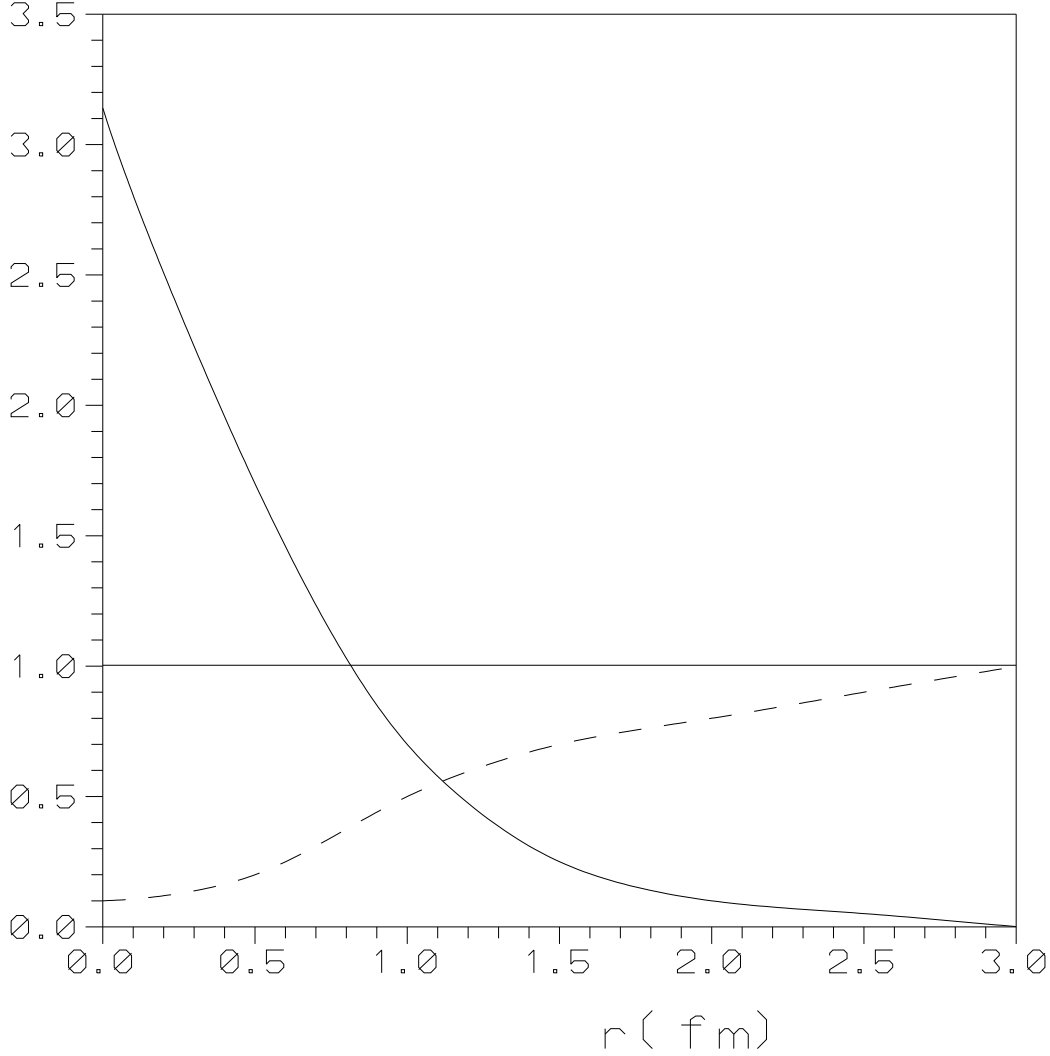}
\ec 
%\centerline{\epsffile{fsg1.eps}}
%\vspace{0.5cm}
\centerline{\parbox{15cm}{\caption{\label{figs1} The radial
dependence of chiral angle $\theta(r)$ - solid line and shape
function $\rho(r)=\exp(-\sigma(r))$ - the dotted lines for the
soliton corresponding to the set of parameters $\fpi=93 MeV$,
$e=4.69$ , $\cg={(122.3 MeV)}^{4}$.}}}
\end{figure}

 As  a starting point we make an attempt to
 describe the pion -- nucleon coupling
constant $\gpinn$ and $N - \Delta$ mass splitting:
$M_{N\Delta}=M_{\Delta} - M_{N}$
  optimizing the only free
parameter of the model - $\cg$. In fact,  the value of $\fpi$
should be fixed to its experimental value,  and the Skyrme
parameter $e$ must be defined by
 Eqs. \re{s6} - \re{s8}
 with $\gpinn={(\gpinn)}^{exp}=13.5$
during the numerical solving procedure of corresponding equations
of motion. The typical solutions
 could be seen from Fig.\ref{figs1}
   where solid and dotted curves display the profile
 function $\theta(r)$ and the shape function
  $\rho(r)\equiv\exp{(-\sigma(r))}$
 respectively.
%\vspace{1.5cm}
%%%%%%%%%%%%%%%%%%%   FIGURE%%%%%%%%%%%%%%
%\vspace{1.5cm}
 It is well known
 \ci{anov} that the Lagrangian \re{s1} with $\cg={(300 MeV)}^4 $
 gives a large value for   the   soliton mass.
 Our precise analysis \ci{ouruz} show that
the gluon condensate
 should be reduced up to the $\cg={(122. 3MeV)}^4$ to
 reproduce
 $\gpinn$ and the masses ($M_\Delta -M_N=293 MeV, M_N=954.5 MeV$) \ci{ouruz}.

     In this case the calculated value of
 axial coupling constant $\ga$, which is  given by the following expression:
 \be
 \ga=-\disc\frac{\pi}{3{e}^{2}}
{\int_{0}^{\infty}}dx\:x^{2}\{\rho^2(\thprim+
 \stwo/x)+
 4[\stwo({\thprim^2}+d)/x+2{\thprim}d]\}
\lab{s21} \ee
 is equal to $1.01$  i.e.  much  better then that in OSM \ci{anw}
$ (\ga^{OSM }=0. 61 ) $. It is clear from Eq. \re{s21} that $\ga$
is very sensitive to the Skyrme parameter $ e $  which is
connected to $\gpinn$ through the equations \re{s6} - \re{s8},
since Goldberger Treiman relation does not work due to $\lhisb$
term.

\begin{figure}[hbt]
%   \vspace{0.5cm}
\bc \epsfysize=14cm \epsfbox{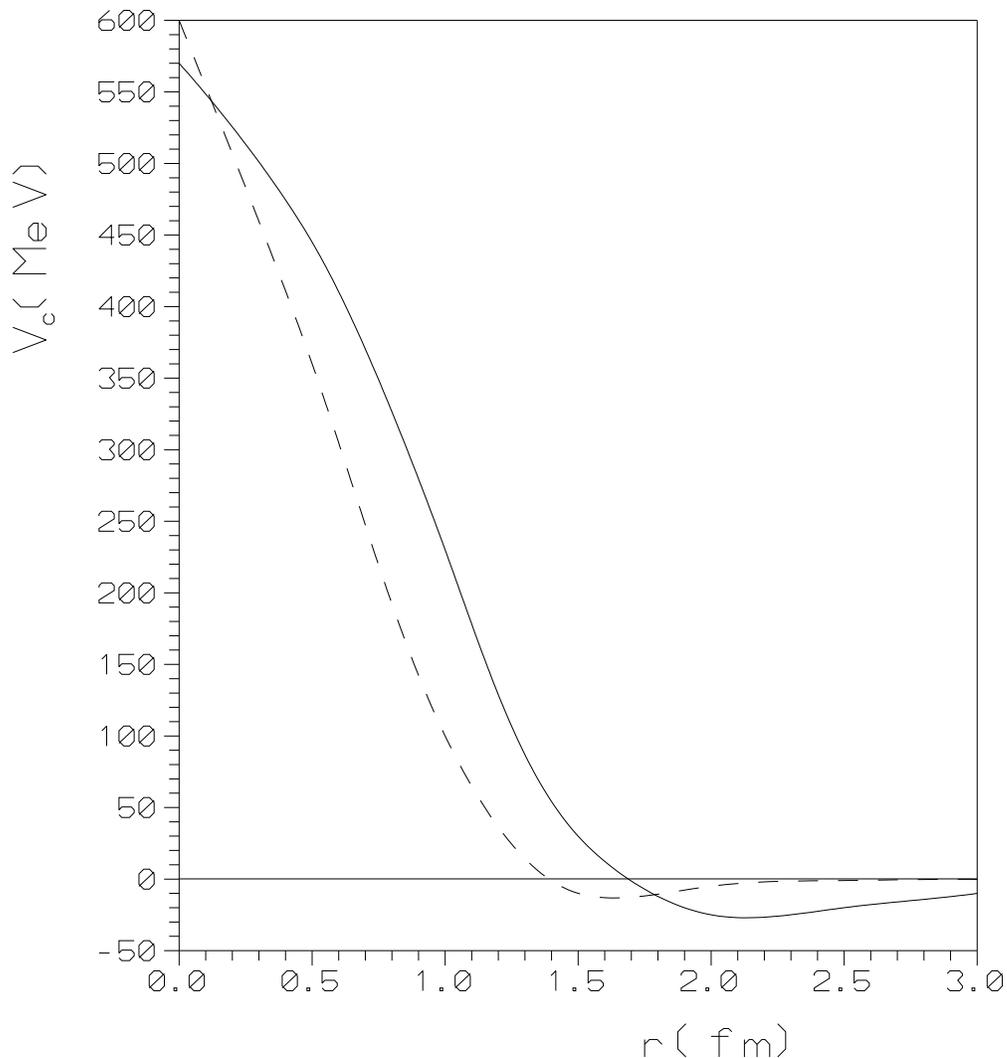} \ec
%\epsfysize=9cm
%   \centerline{\epsffile{fsg2.eps}}
%   \vspace{1.0cm}
   \centerline{\parbox{15cm}{\caption{\label{figs2}
Central isospin-independent potential as a function of
internucleon distance $r$
 calculated from Eqs. \re{s16},\re{s1720} .
The dashed curve is the corresponding interaction component in
the "Paris" potential \ci{paris}.}}}
\end{figure}

 Thus by using the following optimal set of parameters: $\fpi=93 MeV$ ,
 $e=4.69$ , $\cg={(122.3MeV)}^4$ in Eqs. \re{s16},\re{s1720} we calculated
the scalar
 - isoscalar part of central nucleon - nucleon potential -- the solid
 line in Fig.  \ref{figs2}.
%\vspace{1.5cm}
%%%%%%%%%%%%%%%%%%%   FIGURE%%%%%%%%%%%%%%
%\addtocounter{figure}{1}
%\vspace{1.0cm}
  It may be compared with corresponding interaction component
 of the realistic phenomenological  "Paris" potential
  \ci{paris}, displayed there with the dashed  line.
   It is seen from the figure that the attraction is shifted
 to larger separations and much stronger.
 The similar result was obtained in ref.  \ci{meisnn}
 within the framework of realistic pseudoscalar
 - vector chiral Lagrangian where the $\sigma$ - meson exchange
 was  simulated by
 a correlated two-pion exchange between the two solitons.

\begin{figure}[!t]
   \epsfysize=9cm
   \centerline{\epsffile{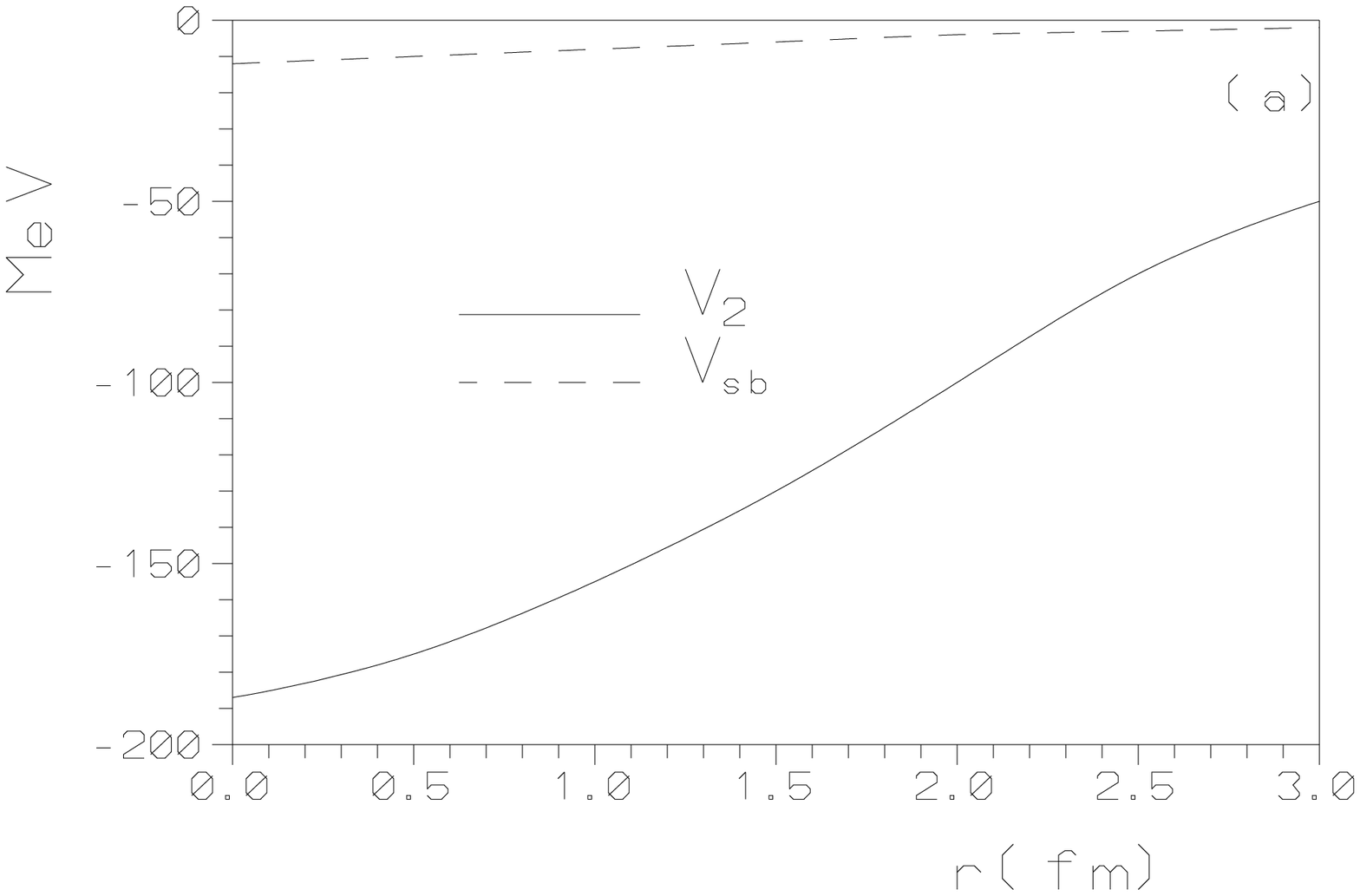}}
   \epsfysize=9cm
   \centerline{\epsffile{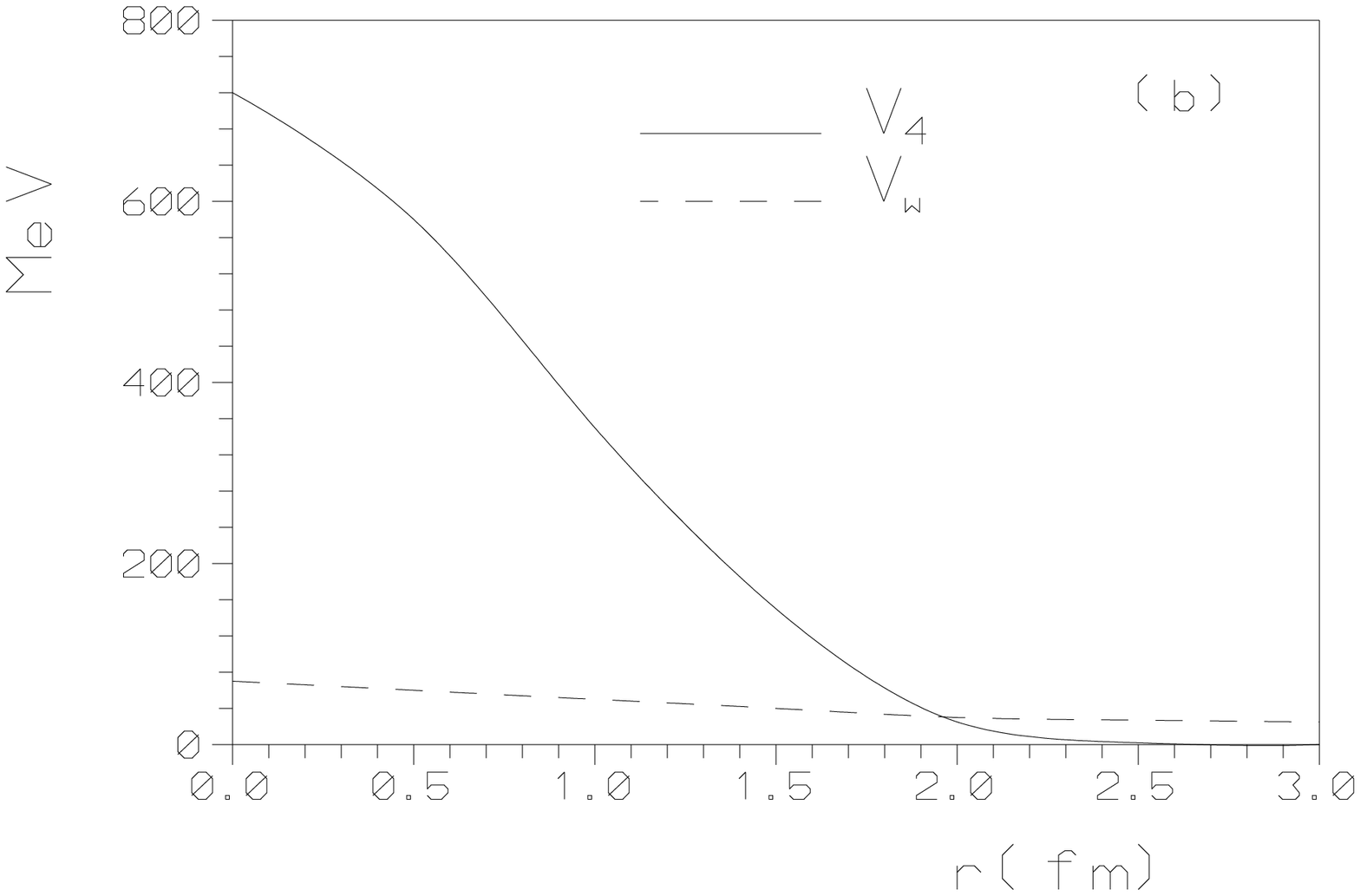}}
\vspace{0.8cm}
 \centerline{\parbox{15cm}{\caption{\label{figs3}
The net contributions to the central NN potential given by the
Eqs. \re{s16}, \re{s1720}. The $V_{2}$,  $V_{sb}$, $V_4$, and
$V_{W}$ curves are the contributions from $\ltwo$,  $\lhisb$,
$\lfora$ and $W_{\sigma}$
 terms respectively.
}}}
\end{figure}

 To discuss this effect in a more
detail we bring in figures Fig.\ref{figs3} (a) and
Fig.\ref{figs3}(b)
 the net contributions from each
term.
%%%%%%%%%%%%%%%%%%%   FIGURE%%%%%%%%%%%%%%
 It is well known that in the OSM the second order derivative term
$\ltwo$ does not contribute to the central potential at all [$
\rho(x)=1 $ in Eqs. \re{s16}, \re{s1720}], whereas the
contribution from $ \lfora$ is strongly repulsive (see
Fig.\ref{figs3}(b)).
 After the inclusion of
scalar dilaton this repulsion may be compensated by an attractive
contribution arising from the kinetic term $\ltwo$ [see
Fig.\ref{figs3}(a)] in the region $ r\ge1.5 fm$.
 As to the contributions
from $\lhisb$ and $W_\sigma$ terms they are not large, especially,
 at short
and medium distances [see  Fig.\ref{figs3}(a) and
Fig.\ref{figs3}(b)].

Note that the relevance of product approximation in the short
range region of the interaction may be doubtful. However, the
considering attraction appears at separations greater than $ 1.2
fm$ where the approximation is valid.

We conclude that  the model we have presented here leads to a
correct values  of $\fpi$, $\gpinn $, $M_{N\Delta}$ and gives a
desired attraction in the central part of $NN$ interactions. On
the other side,  by comparing this result with a result of Yabu
et al. \ci{yabu} where               no attempt had been made
 to optimize
the parameters we may affirm that the attraction
 appears in a natural way in both cases regardless  the
 origin of dilaton.
One may expect that the inclusion of $\omega-$ meson coupling term
to the Lagrangian \ci {rs} and taking into account of deformation
effects  \ci{okadef,otofdef,meissdef} will give a better
description of $NN$ and $\nbn{N}$ interactions.That will be the
subject of the next chapter.
%%%%%%%    END OF SINGAP %%%%%%%%%%%%%%%%%%%%%%%%

%%%%%%%%%%%%%%%%%%%%%%%%%%%%%%%%
%\include{ch3dfm}
%%%%%%  ReMOVE  these 2 lines if you want your fig.s (tabs) to be numbered
%%as chapter.number like V.3
\setcounter{fignum}{\value{figure}}
\setcounter{tabnum}{\value{table}}
\chapter[
The deformation  of the interacting nucleon in the Skyrme model]{}
\setcounter{figure}{\value{fignum}}
\setcounter{table}{\value{tabnum}}
\bc {\Large\bf The Deformation  of the Interacting Nucleon in the
Skyrme Model.} \footnote{The  present chapter is based on
following articles by the author and his collaborators:
\ci{ourplbdef,ourjafdef}} \ec

\section{Introduction}
\indent

 The possible changes in the radius of
 a nucleon in interaction
with another nucleon was investigated in the skyrmion model by
Kalbermann et. al.  \ci{kalberman}.
 It was found that there is a  swelling of the nucleon at intermediate
distances between  the nucleons which were assumed to preserve
their spherical shape. On the other hand Hajduk and Schwesinger
\ci{hs}
%\ (C. Hajduk et al 1986)
considering the skyrmion to be soft for  the deformation showed
that there are several deformed states of rotating skyrmions. In
particular,  besides the ground state of a spherical shape there
exist a degenerate doublet exotic states with the same quantum
numbers of the nucleon ($ s=t=1/2 $) of oblate and prolate
shapes. One may therefore wonder whether the nucleon can change
its shape under the action of strong interactions.

Certainly, several studies have been made of the deformation
effects on the Skyrmions \ci{okadef,otofdef,meissdef}, but these
were carried out to obtain a better understanding of the possible
sources of attraction in the central nucleon - nucleon potential.
As a result it was shown that the deformation effect is very
important and may reduce the central repulsion by about $40\% .$

In the  previous chapter  the problem of missing central
attraction in the     $NN$ - interaction has been investigated
within the model of Andrianov and Novozhilov \ci{anov} which
starts with the Skyrme Lagrangian supplimented by a dilaton
scalar field $\sigma (r)$ so as to satisfy the QCD trace anomaly
constraint. It has been  concluded that this model gives the
desired attraction in the central part of the $NN$ interaction.
The resulting central scalar - isoscalar part of the potential is
in qualitative agreement with the phenomenological one.

In the present chapter we shall concentrate on the effects of
modification of the shape of a nucleon at a
 quantitative level
using skyrmions. We shall use the same Lagrangian \re{s1}-\re{s4}
\ci{anov,migdal}which  has been used in the previous chapter.

\section{The interaction of deformed nucleons}
\indent

As it was shown in the previous chapter, the Lagrangian
${\cl}_{AN}(U,\sigma)$ is a generalization of the well - known
original Skyrme model \ci{anw,skyrme} and takes into account the
conformal anomaly of the QCD.

We have also shown that the Lagrangian   produces
 in a natural and transparent way the intermediate - range
  attraction in the
 central part of baryon - baryon interaction
 even in the product approximation  \ci{oursingap}.
 Here we let the interacting nucleons deform.
    Assume that both skyrmions deform into an ellipsoidal shape
 as they come  close together. To describe this we write the chiral field
 $U$ and the dilaton field $\sigma$ as the nonspherical hedgehog
 form given by :
 \be
 \ba
  U_{0}(\vecr)=\exp{ (i\vec{\tau}\hat{q}\theta(q))}\\
  \sigma(\vecr)=\sigma(q)
\ea \lab{def2} \ee
 where the spatial vector $\vec{q}$ has the components
 ${\deltax}x   $, ${\deltax} y$,${\deltaz} z$ and
 $\hat{q}$ is the unit vector: $\hat{q}=\vec{q}/q$
 with the deformation parameters $\deltax$ and $\deltaz$.
 The profile functions  $\theta$ and $\sigma$ are assumed to be
 the solutions of the Euler - Lagrange equations in the
 spherical case given in \re{s6}.
 The ansatz,  in Eq. \re{def2}, leads to a  modification of
 the static mass,  $\disc{M}^{*}_{H}$,
  and the moment of
 inertia,  $\disc{\lambda}^{*}_{M}$, of the Skyrmion
 \be
 \ba
 \disc{M}^{*}_{H}=[AM_{2}+BM_{4a}+M_{{\chi}sb}+M_{W}]/\eta\\
 \disc{\lambda}^{*}_{M}=[\lambda_2+A\lambda_{4a}]/\eta
 \ea
\lab{def3} \ee where $\eta=\deltax^2\deltaz$,
$A=(2\deltax^2+\deltaz^2)/3$,
$B=\deltax^2(\deltax^2+2\deltaz^2)/3$ and  $M_i$ and $\lambda_i$
denote the relevant contributions from ${\cl}_{i} $ term in
Eq.\re{s1}  for the spherical case $\deltax=\deltaz=1$. As we are
mainly interested in the region where the medium range attraction
takes place - $R\sim 1.25 fm$ ( typical separation between
nucleons in  nuclei)  we restrict ourselves to  the familiar
product ansatz: \be \ba U =
\aone{\uzero}(\vec{X}-\vec{q}/2)\apm{\uzero}(\vec{X}+\vec{q}/2)
{\atwo^{+}}\equiv
\uone\utwo \\
{\rho}=\rho(\vec{X}-\vec{q}/2)\rho(\vec{X}+\vec{q}/2)
\equiv\rhoone\rhotwo \;\; , \;\rho\equiv\exp(-\sigma)\\
\ea \lab{def4} \ee where $\aone$, $\atwo$ are the collective
coordinates of skyrmions to describe their rotational  motion,
 $\vec{q} $ is the vector along $z$ axis: $q_x=0,$
$q_y=0,$ $q_z=q=R\deltaz $ and  $R$ is the  distance between
skyrmions. The static skyrmion - skyrmion potential is defined by
\be V(\vec{R},\deltax,\deltaz)= -{\int}d\vec{X}[{\cl}(\uone\utwo
, \rhoone\rhotwo)-{\cl}(\uone , \rhoone) -{\cl}(\utwo , \rhotwo)]
\lab{def5} \ee

The application  of   the usual projection methods
 developed in \ci{okadef,otofdef,meissdef,jjp}
 %\(A. Jackson et al 1985,
%\T. Otofuji et al 1986, E. Nyman et al 1986,
%\ A. De Pace et al 1986, H. Yabu et al 1985)
to Eq.s $\re{def2} - \re{def5}$ yields the following
representation
 for the
  central scalar - isoscalar part of the nucleon - nucleon interaction :
 \be
\ba
 V_{c}(R,\deltax,\deltaz)=\disc\frac{1}{\eta}
 \lbr V_{{\chi}sb}(q)+V_{W}(q)+
 \deltax^2(V_2(q)+\deltaz^2 V_{4a}(q))+
\nwl +(\deltaz^2-\deltax^2)(V_{2}^{def}(q)+\deltax^2
V_{4a}^{def}(q))\rbr \lab{def6} \ea \ee where the terms $
V_{2}^{def} $ and  $ V_{4a}^{def} $
 are the net contributions
from the deformation effect of  the terms
 $\ltwo$ and $\lfora$ in Eq.s \re{s1}-\re{s3} respectively.
They are given by \be \ba
 V_{2}^{def}(q)=\dsf{(\pi f_\pi)}{e}\ds\int_{0}^{\infty} dx x^2\times
\nwl \times \ds\int_{0}^{1} dt\lbr
\rhoone^2(\rhotwo^2-1)(\beta_1+Y_1t_{1}^{2})+\phi_{1}^{2}t_{1}^{2}
(\rhotwo^2-1)+ \nwl +\rhoone\rhotwo\phi_1\phi_2+(1\leftrightarrow
2)\rbr, \nwl V_{4}^{def}(q)=\dsf{(8\pi
f_\pi)}{3e}\ds\int_{0}^{\infty} dx x^2\times \nwl \times
\ds\int_{0}^{1} dt
\lbr(Y_1x_{1}^{2}+3\beta_1)(Y_2t_{2}^{2}+\beta_2)-Y_1Y_2ZZ_3- \nwl
-\beta_1\beta_2-2\beta_2Y_1t_{1}^{2}+ (1\leftrightarrow 2)\rbr.
\ea \ee with \be \ba \rho(x)=\exp{(-\sigma(x))}, \quad
\phi(x)=d\rho/dx , \nwl \beta(x)=(\sin\theta(x)/x)^2,\quad
Y(x)=[\theta'^{2}(x)-\beta(x)]/x^2, \nwl Z=x^2-q^2/4, \quad
Z_3=x^2t^2-q^2/4, \quad q=\delta_zref_\pi, \nwl
t_{1,2}^{2}=q^2/4+x^2t^2\pm xqt, \quad
 x_{1,2}^{2}=t_{1,2}^{2}+x^2(1-t^2),
\nwl \rho_i\equiv\rho(x_i)\quad \phi_i\equiv\phi(x_i)\quad
Y_i\equiv Y(x_i),
 \beta_i\equiv\beta(x_i).
\ea \ee
 Other terms in Eq. \re{def6} don't include deformation effects explicitly.
They are brought in previous chapter.
\begin{figure}[!hbt]
%   \vspace{0.5cm}
\bc \epsfysize=14cm \epsfbox{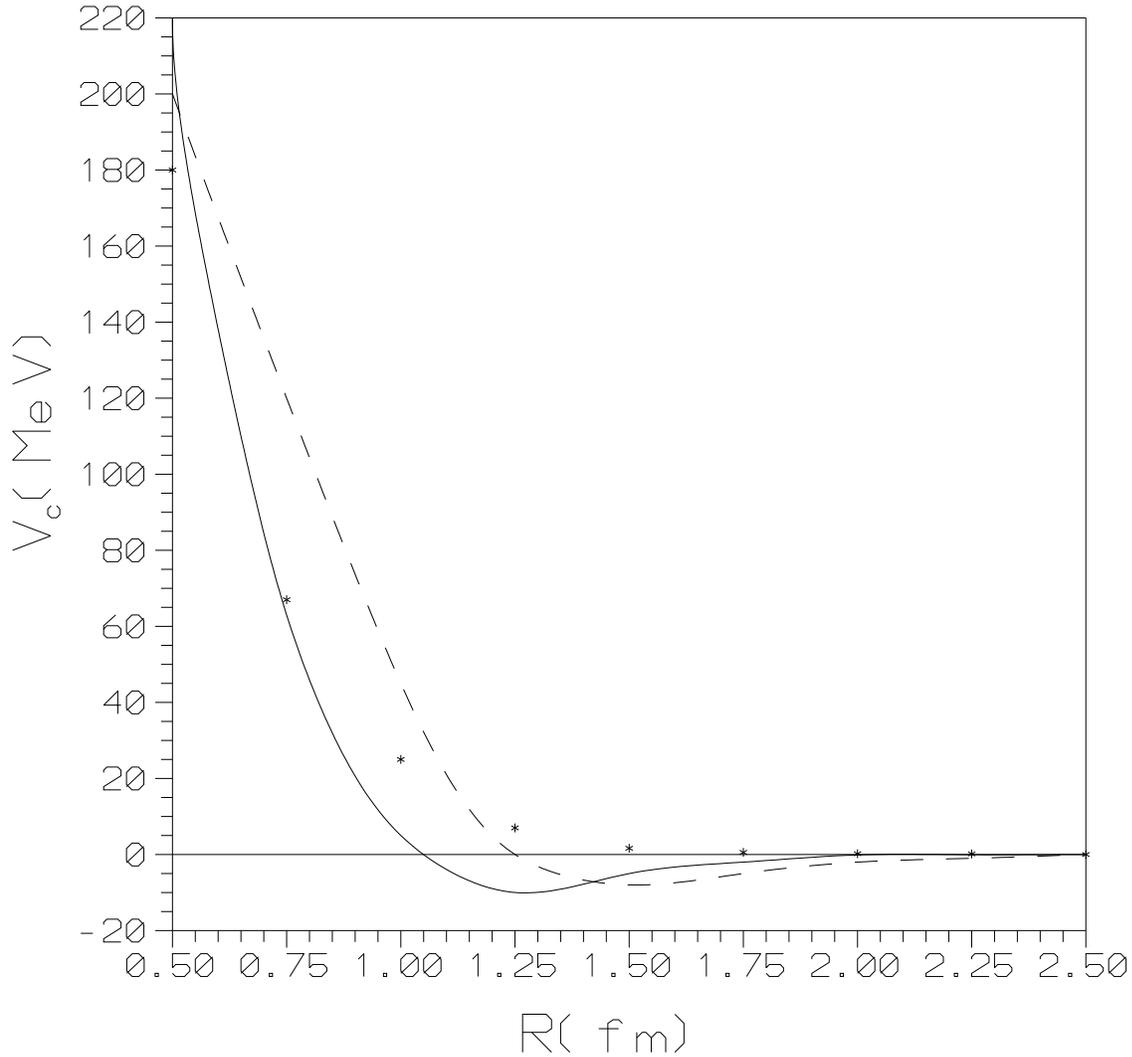} \ec
%  \epsfysize=9cm
%   \centerline{\epsffile{fdf1.eps}}
%   \vspace{1.0cm}
   \centerline{\parbox{15cm}{\caption{\label{figdef1}
Central isospin - independent potential in Eq.\re{def6}
calculated for the cases with  the dilaton field (solid curve)
and without one (dashed curve) as a function of the internucleon
distance $R$ . The dotted curve is the corresponding interaction
component in the "Paris" potential \ci{paris}.}}}
\end{figure}
We shall calculate the deformation parameters
 $\deltax(r) $ and $\deltaz(r)$   by minimizing
 the total static energy of two nucleon system at each separation
 $R$ using Eq.s \re{def2} - \re{def6}.

 The resulting values of the parameters will
be  used to study the changes in the
 shape of the nucleon. Now we  illustrate this
 procedure for the case of the  isoscalar mean square radius
  and the appropriate intrinsic quadrupole moment.
 The normalized isoscalar mean square radius
  along each axis may be defined by
\be { \riso{r_i} }^{*}_{I=0}=\disc\frac{
{\int}d{\vecr}r_{i}^{2}\bzero } {{\int}d{\vecr}\bzero } \lab{def7}
\ee where $i=x,y,z$  and $\bzero$ is the baryon charge
distribution \be
\bzero=\frac{1}{24{\pi}^{2}}{\epsilon}^{ijk}{\Tr}[L_{i}
L_{j}L_{k}] \lab{def8}. \ee The inclusion of the deformation in a
simple way:
 $ r_{i}{\rightarrow}q_{i}/\delta_{i} $ as in Eq. \re{def2}
 yields the following relation between the radius of a free
 spherical nucleon     ${\riso {r}}_{I=0} $ and a deformed one:
 ${ \riso{r_i} }^{*}_{I=0}=\disc\frac{ 1 } {\delta_{i}^{2} }
 {\riso {r}}_{I=0}
 $.
 Therefore the appropriate
  quadrupole moment characterizing the shape of the
  distribution  of
baryon matter is compared  to that of an ellipsoid with axis
$1/\delta_{z}$ and $1/\delta_x$: \be
  Q_{I=0}=3 { \riso{r_z} }^{*}_{I=0}
-{\riso{r}}^{*}_{I=0}=2{\riso{r}}_{I=0}(1/\delzsq-1/\delxsq)\;.
\ee

The explicit formulas  for $Q_{I=1}$ defined by
 $  Q_{I=1}=3 { \riso{r_z} }^{*}_{I=1}
-{\riso{r}}^{*}_{I=1}$
 are rather complicated and may by found elsewhere \ci{ourjafdef}.

\section{Results and discussions}
\indent

 In the numerical calculations we consider the following two
 cases: the Lagrangian with the dilaton and the pure Skyrme model
 when $\sigma=0$ in Eq. \re{s1}.
 \begin{figure}[!hbt]
%   \vspace{0.5cm}
\bc \epsfysize=14cm \epsfbox{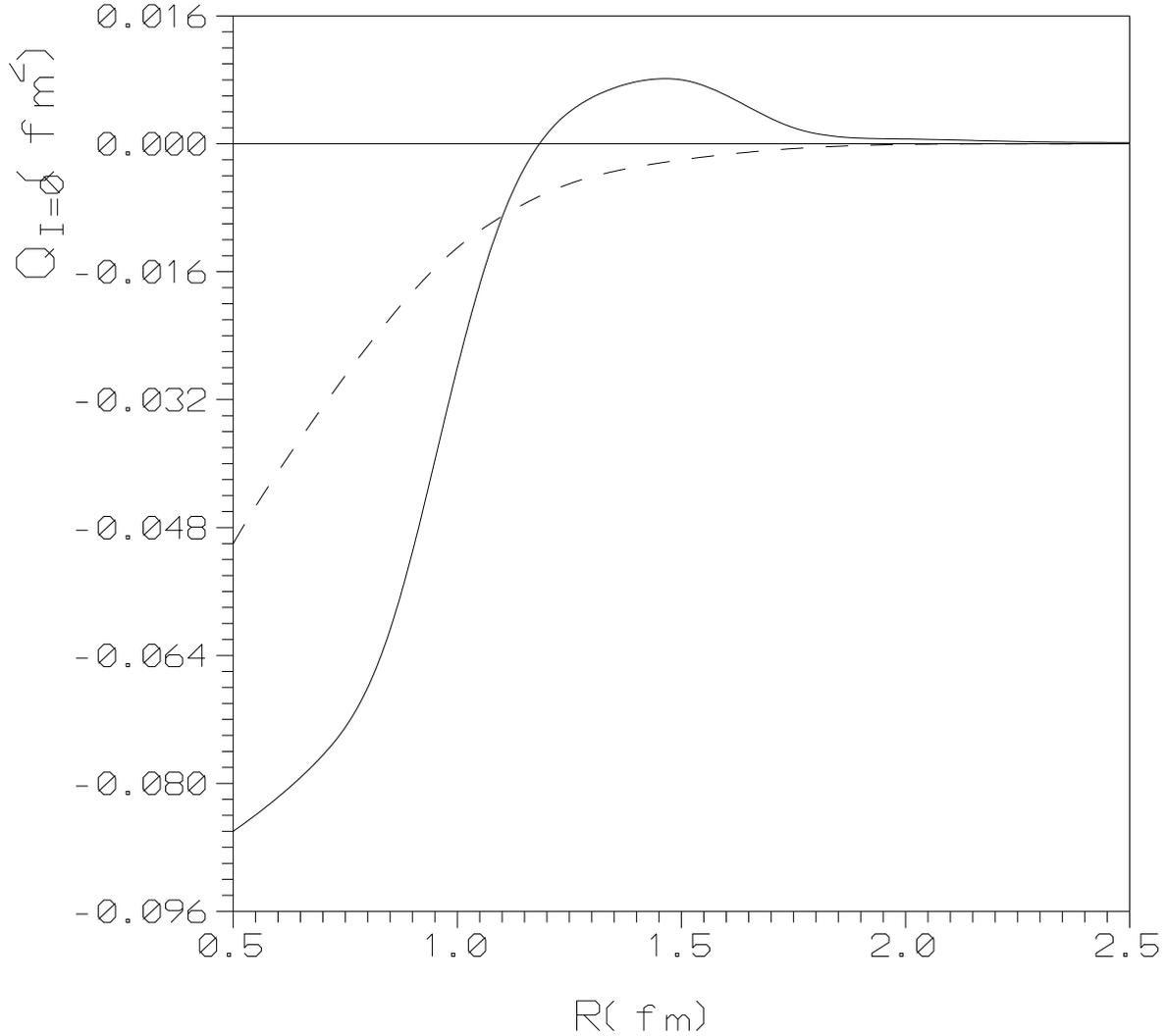} \ec
%   \epsfysize=9cm
%   \centerline{\epsffile{fdf2.eps}}
%   \vspace{1.0cm}
   \centerline{\parbox{15cm}{\caption{\label{figdef2}
The radial dependence of the isoscalar quadrupole moment $\qzero$
of the nucleon.The solid and dashed lines are obtained in the
case with dilaton  and pure Skyrme model respectively.}}}
\end{figure}
In  both cases the parameters $\fpi ,$
 $e$ and $\mpi$ were fixed at the values:
  $\fpi=93 MeV,$  $ e=2\pi ,$  $\mpi=139MeV.$ For the gluon condensate
  we use $\cg=(283 MeV)^4$ as obtained from lattice QCD calculations
\ci{satz}.
%\  (H. Satz 1982) .
The mass of the scalar meson,  $m_{\sigma}$,  defined by $
m_{\sigma}=\sqrt{2\cg}/2\fpi$ is then $ 610 MeV$.

This set of parameters produce the following
  static properties of the nucleon: $M_N=1054 MeV$,
  $\ga=0.65$, $ { \riso{r} }^{1/2}_{I=0}=0.38 fm $  and
  $ { \riso{r} }^{1/2}_{I=1}=0.66 fm $    in the dilaton case.
  No attempt is made here to search for a realistic set of
parameters
  since our interest was mainly in the link between
properties  of $NN$ interaction and the shape of the nucleon.

In Fig.  \ref{figdef1}  the
  central scalar - isoscalar part of the nucleon - nucleon interaction
  has been presented for both cases including deformation effects.
  For a comparison of the corresponding interaction component
 with  the realistic phenomenological  "Paris" potential \ci{paris}
%\  (M. Lacombe et al 1980)
 is also  displayed here ( the dotted  line in Fig. \ref{figdef1}).
   It is clear that  the Lagrangian with the dilaton field
is able to    describe the nucleon - nucleon interaction in the
   intermediate region quite well.
\begin{figure}[!hbt]
%   \vspace{0.5cm}
\bc \epsfysize=14cm \epsfbox{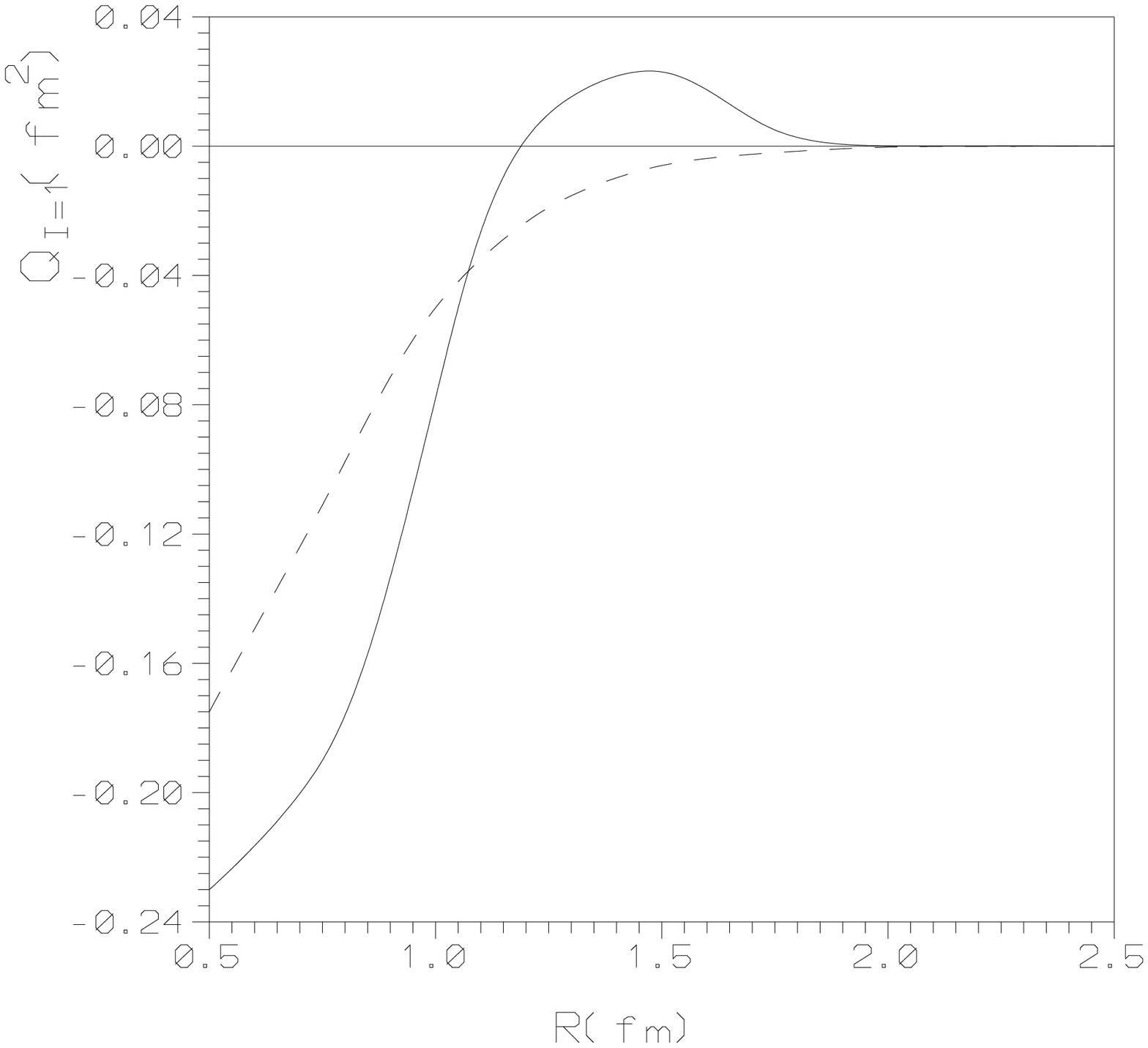} \ec
%  \epsfysize=9cm
%   \centerline{\epsffile{fdf3.eps}}
%   \vspace{1.0cm}
   \centerline{\parbox{15cm}{\caption{\label{figdef3}
The same as in Fig. \ref{figdef2} but for the isovector
quadrupole moment $\qone$.}}}
\end{figure}

    We now turn to the changes in the shapes of   interacting nucleons.
    The intrinsic quadrupole moments
     $\qzero$ and $\qone$ are shown in  Fig.\ref{figdef2}
    and Fig.\ref{figdef3} respectively. In the pure Skyrme model
     when there is no attraction [$V_2=V_{2}^{def}=0$ in Eq.\re{def6}]
     between skyrmions it  becomes oblate  (dashed lines in the
 figures \ref{figdef2}, \ref{figdef3})
     due to  the strong repulsion caused by the $V^4$ terms
     in Eq. \re{def6}. The inclusion of the dilaton leads to the following
     qualitative picture: At  large separations skyrmion is
     obviously in a spherical shape, becomes prolate
      at the intermediate region and deforms to an oblate shape
at small        distances where the repulsion dominates.

      As the nucleons approach each other they change shapes from prolate
      into   oblate       at $R\sim1.2fm$. Comparing  figures
Fig.\ref{figdef2}
      and Fig.\ref{figdef3}
      it  may be noticed
  that the isoscalar intrinsic quadrupole moment $\qzero$
      is much smaller than the isovector one $\qone$ at
      intermediate  separations.
\begin{figure}[!htb]
\bc \epsfysize=14cm \epsfbox{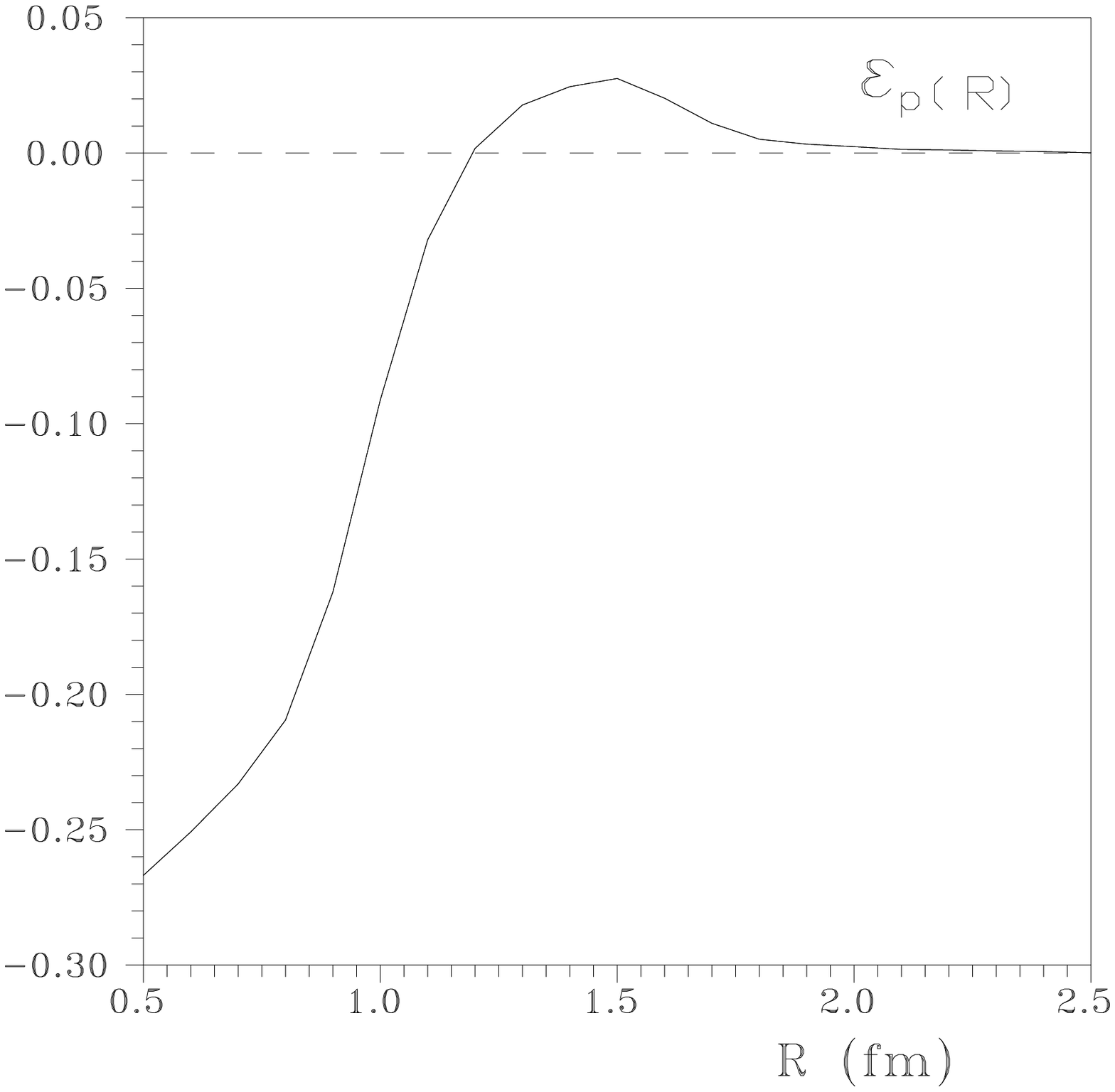} \ec
\centerline{\parbox{15cm}{\caption{\label{figdef4} The
excentricity of the ellipsoidally deformed nucleon  defined as
$\varepsilon=(a^2-b^2)/(a^2+b^2)$,
 where a and b are
the tranverse and conjugate axes of the ellipse. }}}
\end{figure}
     The intrinsic quadrupole moment of the proton  defined by:
\be
    Q_{p}=(\qzero+\qone)/2
\ee
 reaches a  maximum value of
     $Q_{p}=0.016 fm^2$
     at $r\sim 1.5 fm.$ The corresponding excentricity $\varepsilon$ (see
Fig. \ref{figdef4}) reaches the maximum value $\varepsilon\approx
0.02$. Note that for deutron $\varepsilon_D\approx 0.12$, due to
the small mixing of $D$ - state. In this sense, one may conclude
     that the shape of a nucleon in the nuclei is "more spherical" than the
deutron. We expect new data from high - energy electron
     scattering on nuclei to make the situation  clear.

  As a concluding remark we have to underline that the deformed states
    of oblate (prolate) shapes may  not
      necessarily belong to
     the $K=1$ band found in   ref. \ci{hs}
%\ (C.Hajduk et al 1986)
 since for a strongly
      deformed system  the quantization procedure
       used here needs some modifications.
It is a generalization of the well - known original Skyrme model
\ci{anw,skyrme}
%\ (Skyrme 1961, G.S.Adkins et al 1983)
  and takes into account the conformal anomaly
of QCD.

%\include{ch4pirho}
%       ANOMALEOUS DIMENSION \begin{document}
%%%%%%  ReMOVE  these 2 lines if you want your fig.s (tabs) to be numbered
%%as chapter.number like V.3
\setcounter{fignum}{\value{figure}}
\setcounter{tabnum}{\value{table}}
%%%%%%%%%%%%%%%%%%%%%%%%%%%%%%%%%%%%%
\chapter[
The nucleon--nucleon interaction and properties of the nucleon in
a $\pirow$  soliton model including a dilaton field with
anomalous dimension ]{} \setcounter{figure}{\value{fignum}}
\setcounter{table}{\value{tabnum}}
\bc { \Large\bf The nucleon--nucleon interaction and properties of
the nucleon in a $\pirow$  soliton model including a dilaton
field with anomalous dimension }\footnote {The  present chapter
is based on following articles by the author and his
collaborators: \ci{ouruz,ourpirho}} \ec

\section{Introduction}
\indent

Recently  Furnstahl, Tang and Serot (FTS) \cite{fts} have
proposed a new model for nuclear matter and finite nuclei that
realizes QCD symmetries such as chiral symmetry, broken scale
invariance and the phenomenology of vector meson dominance. An
important feature of this approach is the inclusion of light
scalar
 degrees of freedom,  which are given an anomalous scale dimension.
The vacuum dynamics of QCD is  constrained by the trace anomaly
and related low--energy theorems of QCD. The scalar--isoscalar
sector of the theory  is divided into a low mass part that is
adequately described by a scalar meson (quarkonium) with
anomalous dimension and a high mass part (gluonium), that can be
``integrated out'', leading to various couplings among the
remaining fields. The application of the model to the properties
of nuclear matter as well as finite nuclei gave a satisfactory
description. Further developments of the
model~\cite{rhofts,librown} showed that the light scalar related
to  the trace anomaly  can play a significant role not only in
the description of bound nucleons but also in  the description of
heavy--ion collisions. It was also shown that the anomalous
cannot be due to an effect of nuclear density on the trace anomaly
of QCD \ci{rhofts}.

Here a natural question arises: What is the role of this light
quarkonium in the description of the properties of a single
nucleon,  when it is taken into account in topological nonlinear
chiral soliton models,  which are similar to the FTS effective
Lagrangian on the  single nucleon level? In the present Chapter
we introduce a dilaton field with an anomalous dimension into the
$\pirow$--model proposed in ref.~\cite{meisnpa} and investigate
some properties of single nucleon which emerges as a soliton in
the sector with baryon number one $(B=1)$. Note that, analysis of
pion -  nucleon scattering phase shifts made in refs.
\ci{ishida1,ishida2} indicate  the existence of the light scalar
meson.

It is well known that a scalar--isoscalar  meson,
%\footnote{Here
%and in what follows, we call it sigma meson for simplicity.
% Although we assume
%it to be some kind of quarkonium state, its precise dynamical nature
%is of no direct relevance for the following arguments.}
the sigma, plays an important role in the nucleon--nucleon (NN)
interaction especially within one--boson--exchange (OBE)
models~\cite{machadv,machrep}. Note also that, the missing medium
range attraction was a long standing puzzle in Skyrme like models.
In Chapter 2 we have shown that explicit inclusion of a scalar
meson into the Skyrme model produces in a natural way the desired
attraction. However, OBE model includes vector mesons also.
Therefore, it would be quite interesting to investigate the
central part of the NN interaction when the light scalar
($\sigma$), $\rho$
 and $\omega$--mesons
are  taken into account explicitly. This is what will be
considered in present Chapter. In particular, it is important
that when one is to properly describe the intermediate range
attraction in the central NN interaction, the successful
description of the single nucleon properties within the
$\pi\rho\omega$ model should not be destroyed.

We also note that in an soliton approach with explicit regulated
two--pion loop graphs one is able to get the proper intermediate
range attraction.
 In that case, however,
one does not stay within a simple OBE approach any more (as done
here) and also needs to calculate the modifications of the
isovector two--pion exchange to the $\rho$ and so on. For
comparison,
 we mention that recent developments of
the original Bonn OBE potential performed at J\"ulich also
include multi--meson exchanges leading to a renormalization of
various interactions,
 couplings and
cut--off parameters~\cite{juel1,juel2}. The model we investigate
is related closely to the OBE approximation of the NN force.

\section{The $\pirows$  model}
\indent

Including a  $\sigma$--meson by means of the scale invariance and
trace anomaly of QCD into the $\pirow$--model~\cite{meisrep} can
be done in terms of the following chiral Lagrangian of the
coupled $\pirows$ system, \be \ba {\cl}=\dsf{
S_{0}^{2}e^{-2\sigma/d} }{2 } \pal_{\mu}\sigma \pal^{\mu}\sigma
-\dsf{f_{\pi}^{2}e^{-2\sigma/d }} {4}\Tr L_{\mu}L^{\mu}
-\dsf{f_{\pi}^{2}e^{-2\sigma/d} }{2} \Tr[l_{\mu}+
r_{\mu}+ig\vec\tau\vec\rho_{\mu}+ \nwl
\\
+ig\omega_{\mu}]^2
%\quad\\
+\dsf{3}{2}g\omega_{\mu}B^{\mu}-
\dsf{1}{4}(\omega_{\mu\nu}\omega^{\mu\nu}+
\vec\rho_{\mu\nu}\vec\rho^{\, \, \mu\nu})
%\quad\\
+\dsf{f_{\pi}^2m_{\pi}^{2}e^{-3\sigma/d} }{2}\Tr(U-1)- \nwl
\\
%\quad\\
-\dsf{d^{2}S_{0}^2\msig^{2}}{16}[1-e^{-4\sigma/d}(\dsf{4\sigma}{
d}+1)], \\
\label{anom1} \ea \ee where the pion fields are parametrized in
terms of $U=\exp{(i\vec\tau \cdot \vec\pi/\fpi)}$ and
$\xi=\sqrt{U}$, left/right--handed currents are given by
$L_{\mu}=U^+\pal_{\mu}U$, $l_{\mu}=\xi^+\pal_{\mu}\xi$,
  $ r_{\mu}=\xi\pal_{\mu}\xi^+, $
and the pertinent vector meson ($\vec{\rho}, \omega$) field
strength tensors are
$\vec\rho_{\mu\nu}=\pal_{\mu}\vec\rho_{\nu}-\pal_{\nu}\vec\rho_{\mu}+
g[\vec\rho_{\mu}\times\vec\rho_{\nu}]$ and
$\omega_{\mu\nu}=\pal_{\mu}\omega_{\nu}-\pal_{\nu}\omega_{\mu}$.
Furthermore, the topological baryon number current is given by
$B^{\mu}=\varepsilon^{\mu\alpha\beta\gamma} \Tr
L_{\alpha}L_{\beta}L_{\gamma}/(24\pi^2)$.

In  Eq.\re{anom1}, $S_0$ is the vacuum expectation value of
scalar field in free space matter, $\fpi$ is the pion decay
constant $(\fpi=93~{\rm MeV})  $
 and $g=g_{\rho\pi\pi}$
 is determined through the KSFR  relation $g=m/\sqrt{2}\fpi $.
 The model assumes the masses
of $\rho$ and $\omega$ mesons to be equal, $m_\rho=m_\omega=m$.
The mass of the $\sigma$ is related to the gluon condensate in
the usual way~\cite{fts,myransky,jain}
 $\msig=2\sqrt{\cg}/(dS_0)$,
where $d$ is the scale dimension of scalar field ($d>1$). Being
``mapped'' onto the states of a nucleon, the Lagrangian
Eq.\re{anom1} will be similar to the FTS effective Lagrangian.

Nucleons arise as soliton solutions  from the Lagrangian
Eq.\re{anom1} in the sector with baryon number $B=1$ as it has
been explained in Chapter I. To construct them one goes through a
two step procedure. First, one finds the classical soliton which
has neither good spin nor good isospin.
Then an adiabatic rotation of the soliton is performed and it is
quantized collectively.\footnote{We again refer the reader to
ref.~\cite{ourpirho,meisrep} for details.} The classical soliton
follows from Eq.\re{anom1} by virtue of a spherical symmetrical
ans\"atze for the meson fields: \be \ba U(\vec
r)=\exp{(i\vec\tau\hat{r}\theta(r))},  \, \, \, \quad
\rho_i^a=\varepsilon_{iak}\hat r_k\dsf{G(r)}{gr}, \quad
\omega_{\mu}(\vec r)=\omega(r)\delta_{\mu 0},  \, \, \,
\sigma(\vec r)=\sigma(r)\,\,. \lab{anom2} \ea \ee In what follows
we call $\theta(r) $,  $G(r) $,
 $\omega(r) $,  and $\sigma(r)  $
the pion--, $\rho $--,  $\omega $--,  and $\sigma$--meson profile
functions, respectively. The pertinent boundary conditions to
ensure baryon number one and finite energy are, $\theta(0)=\pi,
G(0)=-2,  \omega^{\prime}(0)=\sigma^{\prime}(0)=0,
\theta(\infty)= G(\infty)=\omega(\infty)=\sigma(\infty)=0$. To
project out baryonic states of good spin and isospin,  we perform
a time--independent SU(2) rotation \be \ba U(\vecr,
t)=A(t)U(\vecr)A^{+}(t), \, \,
 \xi(\vecr, t)=A(t)\xi(\vecr)A^{+}(t)\\
 \sigma(\vecr, t)=\sigma(r), \quad
\omega(\vecr, t)=\dsf{\phi(r)}{r}[\vec K \hat r]\\
\vec\tau\cdot \vec\rho_{0}(\vecr,
t)=\dsf{2}{g}A(t)\vec\tau\cdot(\vec K\xi_1(r)+
\hat r\vec K\cdot\hat r \xi_2(r))A^+(t), \\
\vec\tau\cdot\vec\rho_{i}(\vecr, t)=A(t)\vec \tau
\cdot\vec\rho_{i}(\vecr)A^{+}(t) \lab{anom3} \ea \ee with
$2\vec{K}$ the angular frequency of the spinning mode of soliton,
$i\vec\tau\cdot\vec K=A^{+}\dot{A}$. This leads to the
time--dependent Lagrange function \be {\cal L}(t)=\int d\vecr
{\cal L}=-M_{H}(\theta, G, \omega, \sigma) +\Lambda(\theta, G,
\omega, \sigma, \phi, \xi_1, \xi_2)\Tr(\dot{A}\dot{A}^+)~. \ee
Minimizing the classical mass $M_{H}(\theta, G, \omega, \sigma)$
 leads to the coupled differential equations
for $\theta, G, \omega$ and $\sigma $
 subject to the aforementioned boundary conditions.
In the spirit of the large $N_c $--expansion,  one then
extremizes the moment of inertia $\Lambda(\theta, G, \omega,
\sigma, \phi, \xi_1, \xi_2) $
 which gives the coupled differential
equations for $\xi_1$ , $ \xi_2  $ and $\phi $
 in the presence of the background profiles
$\theta,  G,  \omega $ and $\sigma  $.
  The pertinent boundary conditions are
$\phi(0)=\phi(\infty)=0, \, \,$
$\xi_1^{\prime}(0)=\xi_1(\infty)=0,\,\,$
$\xi_2^{\prime}(0)=\xi_2(\infty)=0,\,\,  $ $2\xi_1(0)+\xi_2(0)=2.
$ The masses of nucleon $M_N $ and the mass of $\Delta$ ,
 $M_{\Delta} $,
are then given by $M_N=M_H+3/8\Lambda $ and
$M_\Delta=M_H+3/15\Lambda$.

The electromagnetic form factors  obtained in the usual
way~\cite{meisrep} are: \be \ba G^{S}_E({\vec q}^{\
2})=-\dsf{4\pi m^2}{3g}\disc\il_0^{\infty}
j_0(qr)\omega(r)e^{-2\sigma/d}r^2dr, \nwl
\\
G^{S}_M({\vec q}^{\ 2})=-\dsf{2\pi M_Nm^2}{3g\Lambda}
\il_0^{\infty}\dsf{j_1(qr)}{qr}\phi(r)e^{-2\sigma/d}r^2dr, \nwl
\\
G^{V}_E({\vec q}^{\
2})=\dsf{4\pi}{\Lambda}\disc\il_0^{\infty}j_0(qr) \left\{
\dsf{\fpi^2}{3}[4s_{2}^4+(1+2c)\xi_1+
\xi_2]e^{-2\sigma/d}+\dsf{g\phi\theta^{\prime}s^2}{8\pi^2r^2}\right\}r^2dr,
\nwl
\\
G^{V}_M({\vec q}^{\ 2})=\dsf{8\pi M_N}{3}\disc
 \il_0^{\infty} \dsf{j_1(qr)}{qr}
\left\{ 2\fpi^2[2s_{2}^4-Gc]e^{-2\sigma/d}+
\dsf{3g}{8\pi^2}\omega\theta^{\prime}s^2\right\}r^2dr, \lab{anom4}
\ea \ee where $ s=\sin(\theta),  c=\cos(\theta) $  and
$s_2=\sin(\theta/2)$. The normalization is
$G^{S}_{E}(0)=G^{V}_{E}(0) =1/2$. Similarly,  meson--nucleon
vertex form factors may be calculated \ci{cohen,meisff}. In the
Breit frame one has: \be \ba G^{\pi}(-{\vec q}^{\ 2})= \dsf{8\pi
M_Nf_{\pi}}{3q}({\vec q}^{\ 2}+m_{\pi}^{2})
\disc\il_0^{\infty}j_1( qr)sin(\theta)r^2dr= \nwl
\\
=\dsf{8\pi M_Nf_{\pi}}{3}\disc\il_0^{\infty} \dsf{j_1( qr)}{ qr}
\left[-2\theta^{\prime}c-\theta^{\prime\prime}rc+\theta^{\prime
2}rs+
\dsf{2s}{r}+rm_{\pi}^{2}s\right]r^2 dr, \\
\nwl G^{\rho}_E(-{\vec q}^{\ 2})=\dsf{2\pi}{g\Lambda }\disc
\il_0^{\infty} j_0( qr)
\left[-\xi_1^{\prime\prime}-\dsf{2\xi_1^{\prime}}{r}+ m^2\xi_1
-\dsf{\xi_2^{\prime\prime}}{3}-\dsf{2\xi_2^{\prime}}{3r}
 +\dsf{m^2\xi_{2}^{2}}{3}\right]r^2dr,
\nwl
\\
G^{\rho}_M(-{\vec q}^{\ 2})=-\dsf{8\pi M_N}{3g}
\il_0^{\infty}\dsf{j_1( qr)}{ qr}\left[
-G^{\prime\prime}+2G/r^2+m^2G\right]r^2dr, \nwl
\\
G^{\omega}_E(-{\vec q}^{\ 2})=\disc {4\pi } \il_0^{\infty}j_0(
qr)\left[
\omega^{\prime\prime}+\dsf{2\omega^{\prime}}{r}-m^2\omega\right]r^2dr,
\nwl
 \\
G^{\omega}_M(-{\vec q}^{\ 2})=\dsf{2\pi M_N}{\Lambda}
\il_0^{\infty}\dsf{j_1( qr)}{ qr}\left[
\phi^{\prime\prime}-\dsf{2\phi}{r^2}-m^2\phi\right]r^2dr,
\lab{anom5} \ea \ee where the ``electric'' and ``magnetic''
vector meson--nucleon form factors are connected  to the Dirac
$F_{1}(t) $ and Pauli form factors $F_{2}(t) $ through the
following relations: \be \ba
  G^{i}_{E}(t)=F_{1}^{i}(t)+tF_{2}^{i}(t)/4M_{N}^{2} \;,
\nwl G^{i}_{M}(t)=F_{1}^{i}(t)+F_{2}^{i}(t)$, $(i=\rho, \omega)\;.
\ea \ee
   \section{Results and discussions.}
\subsection{Static and electromagnetic properties of the nucleon.}
\indent

Using the formulas given above we have calculated static and
electromagnetic properties of nucleon. As can be seen from
Eq.\re{anom1}, the Lagrangian has no free parameters in the
$\pirow$ sector. So,  in actual calculations the parameters
$\mpi,  m,  \fpi$ are fixed at their emperical values,
$\mpi=138$~MeV, $m=m_{\rho}=m_{\omega}=770$~MeV,  $\fpi=93$~MeV,
$ g=m/\sqrt{2}\fpi=5.85$.  In the $\sigma$--meson sector there are
in general three  free parameters:  $\msig$, $S_0$ and the  scale
dimension $d$. The latter has been well studied for nuclear
matter calculations~\ci{fts,rhofts}. In  particular,  it was
shown that for $d{\ge}2$  the much debated Brown--Rho (BR)
scaling may be recovered. Therefore,  assuming that there is no
dependence of $d$ on the density,  we shall use the best value
$d=2.6$ found in refs.~\ci{fts,rhofts}. The values for $S_0$ -
were  found to be $S_0=90.6\div 95.6$  MeV ~\ci{fts} .
 So we put  $S_0=\fpi=93$  MeV. The mass of the $\sigma$ or
equivalently the gluon condensate
$\cg=m_{\sigma}^{2}d^2S_{0}^{2}/4=m_{\sigma}^{2}d^2f_{\pi}^{2}/4$
is  uncertain.  We thus consider two cases: $\msig=550$~MeV
and   $\msig=720$~MeV in accordance with recent $\pi\pi$ phase
shift analyses~\ci{ishida1,ishida2} and with OBE values. We
stress again that the precise nature of such a scalar--isoscalar
field is not relevant here, only that it should not be a pure
gluonium state. A summary of static nucleon properties obtained
in both  cases, i.e. with $\msig=550~$MeV   and
$\msig=720$~MeV,  is given in Table \ref{antab1}.
\begin{table}[hbt]
\normalsize \caption{ \large\it
 Baryon properties in the $\pirow$ and $\piros$ models
}
\begin{center}
\begin{tabular}{|l|c|c|c|c|}\hline
           &       $\pirow$  &$\piros$     &$\piros$ & Exp. \\
\hline
$\msig$ [MeV] & $-$              &550          &720& \\
$C^{1/4}_{g}$ [MeV]  & $-$              & 258  & 295  &
     300$\div$400  \\
$B^{1/4}$  [MeV]  & $-$  & 119 & 121  & $-$       \\
$M_N$ [MeV] & 1560     & 1492 & 1511 &        939\\
$\Lambda$ [fm] & 0.88            & 0.88 & 0.88 &    $-$ \\
$M_\Delta-M_N$ [MeV]&    344  & 350  & 350  &        293 \\
$r_H=\langle r_B^2\rangle^{1/2}$ [fm]& 0.5  & 0.5  & 0.5  & $\sim$0.5  \\
$\langle r_E^2\rangle^{1/2}_p$ [fm] & 0.92 & 0.94 & 0.94 &  0.86$\pm$ 0.01\\
$\langle r_E^2\rangle_n$ [fm$^2$] & -0.20& -0.16&-0.16 & -0.119$\pm$0.004\\
$\langle r_M^2\rangle^{1/2}_p$[fm] & 0.84 & 0.85 & 0.85 & 0.86$\pm$ 0.06 \\
$\langle r_M^2\rangle^{1/2}_n$ [fm] & 0.85 & 0.85 & 0.85 & 0.88$\pm$ 0.07 \\
$\mu_p$ [n.m.]        & 3.34 & 3.33 & 3.33 &        2.79\\
$\mu_n$ [n.m.] & -2.58  & -2.53& -2.53&    -1.91 \\
$|\mu_p/\mu_n|$ & 1.29         & 1.30 & 1.30 &        1.46  \\
$g_A$ &0.88                  & 0.95 & 0.95 &  1.26$\pm$0.006\\
$\langle r_A^2\rangle^{1/2}_p$ [fm] & 0.63 & 0.66 & 0.66 &  0.65$\pm$ 0.07\\
\hline
\end{tabular}
\end{center}
\label{antab1}
%\large
\end{table}

One immediately observes that the nucleon mass is again
overestimated. This may not be regarded as a deficiency, since it
is known that quantum fluctuations tend to decrease the mass
substantially.

To estimate the influence of the $\sigma$--meson we also show the
results given by minimal version of $\pirow$ model. As can be
seen from  Table \ref{antab1}, the inclusion of a light sigma
meson into the basic $\pirow$ model just slightly changes the
nucleon mass and its electromagnetic properties. This may be
explained by the fact that
 the role of intermediate scalar--isoscalar meson in gamma--nucleon
interactions is negligible. In contrast,  the presence of the
sigma--meson leads to an enhancement of axial coupling constant
as it was first observed in ref.~\ci{anov}. Although, the
physical mechanism of this change is not clear, it   may be
understood as  mainly due to modification of meson profiles. This
is in marked contrast to the inclusion of pion loop effects, which
tend to lower the axial coupling even further~\cite{zwm}. In the
present model $\ga=0.88$ for the $\pirow$ and $\ga=0.95 $ for the
$\pirows$ model, respectively. One may expect that an appropriate
inclusion of $\sigma$ meson into a more complete version of the
basic $\pirow$ model might give the desired value $\ga=1.26$.
%  The normalized axial form factor
%$G_{A}(q^2)/G_{A}(0)$ is shown in Fig.~1.

\subsection{Meson--nucleon form factors.}
\indent

One of the usual ways to calculate the meson--nucleon interaction
potential within topological soliton models is the so--called
product ansatz~\ci{yabu}. Within this approximation, the
two--Skyrmion potential as a function of the relative angles of
orientation between the Skyrmions has a compact form,  and the
extraction of the NN potential by projection onto asymptotic
two--nucleon states is  straightforward. This procedure gives
only three nonvanishing channels: the central,  spin--spin and
tensor potential.
At large and intermediate distances, the latter two
 compare well  with e.g. the
phenomenological Paris potential~\ci{paris}.  The major inadequacy
found in such type of calculations is the lack of an intermediate
range attraction in central potential. Although many remedies
have been proposed,  this result may not be genuine for  Skyrme
like models. In fact, the product ansatz, which is not a solution
to the equations of motion, can only be considered accurate at
large distances,  and the failure of these calculations to
reproduce the central range attraction may simply be the failure
of the product approximation to provide an adequate approximation
to the exact solution.\footnote{An early study giving credit to
this line of reasoning can be found in ref.\cite{UKM}.} Although
the lack of central attraction may be recovered by the inclusion
of a scalar--isoscalar meson, the inherent ansatz dependence of
the trial configuration remains as a major shortcoming of product
approximation~\ci{jjp}.

On the other hand there is another natural way which was first
used by Holzwarth and Machleidt~\ci{holzmach1}.  They proposed to
calculate $\vnn$ within OBE model taking coupling constants and
meson--nucleon form factors from a microscopical model such as
the Cloudy Bag model or the Skyrme model. It was shown that the
Skyrme form factor is a soft pion form factor that is compatible
with the $\pi N$ and $NN$ systems. We shall use this strategy to
investigate the $NN$ potential within present model.

\begin{figure}[!htb]
%\vspace{0.5cm}
\bc \epsfysize=14cm \epsfbox{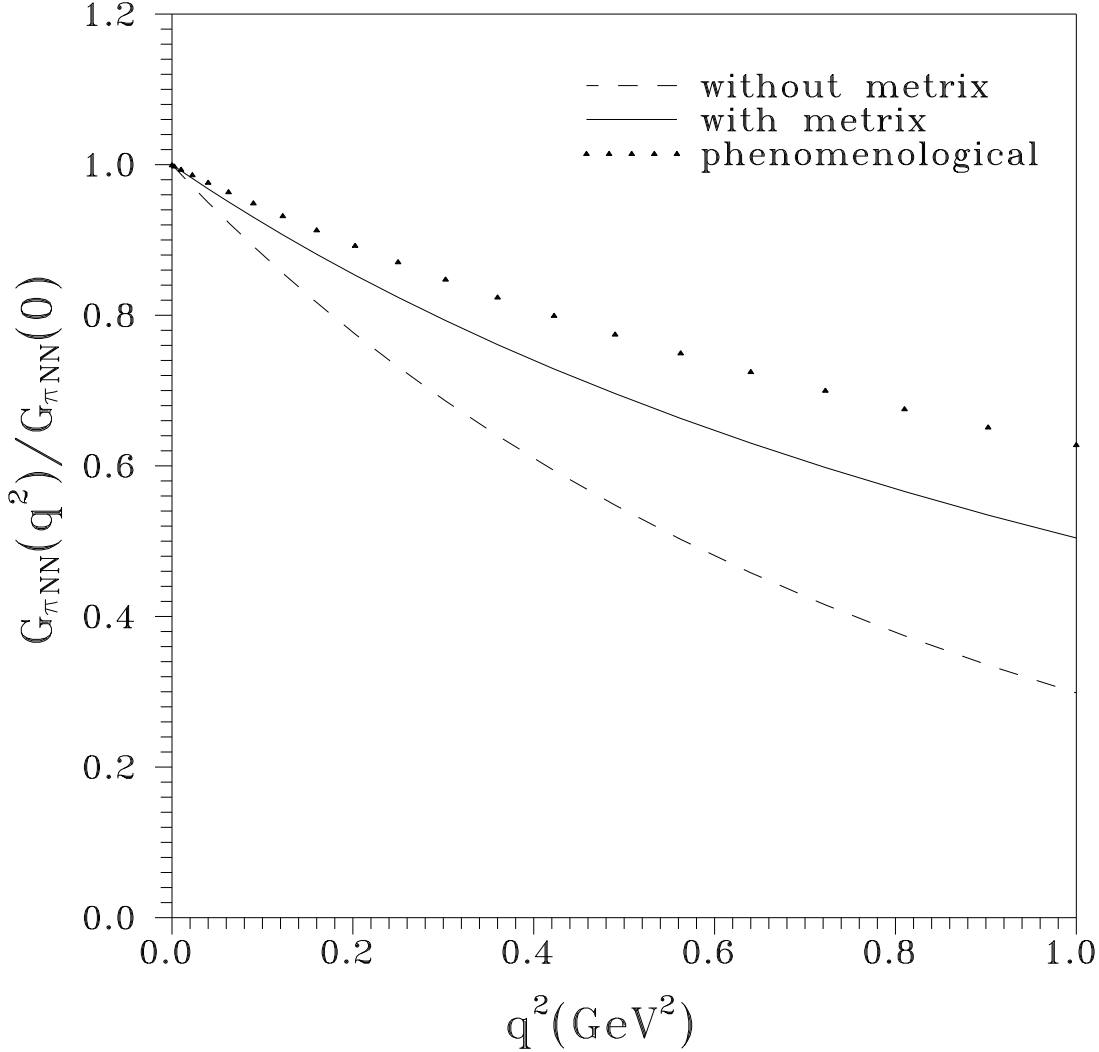} \ec
   \centerline{\parbox{15cm}
   {\caption{\label{figan1}
The normalized $\pi NN$  form factor in the $\pi\rho\omega\sigma$
model ($\msig=720$ MeV). The solid line represents the form
factor when the metric factor is included (Eq. \re{anom6}), while
the dashed line gives the result with no metric factor as in Eq.
\re{anom8}.
 The
dotted line is a monopole form factor with
$\Lambda_{\pi}=1300\,$MeV.
  }}}
\end{figure}

The meson nucleon form factors given by well known procedure,
proposed first by Cohen~\ci{meisff} are given in Eq.\re{anom5}.
Although they were derived in a microscopical and consequent way,
these form factors could not be directly used in standard OBE
schemes. The reason is that the OBE schemes
 ~\ci{machrep} in momentum space  use form factors
defined for fields propagating on a flat metric, whereas the
definition of form factors in Eq.\re{anom5} involve a nontrivial
metric. Hence, before using the latters in OBE scheme one should
modify the procedure by redifining meson fields. The modification
for pion nucleon form factors in $\pirow$ model is clearly
outlined in refs. ~\ci{holzmach1,holzmach2}. Now, applying this
procedure in the Lagrangian in Eq. \re{anom1} we get the
following $\pi NN$ form factor: \be \ba G^{\pi}(-{\vec q}^{\ 2})=
\dsf{8\pi M_Nf_{\pi}}{3q}({\vec q}^{\ 2}+m_{\pi}^{2})
\disc\il_0^{\infty}j_1( qr)\sqrt{M_T(r)}\sin(\theta)r^2dr= \nwl
\\
=\dsf{8\pi M_Nf_{\pi}}{3}\disc\il_0^{\infty} \dsf{j_1( qr)}{ qr}
\lbr
 -2F^{\prime}(r)-rF^{\prime\prime}(r)+\dsf{2F(r)}{r}+
rm_{\pi}^{2}F(r)\rbr
r^2 dr, \\
\lab{anom6} \ea \ee where \be \ba
M_T=\left[1+2\tan^2(\theta/2)\right] e^{-2\sigma/d} , \nwl
F(r)=\sqrt{3+\cos^2(\theta(r))-4\cos(\theta(r))} e^{-\sigma/d} .
\lab{anom7} \ea \ee The influence of metric factor $M_T$ to the
pion--nucleon form factor is illustrated in Fig. \ref{figan1}.

 It is seen that, without the inclusion
of   $M_T$  the form factor is softer than in OBE models (the
dashed line in Fig. \ref{figan1}), while its inclusion via Eqs.
\re{anom6}, \re{anom7} gives a  behavior closer to OBE models. In
fact, a monopole approximation at small $q^2=t$ of the normalised
form factor
 $G^{\pi}(t)/G^{\pi}(0)
\approx\Lambda_{\pi}^{2}/(\Lambda_{\pi}^{2}-t)$ gives
$\Lambda_{\pi}=860 $ MeV and  $\Lambda_{\pi}=1100 $ MeV for
$M_T=1$ and $M_T\not=1$ respectively, compared to its emperical
fit : $\Lambda_{\pi}^{OBE}=1300\,$MeV (dotted line in Fig.
\ref{figan1}). Note, however, that our result for $\Lambda_{\pi}$
including the metric is  in line with recent coupled--channel
calculations of the J\"ulich group~\cite{RB}. There, a monopole
form factor with $\Lambda_{\pi} \simeq 800\,$MeV is obtained.

Introducing a flat metric requires a canonical form for the
kinetic part of the Lagrangian, which determines the dynamics of
the field fluctation. The kinetic term of the scalar meson in Eq.
\re{anom1} \be {\cal L^{\mbox{kin}}_{\sigma}}=
S_{0}^{2}e^{-2\sigma/d}  \pal_{\mu}\sigma \pal^{\mu}\sigma/2 \ee
 can be easily rewriten in a  usual way:
\be {\cal L^{ \mbox{kin}  }_{\sigma}}=
\pal_{\mu}\sigmat\pal^{\mu}\sigmat/2 \ee by the following
redifinition of the basic sigma field: \be
\sigmat(r)=S_0d[1-e^{-\sigma(r)/d}]. \ee
 Now the new field
$\sigmat$ may be identified with the real sigma field. Clearly
this redifinition does not change the nucleons static properties
given in Table~\ref{antab1}. Note also that, using the above
redinition in the last term of Eq. \re{anom1}, one may easily
conclude that $m_{\sigmat}=m_{\sigma}$. The appropiate
sigma--nucleon form factor is given by \be G^{\sigma}(-{\vec
q}^{\ 2})=-4\pi\disc\il_0^{\infty} j_0(
qr)\left[\sigmat^{\prime\prime}+\dsf{2\sigmat^{\prime}}{r}-
m_{\sigma}^2\sigmat\right]r^2dr, \lab{anom8} \ee and may be used
in OBE models. We have not introduced any metric factors in the
form factors of the heavier mesons since these should play a
lesser role than in the case of the pion.

 For small values of
the squared four--momentum transfer $t$ each form factor can be
parametrized in monopole form:
 \be
G_{i}(t)=g_{i}(\Lambda_{i}^{2}-m_{i}^{2}
)/(\Lambda_{i}^{2}-m_{i}^{2} ) \ee $(i=\pi, \rho, \omega,
\sigma)$. We present in Table~\ref{antab2}
 the range parameters
(cut--offs) and the coupling constants of the resulting
meson--nucleon dynamics.

\begin{table} [hbt]
%\normalsize
\caption{ \large\it Meson--nucleon coupling constants and
cut--off parameters of meson--nucleon form factors. The
$\Lambda_i(i=\pi, \rho, \omega\sigma)$ are cutoff parameters in
equivalent monopole fits $1/(1-t/\Lambda_i^2)$ to the normalized
form factors $G_{iNN}(t)/G_{iNN}(0)$ around $t=0$. The empirical
values are from OBE potential fit \ci{machrep} }
%\vspace{1cm}
\begin{center}
\begin{tabular}{|l|c|c|c|c|}\hline
           &         $\pirow$      &$\piros$(550)  &$\piros$(720)& OBE/Emp.\\
\hline $G_{\pi NN}(0)$ &
14.74& 13.97 & 14.17 & 13.53\\
$G_{\sigma NN}(0)$ &
$-$ & 6.2   & 6.19  & 9.1 (12.41) \\
$F_1^\rho(0)$ &
2.55 & 2.76 & 2.68 & 2.24\\
$F_2^\rho(0)$ &
14.33 & 15.01 & 14.67 & 13.7  \\
$F_2^\rho(0)/F_1^\rho(0)   $ &
5.6      & 5.43  &5.47   & 6.1  \\
$F_1^\omega(0)$ &
8.78 & 10.73& 10.15& 11.7 \\
$F_2^\omega(0)$ &
$-2.15$ & $-2.78$ & $-2.65$ & 0 \\
$F_2^\omega(0)/F_1^\omega(0)   $ &
-0.24     &-0.25  &-0.26   & 0  \\
$\Lambda_\pi$ (GeV) &
1.2  & 1.1  & 1.1  & 1.3$\div$2.0 \\
$\Lambda_\sigma$ (GeV) &
$-$ & 0.59 & 0.60 & 1.3$\div$2.0    \\
$\Lambda^{\rho}_{1}$ (GeV) &
0.62   & 0.63 & 0.63 & 1.3 \\
$\Lambda^{\rho}_{2}$ (GeV) &
0.92   & 0.92 & 0.92 & 1.3 \\
$\Lambda^{\omega}_{1}$ (GeV) &
0.95     & 0.89 & 0.91 & 1.5 \\
$\Lambda^{\omega}_{2}$ (GeV) &
1.12   & 0.86 & 0.89 & - \\
\hline
\end{tabular}
\end{center}
\label{antab2}
%\large
\end{table}

 One can see that
there is no interference between $\sigma$--meson nucleon and e.
g. the pion--nucleon coupling constants. In other words,  the
inclusion of $\sigma$--meson does not significantly affect
meson--nucleon form factors that had been given by the $\pirow$
model. As it is seen from Table~\ref{antab2} the values for
meson--nucleon coupling constants are close to their emperical
values (in some cases obtained by OBE model fits). This is one of
the main advantages of the inclusion of a scalar--isoscalar meson
as done in the present approach.

\begin{figure}[!hbt]
\bc \epsfysize=14cm \epsfbox{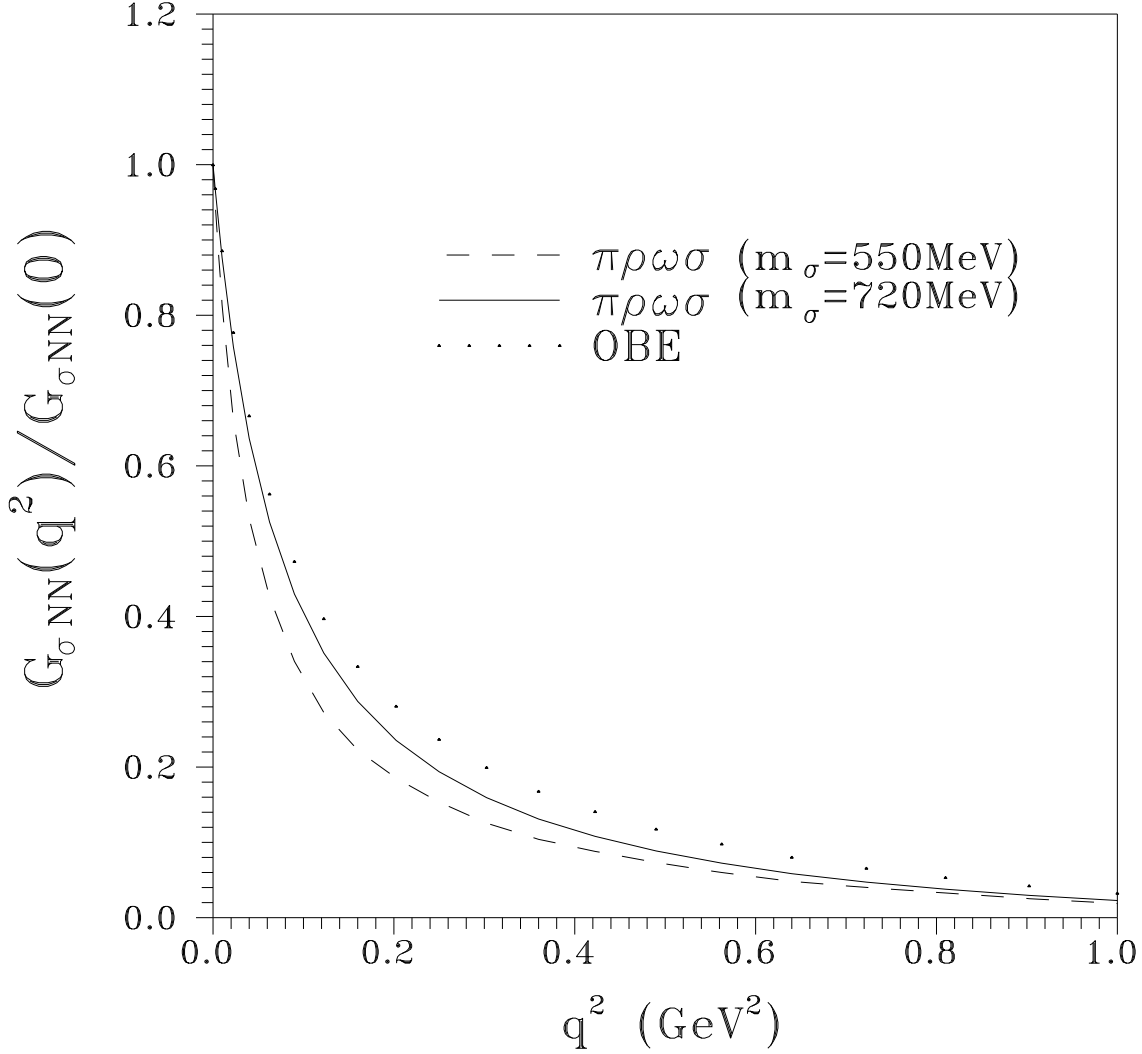} \ec
   \centerline{\parbox{15cm}
   {\caption{\label{figan2}
The sigma - nucleon form factor
 $G_{\sigma NN}(\vec q^{\,2})/G_{\sigma NN}(0)$.
The dashed and solid lines are for $m_{\sigma}=550$~MeV and
$m_{\sigma}=720$~MeV, respectively. Typical OBE monopole fits with
 $\Lambda = 1.3 \ldots 2$ GeV are shown by
the band enclosed by the dotted lines.}}}
\end{figure}

The corollary of the present model is that it gives  significant
information on the $\sigma$--nucleon interaction. As it is seen
from the Table~\ref{antab2}, the value for $ g_{\sigma NN}$ and
the cut--off parameter of sigma --nucleon vertex
$\Lambda_{\sigma}$
 are  smaller than their OBE prediction
 $\Lambda_{\sigma}^{OBE}\approx 1300\div 2000$ MeV.
    This contrast is
evidently seen from the  Fig. \ref{figan2}, where
$G^{\sigma}(t)/G^{\sigma}(0)$ for two cases : $\msig=720 $MeV
and  $\msig=550 $MeV is presented with the solid and dashed
 lines respectively. The band enclosed by the dotted lines refers to the OBE
monople form factor  with $\Lambda_{\sigma}^{OBE}=1300\ldots
2000\,$MeV.
  One can  conclude that the present model
%%%%% note if the cutoff gets smaller, the form factor is softer! %%%%
gives a softer $\sigma NN$ form factor than it obtained by OBE.
As it had been noticed before,  the  $t$--plane for each form
factor has a cut along the positive real axis extending from
$t=t_0$ to $\infty$.  The cut for the $\sigma$--nucleon vertex
function starts at $t_0=4\mpi^{2} $ reflecting the kinematical
threshold for the $\sigma\rightarrow\pi\pi$ channel. More
precisely this result follows from the asymptotic behavior of the
meson profiles: For $r\rightarrow\infty$, we have
$\theta(r)\sim\exp(-\mpi r)/\mpi r$,
 $ \sigma (r)\sim\theta ^2 (r)$, which are derived from the
equations of motion.

\subsection{Nucleon -  Nucleon  central interaction}
\indent

Once the vertex function of the corresponding meson--nucleon
interaction has been found, its appropriate contribution to the $
NN$ interaction may be easily calculated by using well known
techniques from OBE.  The detailed formulas are given
elsewhere~\ci{meisnn,machrep}.
\begin{figure}[!htb]
\bc \epsfysize=14cm \epsfbox{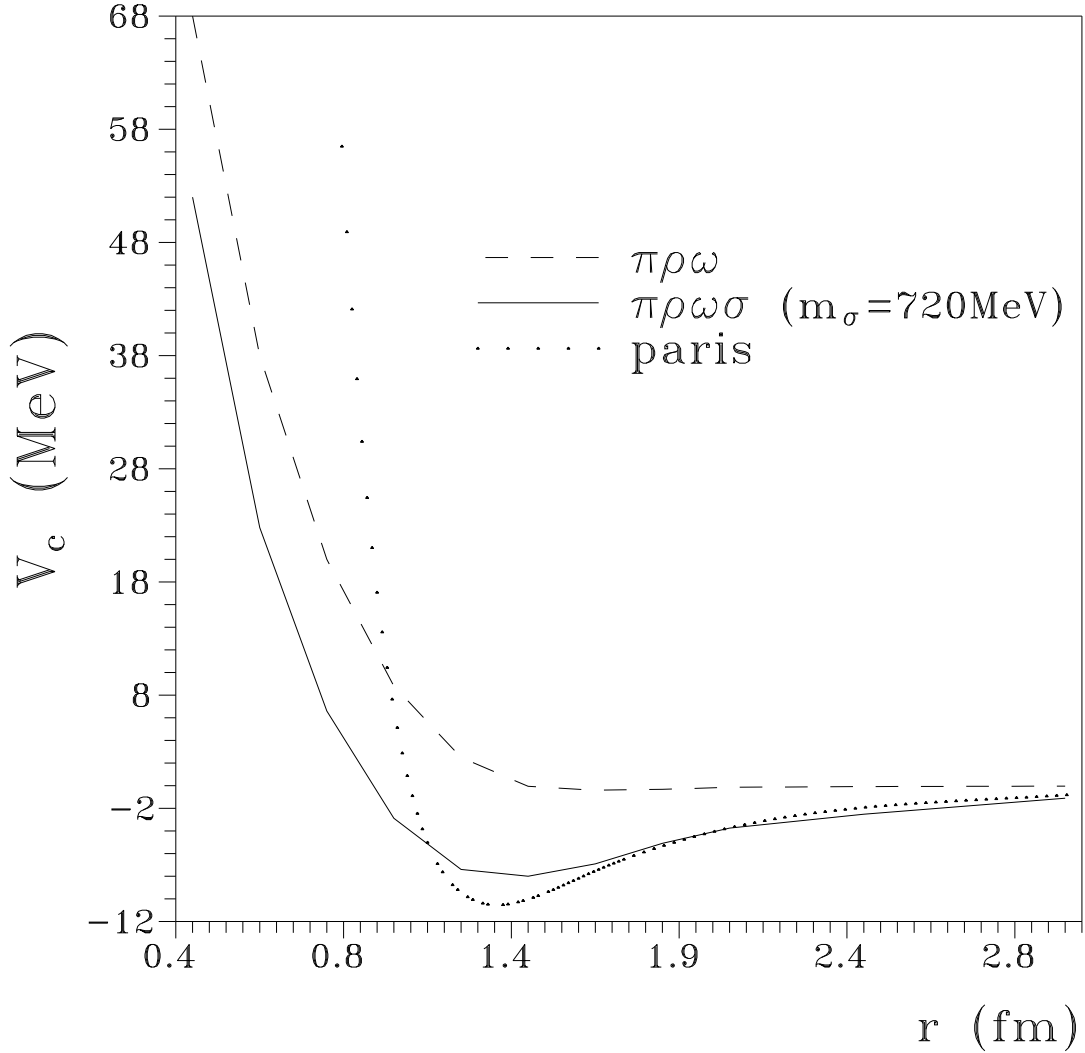} \ec
   \centerline{\parbox{15cm}{\caption{\label{figan3}
The central potential in the $S=1$, $T=0$ state for
$\pi\rho\omega$ and $\pi\rho\omega\sigma$ models (dashed and
solid lines, respectively). No contribution  from two--meson
exchange has been taken into account. The dotted line corresponds
to the Paris potential~{\protect \ci{paris}}.
  }}}
\end{figure}
In particular, the contribution of the $\sigma$--meson exchange
to the central potential is given by \be V^{c}_{\sigma}(r)=\disc
\il_0^{\infty}\frac{k^2 dk}{2\pi^2} \frac{G^{2}_{\sigma NN}
(k^2)}{k^2+\msig^2} \, j_0(kr)~. \ee The central $NN$ potential
in  the $T=0,  S=1 $ state (the deuteron state) is presented in
Fig. \ref{figan3} in comparison with Paris potential. Our
prediction is in  good agreement with the emperical one. Note
that the desired attraction  in the central
 $V_{NN}$ has before been obtained in the $\pirow$ model by means of
two--meson exchange~\ci{meisnn}.

In conclusion it should be noted that we do not intend to
describe (cover) all $NN$ phase shifts staying only in the
framework of the present model. Besides other mesons, which are
usually included in OBE picture, the full model should also take
in account e.g. $N\Delta\rho$ couplings. In addition the $2\pi$
exchange and its strong mixing with $\sigma$ meson exchange (see
e.g. ref.\ci{birse}) should be considered. Another reason
possible which limits the accuracy of $NN$ phase shifts in the
present model is that the $\sigma$NN coupling is not sensitive to
the mass of sigma (see Table~\ref{antab2}) as it is in the OBE
phenomenology.
 In fact, even
when the $2\pi$ exchange is disregarded , the pure OBE model has
to consider two types of sigmas with nearly the same masses but
with quite different coupling constants. So, we refrain from
performing direct calculations of $NN$ phase shifts in the
present model. Instead, we  point out that    the meson--nucleon
form factors found in the present model could be useful in a
wider context of calculations of nucleon--nucleon observables
(phase shifts, deuteron properties etc) and may give  more
information on meson--nucleon and nucleon--nucleon dynamics.

\section{Summary}
\indent

To summarize this chapter, here we have developed a topological
chiral soliton model with an explicit light scalar--isoscalar
meson field, which plays a central role in nuclear physics, based
on the chiral symmetry and broken scale invariance of QCD. We
have shown that for the single nucleon properties the successfull
description of the electromagnetic observables of the $\pirow$
model is not modified and even the value for the axial--vector
coupling is somewhat improved. In the two--nucleon sector, this
extended $\pirows$ Lagrangian leads to the correct intermediate
range attraction in the central potential and a  soft $\sigma NN$
formfactor for   both values of sigma meson mass
 $\msig=550$ MeV and   $\msig=720$ MeV.
% %%%%%%%%   end OF ANOM %%%%%%%%%%%%%%%%%%%%%%%%%

%\include{ch5npa}
\renewcommand{\fpi}{F_\pi}
%%%%%%  ReMOVE  these 2 lines if you want your fig.s (tabs) to be numbered
%%as chapter.number like V.3
\setcounter{fignum}{\value{figure}}
\setcounter{tabnum}{\value{table}}
%%%%%%%%%%%%%%%%%%%
\chapter[
Density dependence of nucleon properties  in nuclear matter. ]{}
\setcounter{figure}{\value{fignum}}
\setcounter{table}{\value{tabnum}}
\bc { \Large\bf Density Dependence of Nucleon Properties  in
Nuclear Matter. }\footnote {The  present chapter is based on
following articles by the author and his collaborators:
\ci{ourjafmed,ourjafmedff,ourprcmed, ournpamed,ourfinite}} \ec

\section{Introduction}
\indent

It is well  established that in - medium nucleon - nucleon (N-N)
cross sections manifest in heavy ion collisions need not to be
the same as their free space values \ci{cugnon,alm}. The origin
of these changes must be reflected in modifications of the N - N
interaction which is predominantly one-boson exchange. The
parameters in meson exchange models are meson nucleon coupling
constants and their masses. To avoid divergencies in loop
integrals the meson - nucleon vertices are modified by form -
factors which effectively provide cut-off parameters. All these
parameters would change in the nuclear medium.

Even for the free N-N interaction the boson exchange models use a
phenomenological ansatz for vertex form factors. Recently
Holzwarth  and Machleidt  \ci{holzmach1}  have shown that among
the QCD inspired pion - nucleon form factors the Skyrme form
factor \ci{meisff} is the most preferable: it can describe well
both the pion
 - nucleon and nucleon - nucleon systems.

The purpose of this Chapter is to investigate the role of the
medium in modifying the properties like meson nucleon coupling
constant, meson-nucleon form factors and the axial vector form
factor and the coupling constant. The calculations provide a
consistent approach to modification of these properties in a
modified Skyrme model that may be valid in a nuclear medium. The
$\sigma$-meson is introduced as a dilaton field to satisfy scale
invariance. Even though it is well known that scale invariance is
badly broken, this provides a way to introduce the $\sigma$-meson
which is so essential for an N-N interaction that is consistent
with the two-body scattering data and the bound deuteron.
%In sections II and III the model will be presented; in
%section IV the strong vertex form factors are derived and
%in sections V and VI the renormalization of masses and coupling constants
%respectively will be discussed. The summary and discussion of the
%present approach are presented in the last section.

\section{The modified Skyrme Lagrangian}
\indent

In ref.s  \ci{ourjafmed,ourprcmed} a medium modified Skyrme
Lagrangian was proposed  and applied to study the static
properties of nucleons embedded in the nuclear medium. It gave a
good description of changes in nucleon mass and its size in the
medium. Here we shall outline the basic features of the
Langrangian and then extend it by including the  dilated $\sigma$
meson field in order to satisfy the scale invariance. It is
well-known that scale invariance is badly broken, this symmetry
is retained here to investigate its consequences.

Our basic assumption in modifying of the Skyrme Lagrangian \be \ba
{\cal L}_{\mbox{sk}}=\dsf{F_{\pi}^2}{16}
\Tr(\partial_{\mu}U)(\partial^{\mu}U^+)+
\dsf{1}{32e^2}{\Tr}[U^+{\partial_{\mu}U}, U^{+}\partial_{\nu}U]^2-
\nwl
\\
\quad \quad

-\dsf{F_{\pi}^2m_{\pi}^2}{16}{\Tr}(U^{+}-1)(U-1), \lab{npa21} \ea
\ee where $F_{\pi}$ and $e$ are the parameters of the model, was
that, in the lowest order expansion of the chiral field $
U=\exp(2i\vec\tau\vec\pi /F_{\pi})\approx 1+2i (\vec \tau\vec
\pi)/{F_{\pi}} -2{{\vec{\pi}}^2}/\fpi^2 + \dots $ the
appropriate  medium modified Skyrme Lagrangian $\lsk^{*}$
 should give
the well known  equation for the pion field \ci{polar}: \be
\partial_{\mu}\partial^{\mu}\pi+(m_{\pi}^2+\hat \Pi)\pi=0,
\lab{npa22} \ee where $\mpi$ is the pion mass and $\hat \Pi$ is
the self energy operator. This may be achieved simply by
including $\hat \Pi$  in the pion mass term of the Skyrme
Lagrangian: \footnote{Here and below the asterisk indicates the
medium modified operators in quantities.} \be \lhisb^*=
-\dsf{F_{\pi}^2m_{\pi}^2}{16}{\Tr}[(U^+-1)(1+\hat
\Pi/m_{\pi}^2)(U-1)]. \lab{npa23} \ee

In general,  the operator  $\hat \Pi=\hat \Pi_s+\hat \Pi_p$ in
Eq.  \re{npa23} acts both on the center of mass coordinate, R,
and on the internal coordinate, $\vec r$, of the Skyrmion. Only
homogenous nuclear matter is considered. The translational
invariance of the medium and that of the basic Skyrme Lagrangian
\re{npa21} makes the coordinate dependence of $\hat \Pi$ simpler:
$\hat \Pi\equiv \hat \Pi(\vec r-\vec R)$ which is also relevant
for a moving Skyrmion. Further we assume that the Skyrmion is
placed at the center of the nucleus i.e. $\vec R = 0$.

The $P$ - wave part of the pion self energy $\hat \Pi_{\Delta}$
is dominated by the $P_{33}$ resonance. A simple model used in
practical calculations is the delta - hole model which
concentrates on the  pion - nucleon - delta interaction ignoring
the nucleon particle - hole excitations. In momentum space the  $
\hat\Pi_{\Delta}$ is given by $\hat \Pi_{p}(\omega,\vec k)= \hat
\Pi_{\Delta}(\omega,\vec k)= -{\vec k^2\chi(\omega,\vec
k)}/(1+\gzero\chi(\omega,\vec k)) $ \ci{polar} where $\gzero$ is
the Migdal parameter which accounts for the short range
correlations, also known as the Ericson -  Ericson -  Lorentz -
Lorenz effect. The pion susceptibility $\chi$  has nearly linear
dependence on the nuclear density $\rho$: \be \chi(\omega,\vec
k)\approx\dsf{8}{9}\left(\dsf {f_{\pi N\Delta}}{\mpi}\right)^2
\dsf{\rho\omega_{\Delta}} {\omega_{\Delta}^2
-\omega^2}\exp(-2\vec k^2/b^2), \ee where $f_{\pi N\Delta}\approx
2 f_{\pi NN}$  ( $f^2_{\pi NN}/4\pi\approx 0.08$) is the coupling
constant, $\omega_{\Delta}\approx \vec
k^2/2M_{\Delta}+M_{\Delta}-M_N$ and $b$ $(b\approx 7\mpi)$ is the
range of the vertex form factor that is chosen to have a Gaussian
form. For the pions bound in the nuclear matter $\omega$ is small
$0\le\omega\le\mpi$, so in the lowest order $\hat \Pi_{\Delta}$
has the form: \be \disc{\Pi_\Delta}(\omega,\vec{k})\approx - \vec
k^2\chi_{\Delta}-\dsf{{\vec
k}^2\omega^2\alpha_{rt}}{\omega_{\Delta}^2}, \lab{npa25} \ee where
\be \ba \chi_{\Delta}={4\pi c_0\rho}/{(1+4\pi\gzero\rho)}, \nwl
\alpha_{rt}={\chi_{\Delta}}/{(1+4\pi\gzero c_0\rho)},\nwl
c_0={8}{f_{\pi N\Delta}^2}/{9m_{\pi}^2}{4\pi \omega_{\Delta}}\;\;.
\ea \ee In the static case $(\omega\rightarrow 0)$ it coincides
with the Kisslinger optical potential \ci{polar}  which was used
in ref. \ci{ourprcmed}.

Then in coordinate space the following symmetrized Lagrangian may
be obtained \be \ba \lsk^*=\dsf{\fpi^2}{16}\Tr
\left(\dsf{\partial U}{\partial t} \right)\left(\dsf{\partial
U^+}{\partial t}\right)-\dsf{\fpi^2}{16} (1-\chi_{\Delta})
\Tr(\vec \nabla U^+)(\vec \nabla U)+\\
\quad \\
+\dsf{\fpi^2}{16}\dsf{\alpha_{rt}}{\omega_{\Delta}^2}
\Tr\left(\dsf{\partial }{\partial t} \dsf{\partial U^+ }{\partial
r} \right) \left(\dsf{\partial }{\partial t} \dsf{\partial  U}{
\partial{r}}\right)
+\dsf{\fpi^{2} m_{\pi}^{*2}    }{16}   \Tr (U+U^+-2)+\\
\quad\\
+\dsf{1}{32e^2}{\Tr}[U^+{\partial_{\mu}U},
U^{+}\partial_{\nu}U]^2 , \lab{npa26} \ea \ee where
$m_{\pi}^{*2}=\mpi^2(1+\Pi_S(\rho)/\mpi^2)$, the effective pion
mass arises from the $S$ - wave part of the self energy
$\Pi_S(\rho) $. Since the operators in nuclear matter do not
satisfy the Lorentz invariance there  are two different effective
coupling  constants  in the kinetic terms of Eq. \re{npa26}
\footnote{The similar second order derivative terms had been
suggested in ref \ci{wirsba}. }.
 The third term with "mixed" derivative in the Lagrangian
in Eq. \re {npa26} clearly vanishes in
 free space  ($\alpha_{rt}=0$  if $\rho=0$)
and contributes mainly to the moment of inertia, $ I $, of the
skyrmion. We shall consider this contribution in Sect. 5 of the
present Chapter by estimating the N$\Delta$ mass splitting.

\section {Inclusion of scalar meson}
\indent

Recently arguments have been given in favor  of the empirical
evidence for  a scalar-isoscalar meson, $\sigma$. For example the
isosinglet resonance  with a mass $m_{\sigma}=553.3\pm 0.5$ MeV
and width $\Gamma=242.6\pm 1.2 MeV$ \ci{ishida1,ishida2} found in
the recent  $\pi \pi$  phase shift analyses is  believed to have
the properties of the $ \sigma$ meson   that is essential for the
N-N potential in OBE  models to explain the NN-scattering data.

However  the $\sigma $  meson   as a chiral partner  of the pion
was originally  excluded from the Skyrme type  Lagrangians. The
only way of  including $\sigma$ meson  in the Lagrangian is by
means of a dilaton field  appropriate to the scale transformations
$x^{\mu}\rightarrow e^{\mu}x^{\mu}$. This procedure has been
outlined in  Chapter 1.

Now a skyrmion imbedded  in nuclear matter is considered. When a
dilaton field is introduced into the effective Lagrangian the
dilaton potential $V({\sigma})$ would be modified by a $\sigma$
field generated by the medium itself. While several attempts have
been made \ci{kalfluid}, the correct nature of this modification
is still poorly understood. On the other hand it is well known
that the gluon condensate $C_g$ as well as the quark condensate
$\langle\bar{q}q\rangle$  decrease in the medium due to the
partial restoration of the chiral symmetry. The present study is
restricted to considering the medium modification of the dilaton
potential, Eq. \re {vsig}, by taking into account mainly the
change of $C_g$ i.e. $C_g \rightarrow C_g^*$.

Thus putting all these considerations together the following
Lagrangian is proposed for homogeneous nuclear medium in the
static case \be \ba \lsk^*=\dsf{\fpi^2\alpha_p\chi^2}{16}\Tr \vec
L_i^2+ \dsf{1}{32e^2}{\Tr}[\vec L_i,\vec L_j]^2+
\dsf{\fpi^2m_{\pi}^{*2}}{16}\chi^3\Tr (U+U^+-2)-\\
\quad\\
-\dsf{\fpi^2}{8}(\vec\nabla\chi)^2+
\dsf{C_g^*}{24}\left[1-\chi^4-\dsf{4}{\veps}
(1-\chi^{\veps})\right], \lab{npa32} \ea \ee where
$\chi(r)=e^{-\sigma(r)}$, $\alpha_p=1-\chi_{\Delta}$,
 $\vec L_i=U^+\partial_iU$.
Note that the Skyrme parameter $e$ coupled to the fourth
derivative term  remains unchanged since this term is related to
the exchange of a very heavy $\rho$ - meson with mass
$m_{\rho}^*=m_{\rho}\rightarrow \infty$ \ci{bhaduri}.

There may be two alternative approaches to applying this
Lagrangian in nuclear physics. In Quantum Hadrodynamic models,
the $\sigma$ field plays the role of an external field modifying
the properties of a soliton \ci{kalfluid} or a bag
\ci{saitothomas} which moves in the background generated by the
medium. In the present model this approach would mean that
$\sigma\equiv \sigma(R)$ and $U\equiv U(r)$ \ci{kalfluid} that
makes the scale invariance doubtful. In contrast, the mean field
approximation is not used for the $\sigma$ field. Instead, the
$\sigma$-field is strongly coupled to the nonlinear pion fields
so as to generate the soliton . We assume that
$\sigma\equiv\sigma(r-R)$ and  $U=U(r-R)$ for a moving skyrmion.
In ref.  \ci{ourprcmed} a similar approximation (Eq. \re {npa32}
with $\sigma=0)$ was  used to estimate the medium modified static
properties of the nucleon and found a well known behavior of the
nucleon mass $M_N^*/M_N<1$ and its size $R^{*}_{N} / R_{N} > 1 $.
In the next section we shall investigate in detail the dynamical
properties of the meson - nucleon system.

\section{Meson - nucleon form factors and Goldberger - Treiman relation}
\indent

The semiclassical procedure for calculating the meson - nucleon
vertex form - factors in a topological chiral effective
Lagrangian \ci{cohen,meisff} is well-known. In fact, the results
of  more accurate methods \ci{verch,uehara}, based on the correct
quantization of the fluctuating chiral fields nearly coincide
with the original result that was given by Cohen et al.
\ci{cohen}. Using the ansatz $U(\vec r ,t)=A(t)U_0(\vec r-\vec
R(t))A^+(t)$
 and defining the pion field as
\be \pi_{\alpha}(\vec r)=-\dsf{iF_{\pi}}{4}\Tr
[\tau_{\alpha}AU_0(\vec r- \vec R)A^+] \lab{npa41} \ee the
following expressions are obtained \be \ba G^{*}_{\pi
NN}(q)=\dsf{4\pi M_N^*F_{\pi}\alpha_p(\vec q^{\ 2}+
m_{\pi}^{*2}/\alpha_p)}{3q}\int\limits_0^{\infty} j_1(qr)
\sin(\theta)r^2dr=\\
=\dsf{4\pi M_N^*F_{\pi}\alpha_p}{3}\int\limits_0^{\infty}
\dsf{j_1(qr)}{qr}S_{\pi}(r)r^3dr \lab{npa42} \ea \ee for the pion
nucleon form factor and \be \ba G^{*}_{\sigma NN}(q)=2\pi
F_{\pi}(\vec q^{\ 2}+m_{\sigma}^{*2}) \ds\int\limits_0^{\infty}
j_0(qr)\sigma(r)r^2dr= \nwl \quad \quad =2\pi
F_{\pi}\int\limits_0^{\infty}j_0(qr)S_{\sigma}(r)dr \lab{npa43}
\ea \ee for the sigma nucleon form factor respectively. The
details and the explicit expressions for the source functions
$S_{\pi}(r)$ and $S_{\sigma}(r)$
 are given in the Appendix A.
Here we note that, in the chiral soliton models formulas for
meson nucleon form - factors are mainly determined by the
quantization scheme rather  than by  details of the Lagrangian.
The latter manifests itself through equations of motion whose
solutions $\theta(r)$ and $\sigma(r)$ with spherically symmetric
 ansatz are
$U_0=\exp({i\vec{\tau}\vec{n}\theta(r)})$, $\vec{n}=\vec{r}/r$,
$\sigma(\vec r)\equiv\sigma(r)$ that should be used in Eq.s \re
{npa42}, and \re {npa43}. The effective mass of the nucleon
$M_N^*$ is given by \be M_N^*=M_H^*+\dsf{3}{8I^*} \lab{npa44} \ee
where $M_H^*$ is the mass of the classical hedgehog soliton and
$I^*$ is the moment of inertia  of the spinning mode.

The effective pion decay constant $f_{\pi}^*$ should be
understood before considering the Goldberger - Treiman (GT)
relation. Medium renormalized pion - decay constant $f_{\pi}^*$
can be naturally defined by the PCAC relation: \be \vec \nabla
\vec A_{\alpha}(x)=f_{\pi}^*m_{\pi}^{*2}\pi_{\alpha}(x)
\lab{npa45} \ee The medium modified axial coupling constant
$g_A^*$ measures the spin - isospin correlations in a nucleon,
embedded in a medium and is defined as the expectation value of
the space component of the axial current $A^{i}_{\alpha} $
 in the nucleon state at zero momentum transfer
\ci{skpi1,skpi2}: \be \lim_{q\rightarrow
0}\bra{N(\vec{P'})}A^{i}_{\alpha}(r)\ket{N(\vec{P})}=
\dsf{2}{3}\lim_{q\rightarrow 0}G^{*}_{A}(\vec{q}\ ^2)\bra{N}
\frac{\sigma_{i}\tau^{\alpha}}{2}\ket{N}\exp(i\vec{q}\vec{r})
\lab{npa46} \ee where $\vec{q}=(\vec{P'} - \vec{P})$ and
$G_A^*(\vec{q}\ ^2)$ is the axial form factor of nucleon,
$G_A^*(0)=\gastar$. Here $\sigma_{i}$ is the component of the
nucleon spin. Due to the semiclassical quantization prescription,
Eq. \re {npa41}, the matrix element of the pion field evaluated
between nucleon states is given by: \be
\bra{N(P')}\pi_{\alpha}(\vec{r}-\vec{R})\ket{N(P)}=
\frac{\fpi\exp{(i\vec{q}\vec{r})}}{6}{\int}d\vec{x}
e^{-i\vec{q}\vec{x}}\sin(\theta)
\bra{N}\sigma_{\alpha}(\vec{\tau}\hat{x})\ket{N} \lab{npa47} \ee
Evaluation of the matrix elements between nucleon states for both
sides of Eq. \re {npa45} yields \be \gastar=\dsf{4\pi\fpi
f_{\pi}^{*}m_{\pi}^{*2}}{9} \int\limits_0^{\infty}
\sin(\theta)r^3dr \lab{npa48} \ee that was originally derived in
\ci{workman} for a free particle. By comparing this equation with
the expression for pion nucleon coupling constant:
$\gpinnstar=G^{*}_{{\pi}NN}(q^2)|_{q=0} $
 given by Eq. \re {npa42}
the following  medium modified Goldberger - Treiman relation is
realized \be \disc{g_{\pi NN}^{*}}f_{\pi}^{*}=\gastar \mn^* \quad
. \lab{npa49} \ee Although this relation has been suggested
earlier in ref.s
 \ci{meissmed,br} it has not been proved yet.
On the other hand $\gastar$ and the axial form factor $G_A(q^2) $
may be calculated directly from the Lagrangian in Eq. \re {npa32}
in terms of the Noether currents \ci{skpi1,skpi2}. This gives
$$
G_A(q^2)=
-4\pi\int\limits_0^{\infty}[j_0(qr)A_1(r)+\dsf{j_1(qr)}{qr}A_2(r)]r^2dr,
$$
with \be \ba
A_{1}(r)=\dsf{s_2}{8r}\left[e^{-2\sigma}\alpha_pF_{\pi}^2+
\dsf{4}{e^2}\left(\theta^{\prime 2}+d\right)\right],\\
\quad \\
A_{2}(r)=-A_{1}(r)+\dsf{\theta^{\prime}}{4}\left(\fpi^2\alpha_p
e^{-2\sigma}+\dsf{8d}{e^2}\right) \lab{npa410} \ea \ee and \be
g_A^*=-\dsf{\pi}{3}
\int\limits_0^{\infty}\left\{e^{-2\sigma(r)}F_{\pi}^2\alpha_p
\left(\theta^{\prime}+\dsf{s_2}{r}\right)+\dsf{4}{e^2}
\left[(\theta^{\prime 2}+d)\dsf{s_2}{r}+
2\theta^{\prime}d\right]\right\} r^2dr \lab{npa411} \ee where
$d=\sin^2(\theta)/r^2$, $s_2=\sin(2\theta)$, and $\alpha_p$
 is defined in Eq. \re {npa32}.
The renormalized pion decay constant $f_{\pi}^*$ is obtained by
combining the results given in Eq.s \re{npa49} and \re{npa411}.

\section{ Renormalization of hadron masses}
\indent

Before going to a quantitative analyses of the medium effects we
fix the following set of parameters for free space:
$F_{\pi}=2f_\pi=186MeV$, $m_{\pi}=139MeV$, $C_g=(260MeV)^4$.
 This gives a good description  of the
sigma meson properties $m_{\sigma}=550MeV$,
$\Gamma_{\sigma}=251.2MeV$, that may be compared with their
experimental values obtained from the recent $\pi\pi$ phase shift
analysis \ci{ishida1,ishida2}. The Skyrme parameter $e$ has been
adjusted to reproduce the pion nucleon coupling constant: $g_{\pi
NN}=13.5$ for $ e=4.05$. The well-established fixed parameters of
the $P$ - wave pion self energy in  Eq. \re {npa25} are used in
the pion sector: $\gzero=0.6$, $c_0=0.13m_{\pi}^{-3}$
\ci{ourprcmed,polar}.

Now, the medium dependence of the input parameters may be
considered. The possible renormalization of the Skyrme parameter
$e$ cannot be studied in the present approach unless the $\rho$
meson is included in the Lagrangian explicitly. So we take
$e^*=e$.

We adopt the following parametrization of $m_{\pi}^*$ \be
m_{\pi}^*=m_{\pi}\sqrt{1+\hat{\Pi}_{s}(\rho)/m_{\pi}^{2} }=
m_{\pi}\sqrt{ 1-4{\pi}b_{0}\rho\eta/m_{\pi}^{2}} \ , \lab{npa51}
\ee where $\eta=1+\mpi/\mn $ and $b_0$ is an effective S - wave
$\pi - N$ scattering length. It is anticipated \ci{ourprcmed}
that the results will not be too much sensitive to the value of
$b_0$.

The only  input parameter in the scalar meson sector is $C_g^*$.
The medium renormalization of the gluon condensate $C_g^*$, in
contrast with the renormalization of the quark condensate
$\langle\bar{q}q\rangle$
 \ci{birserev} and meson masses \ci{ko},
is poorly known. However in the present approach \re{npa32}
$C_g^*$  may be determined  by $m_{\sigma}^*$  through the
equation \be
 C_g^*=\dsf{ 3\fpi^2m_{\sigma}^{*2} N_f}{4(4-\veps)}.
\lab{npa52} \ee Various approaches \ci{saitothomas,meissmed, br}
show that $m_{\sigma}^*$, has a linear density dependence. The
following parametrization \be
\dsf{m_{\sigma}^*}{m_{\sigma}}=1-0.12\dsf{\rho}{\rho_0}
\lab{npa53} \ee is adopted here. It is consistent with the one
obtained in the Quark coupling model (QCM)  of framework
\ci{saitothomas}.

The results for the static properties of hadrons are presented in
Table \ref{npatab1}. The  second ($m_{\pi}^*$) and the third
$(m_{\sigma}^*)$ columns  of  the table should be considered as
input data for they were taken from other models
\ci{polar,saitothomas}. The medium renormalized gluon condensate
$C_g^*$ is calculated from Eq.s \re {npa52} and \re{npa53} with
$N_f=2$, $\varepsilon=8N_f/(33-2N_f)$. The change in the gluon
condensate is small $\sim 5\%$  at  normal nuclear matter
density, $\rho_0=0.5m_{\pi}^3$. The stiffness of the gluon
condensate  as a consequence of the lack of scale invariance of
QCD has been shown by Cohen \ci{cohencg} who found that the fourth
root of the condensate might be altered by no more than $4\%$
\ci{birserev}.

%%%%%%%%%%%%%%%%%%%%%%%%%%%%%%%%%%%%%%%%%%%%%%%%%%
\begin{table}[tbp]
\caption{\it Density dependence of hadrons properties. All
quantities are in $MeV$.} \bc
\begin{tabular}{cccccccc}
$\rho/\rho_0$&$ m_{\pi}^*$&$m_{\sigma}^*$&$(C_g^*)^{1/4}$&$M_N^*$&
$\Gamma_{\sigma\rightarrow\pi\pi}^*$&$\delta M_{N\Delta}^a$&
$\delta M_{N\Delta}^b$\\
\hline
0.0&139.00&550.1&260.70&1413&251.2&283.7&283.7\\
0.5&144.90&513.8&251.06&1271&88.7 &200.1&238.1\\
1.0&149.06&493.8&246.12&1157&34.6 &161.9&205.4
\end{tabular}
\ec \lab{npatab1}
\end{table}
%\large
%%%%%%%%%%%%%%%%%%%%%%%%%%%%%%%%%%%%%%%%%%%%%%%%%%

The main contribution to the $\sigma\pi\pi$ vertex,
 and  hence, to the decay width of $\sigma $ meson at the tree level:
$ \Gamma_{\sigma\rightarrow \pi\pi}={m_{\pi}^{*3}x^3\alpha_p^2
\sqrt{1-4x^2}(1-2x^2)^2}/{4\pi F_{\pi}^2}, $ where
$x=m_{\pi}^*/m_{\sigma}^*$, arises from the first term of the
Lagrangian in Eq. \re {npa32}. The Table \ref{npatab1} shows that
the  width
 $\Gamma_{\sigma\rightarrow \pi\pi}$ is decreased significantly
in the medium. This stimulates an interest to observe the
$\sigma$ mesons in nuclei by experiments proposed recently by
Kunihiro et el. \ci{kunihiro}.

Now medium effects on the mass of nucleon $M_N^*$ and $N\Delta$
mass splitting $\delta M_{N\Delta}= M_{\Delta}^*-M_N^*$ will be
considered. In general, it is almost impossible to reproduce
simultaneously the experimental values of
 masses and coupling
constants within the Skyrme model even for a free particle. Since
dynamics is the main interest, the set of parameters, was chosen
so as to reproduce the pion nucleon coupling constant:
$\gpinn=13.5$. It is clear from Table \ref{npatab1}
 that the free space value of the nucleon
mass $M_N$ is slightly large $\mn=1413 MeV$, whereas $\delta
M_{N\Delta}$
 is reproduced
rather well $\delta M_{N\Delta}=284MeV$ ($\delta
M_{N\Delta}^{\mbox{exp}}=293 MeV)$.
 The effective mass of the nucleon $\mn^*$
 in normal nuclear matter density is
decreased by a factor of $ \mn^*/\mn=0.82  $ which is in a good
agreement with the estimates based on QCD sum rules
$M_N^*(QCD)=680\pm 80MeV$  i.e.
 $\mn^*/\mn=0.72\pm0.09 $ \ci{furnstahl}.
We underline that the mass of the nucleon should be treated as
the mass of the baryon which  emerges as a soliton in the  sector
with baryon  number one $(B=1)$.

The study of $N\Delta $ mass splitting gives  a chance to
estimate the contribution from the "mixed derivative" term - the
third term on the r.h.s of Eq. \re {npa26}: \be {\cal L}_{rt}=
\dsf{\fpi^2}{16}\dsf{\alpha_{rt}}{\omega_{\Delta}^2}
\Tr\left(\dsf{\partial }{\partial t} \dsf{\partial U^+ }{\partial
r} \right) \left(\dsf{\partial }{\partial t} \dsf{\partial  U}{
\partial{r}}\right) \lab{npa54}. \ee

 Actually, owing to the canonical quantization
$\delta M_{N\Delta} $ is related to the moment of inertia $I$ by:
$ \delta M_{N\Delta}=M_{\Delta}^*- M_{N}^*=3/2(I^{*}_{0}+I_{rt}),
$ where  $I_{rt}$  is the net contribution from ${\cal{L}}_{rt}$
(clearly $I_{rt}=0$ in free space). The explicit expressions for
 $I^{*}_{0}$ and $I_{rt}$ are given in the Appendix A.
In Table \ref{npatab1} are shown  $\delta M_{N\Delta}$ that has
been   calculated  with the inclusion of $I_{rt}$ (denoted here
$\delta M_{N\Delta}^a$) and without the inclusion of
$I_{rt}(\delta M_{N\Delta}^b)$. In the nuclear medium the
${\cal{L}}_{rt}$ term leads to an enhancement of  the moment of
inertia  decreasing $\delta M_{N\Delta}$ significantly. Even
without the term ${\cal{L}}_{rt}$ in the Lagrangian the shift of
$\delta M_{N\Delta}$ from its free value $\delta
M_{N\Delta}^*/\delta M_{N\Delta}=0.75$ is larger than that
obtained by Meissner \ci{meissmed} in a medium modified chiral
soliton model based on the Brown Rho (BR) scaling law ($\delta
M_{N\Delta}^*/\delta M_{N\Delta}=0.87)$.

\section{Renormalization of coupling constants and form - factors}
\indent

The axial - vector exchange currents
 must be considered in order to investigate the medium
effects on $g_A$ and on the axial form factor $G_A(q^2)$. However
it is known that the bulk of exchange current effects arise from
the $\triangle$-hole contributions. Including such $\triangle$-h
effects would imply that bulk of the exchange currents effects
are included in the effective $g_{A}$. In a heavy nucleus it is
meaningful to take these axial exchange operators into account as
corrections to the \underline{effective} axial current operator
of a single nucleon $\vec A_{\alpha}=-g_A^*\vec
\sigma\vec\tau_{\alpha}$. So $g_A^*$ in Eq. \re {npa411} may be
considered as an effective axial coupling constant modified by
the medium polarization and screening effects since we are
considering an effective one body problem of a nucleon embedded
in the nuclear medium.

The second column of Table \ref{npatab2} displays a well known
quenching behavior of $\ga$ that is  mainly caused by a
  factor $\alpha_p=1-\chi_{\Delta} < 1 $ in the
first term of Eq. \re {npa32}. Note that the same set of input
parameters $\fpi, e, c_0$ (but without dilaton field $\sigma=0) $
gives the desired ratio $\gastar/\ga=0.8 $ \ci{ourprcmed}. The
present calculations show that the inclusion of the scalar meson,
which induces an additional attraction,  prevents larger
quenching: $\gastar/\ga=0.9 $.

%%%%%%%%%%%%%%%%%%TABLE 2%%%%%%%%%%%%%%%%%%%%%%%%%%%%%%%%%%%%%
\normalsize
\begin{table} [tbp]
\caption{\it Coupling constants and cut-off parameters at finite
density. (All values are normalized to their free space ones.)}
\bc
\begin{tabular}{cccccccc}
$\rho/\rho_0$& $g_{A}^*/g_{A}$& $g_{\pi NN}^*/g_{\pi NN}$&
$g_{\sigma NN}^*/g_{\sigma NN}$& $f_{\pi}^*/f_{\pi}$&
$\Lambda_{\pi}^*/\Lambda_{\pi}$&
$\Lambda_{\sigma}^*/\Lambda_{\sigma}$&
$r_a^*/r_a$\\
\hline
0.5&0.96&0.91&0.88&0.98&0.70&0.90&0.93\\
1.0&0.92&0.80&0.78&0.94&0.56&0.84&0.55
\end{tabular}
\ec \lab{npatab2}
\end{table}
%\large
%%%%%%%%%%%%%%%%%%%%%%%%%%%%%%%%%%%%%%%%%%%%%%%%%%%%%%%%%%%%%%%%%

In nuclei, a nucleon polarizes the medium in its vicinity. This
leads to a screening effect that reduces the effective pion
nucleon coupling strength. In the present approach the screening
mechanism may be described as being due to virtual $\Delta h$
excitations that have been taken into account by the self -
energy $\disc{{\hat\Pi}_\Delta}$ term in Eq. \re {npa25}. At
normal  nuclear matter density the renormalization of $g_{\pi
NN}$ amounts to a reduction of $25\%$ of the coupling strength.
This is sufficient to explain the quenching of the Gamov - Teller
strength in heavy nuclei \ci{khannatowner}. Furthermore this is
consistent with a general argument based on Ward-Takahashi
relations \ci{xzh}.

The effective pion decay constant $f_{\pi}^*$  is obtained by
using the GT relation  \re {npa49}. Considering the ratios
$f^{*}_{\pi}/f_{\pi}$ (Table \ref{npatab2})
 and $M_N^*/\mn$ (Table \ref{npatab1})
it may be concluded that they both decrease in the nuclear medium.
It would be interesting to compare these ratios with well known
Brown-Rho scaling law \ci{br} $f^{*}_{\pi}/f_{\pi} =\mn^*/\mn$
 predicted by
a simple Skyrme model.   As it is seen from Tables \ref{npatab1}
and \ref{npatab2} in our model $f^{*}_{\pi}/f_{\pi}
\neq\mn^*/\mn$. Clearly, this deviation from  Brown-Rho scaling
law is caused by the the additional terms including  the  dilaton
field.

%  REVISED IN NPA
%
% which has been derived in a simple Skyrme model,
% does not hold in the present approach. Due to  the change in mass is larger
%than that of the pion decay coupling constant.
%
%It is well known that
%if meson masses   decrease at the same rate in accordance
%with the BR scaling law then the nuclear saturation cannot be reached
%at the empirical nuclear matter density \ci{birserev,rappmach}.
%Even modest changes in meson masses or coupling constants produce large
%changes in the nuclear saturation.
%For instance a decrease of the sigma meson mass with density
%can prevent nuclear matter from saturation. Therefore
%new mechanisms for saturation, such as a decrease in the
%$\sigma N$ coupling constant, have to be considered.
%The ratio of $\gsnnstar$ to its free value  is presented in Table 2.
%The changes
%in $\gpinn$ and $\gsnn$ are nearly the same. Both are reduced
%in the medium by $\approx 25\%$ at $\rho=\rhozero$.

The masses of the omega and  sigma mesons are supposed to decrease
by the scaling law.There are experimental indications that this
is true.In the same way the $\sigma -  N$ coupling constant is
expected to decrease. The ratio of $\gsnnstar$ to its free value
is presented in
 Table \ref{npatab2}.
The changes in $\gpinn$ and $\gsnn$ are nearly the same. Both are
reduced in the medium by $\sim 25\%$ at $\rho=\rhozero$.
%%%%%%%%%%%%%%%%%%%%%%%%%%%%%%%%%%%%%%%%%%%%%%%%%%%%%%%%%%%%%%
\begin{figure}[!htb]
\bc \epsfysize=14cm \epsfbox{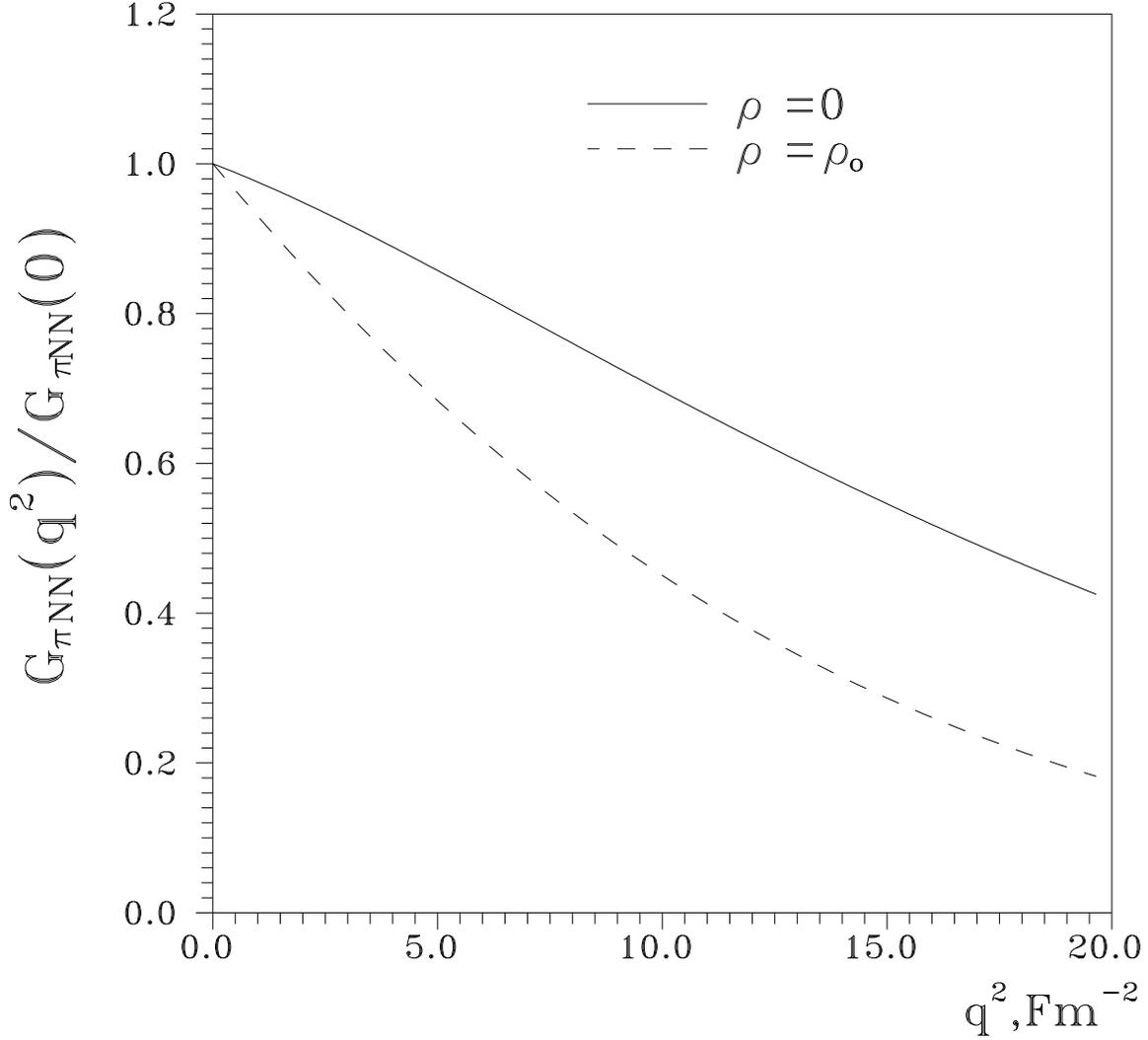} \ec
   \centerline{\parbox{15cm}
   {\caption{\label{npafig1}
 The $\pi NN$ form factor.The solid and dashed curves
give results for the free space ($\rho=0$) and the nuclear matter
($\rho=\rho_0)$ respectively.}}}
\end{figure}

\begin{figure}[htb]
\bc \epsfysize=14cm \epsfbox{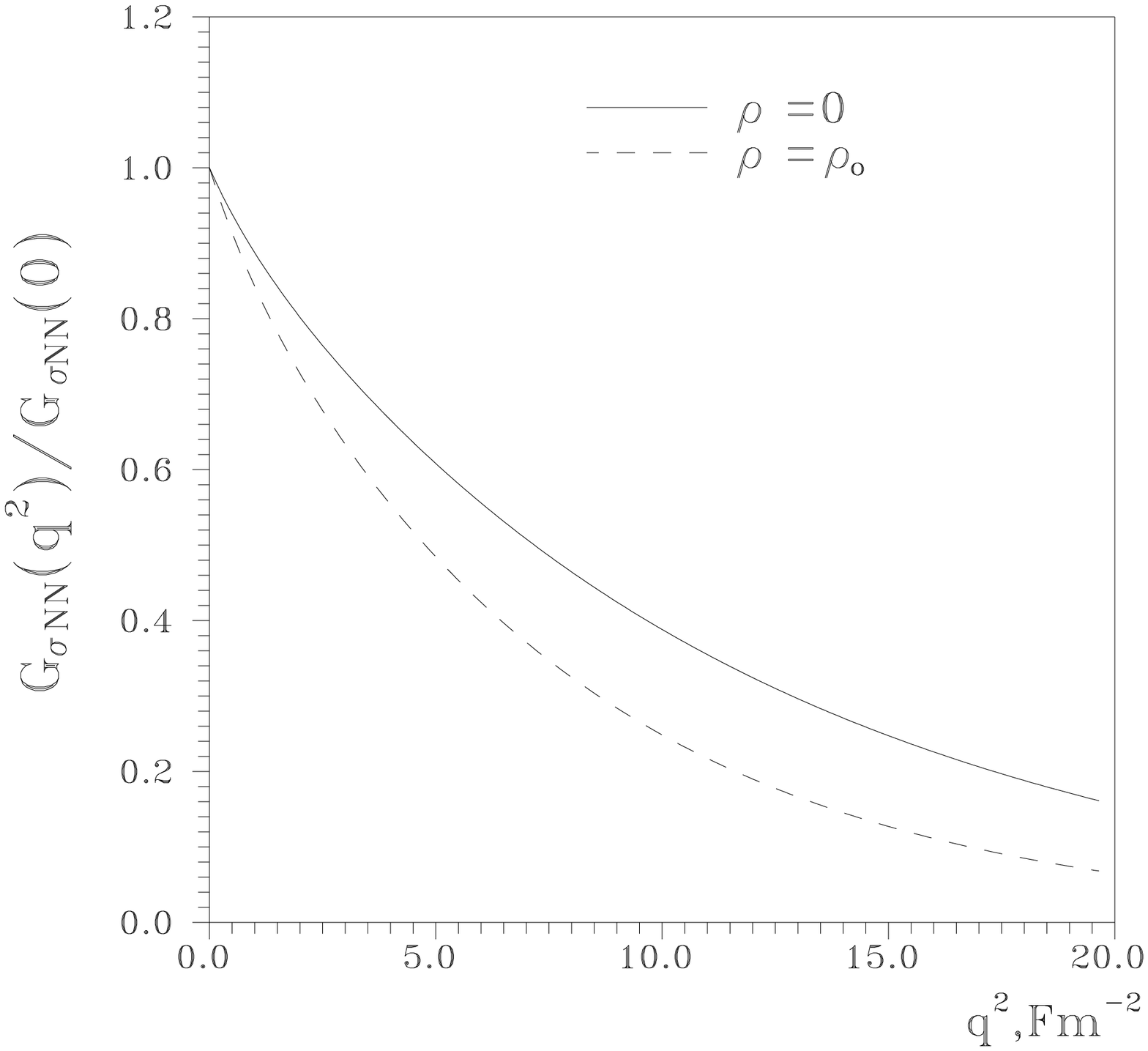} \ec
   \centerline{\parbox{10cm}
   {\caption{\label{npafig2}
The $\sigma NN$ form factor. Solid and dashed curves are for
$\rho = 0$ and $\rho = \rho_0$ respectively. }}}
\end{figure}

%%%%%%%%%%%%%%%%%%%%%%%%%%%%%%%%%%%%%%%%%%%%%%%%%%%%%%%%%%%%%

In Figs. \ref{npafig1}  and \ref{npafig2},
 the renormalized
 $\pi NN$  and  $\sigma NN$ vertex
form factors respectively at $\rho=\rho_0$ (dashed lines) in
comparison with these in the free space (solid lines) are
displayed. Appreciable quenching of both form factors is
observed. At small momentum transfer these can be parametrized by
a monopole form i.e. $G_{\pi NN}(\vec q\ ^2)=\gpinn/(1+\vec q\
^2/\Lambda_{\pi}^2)$ and $G_{\sigma NN}(\vec q\ ^2)=\gsnn/(1+\vec
q\ ^2/\Lambda_{\sigma}^2)$. Table \ref{npatab2}
 shows that the cut off parameter $\Lambda_{\pi}$ is decreased
significantly at $\rho=\rho_0$. Relatively small changes in
$G_{\sigma NN}(\vec q\ ^2)$ seem to be caused by a stiffness of
the $ \sigma $-field or equivalently
  $C_g^*$  \ci{birserev}.

\begin{figure}[htb]
\bc \epsfysize=14cm \epsfbox{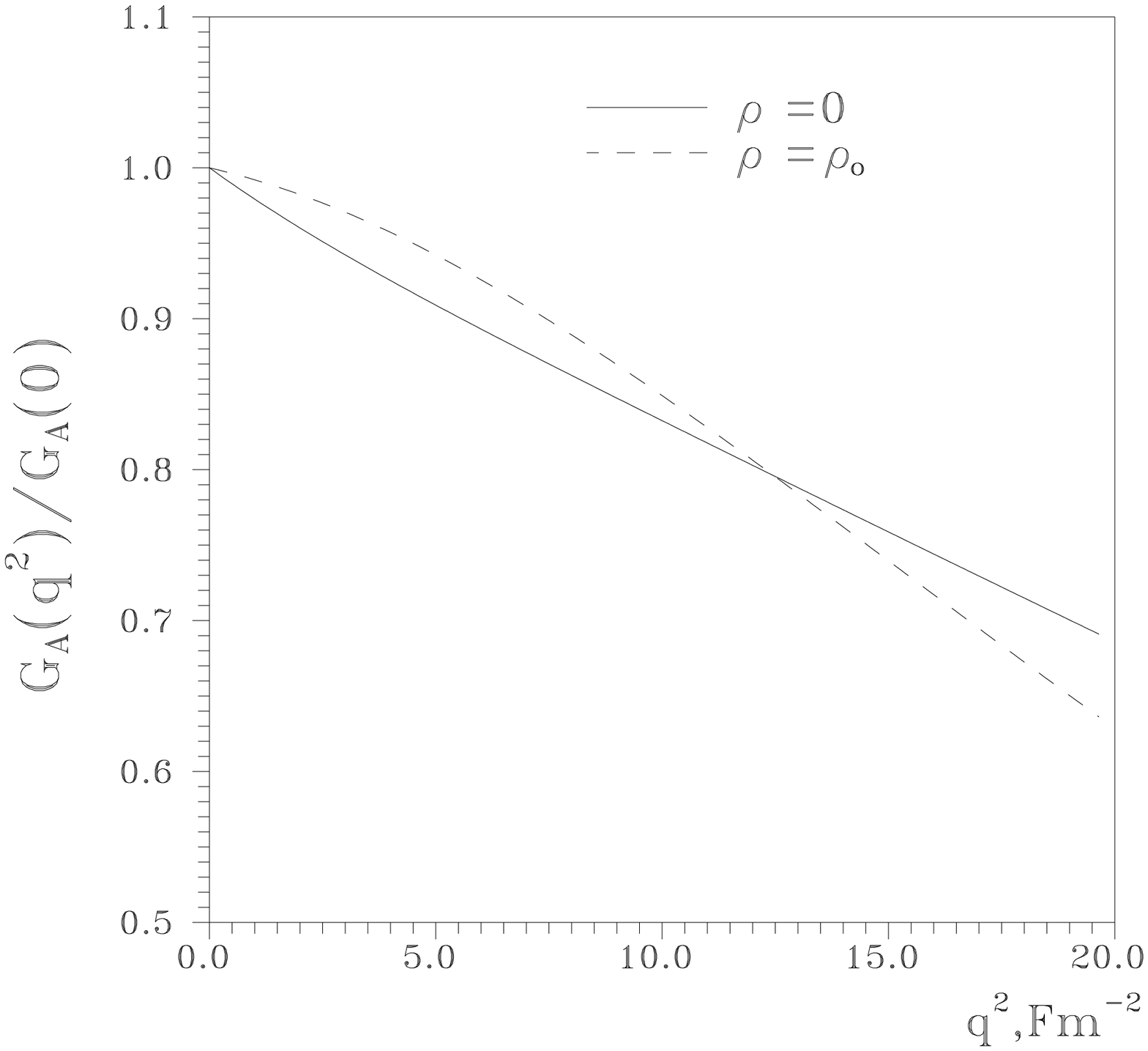} \ec
   \centerline{\parbox{15cm}
   {\caption{\label{npafig3}
The axial form factor. Solid and dashed curves are for $\rho = 0$
and $\rho = \rho_0$ respectively. }}}
\end{figure}

The nucleon axial form factor $G_{A}(\vec{q}^{\
2})/G_{A}(0)$calculated for $\rho=0$ and
 $\rho=\rhozero$  is presented in Fig. \ref{npafig3} with solid and
dashed curves respectively. It is seen that, the modification of
$G_{A}(\vec{q}^{\ 2})$ is not as simple as that of $G_{\pi
NN}(\vec{q}^{\ 2})$. The medium leads to a quenching of the meson
nucleon form factors  over a range of $\vec{q}^{\ 2}$, while the
quenching of $G_{A}(\vec{q}^{\ 2})$ takes place at $\vec{q}^{\ 2}
=0$ and $\vec{\ q}^{2} > 10 fm^{-2}$.

%%%%%%%%%%%%%%%%%%%%%%%%%%%%%%%%%%%%%%%%%%%%%%%%%%%%%%%%%%%%%%%%%%%%

%%%%%%%%%%%%%%%%%%%%%%%%%%%%%%%%%%%%%%%%%%%%%%%%%%%%%%%%%%%%%%%%%%%

\begin{figure}[htb]
\bc \epsfysize=14cm \epsfbox{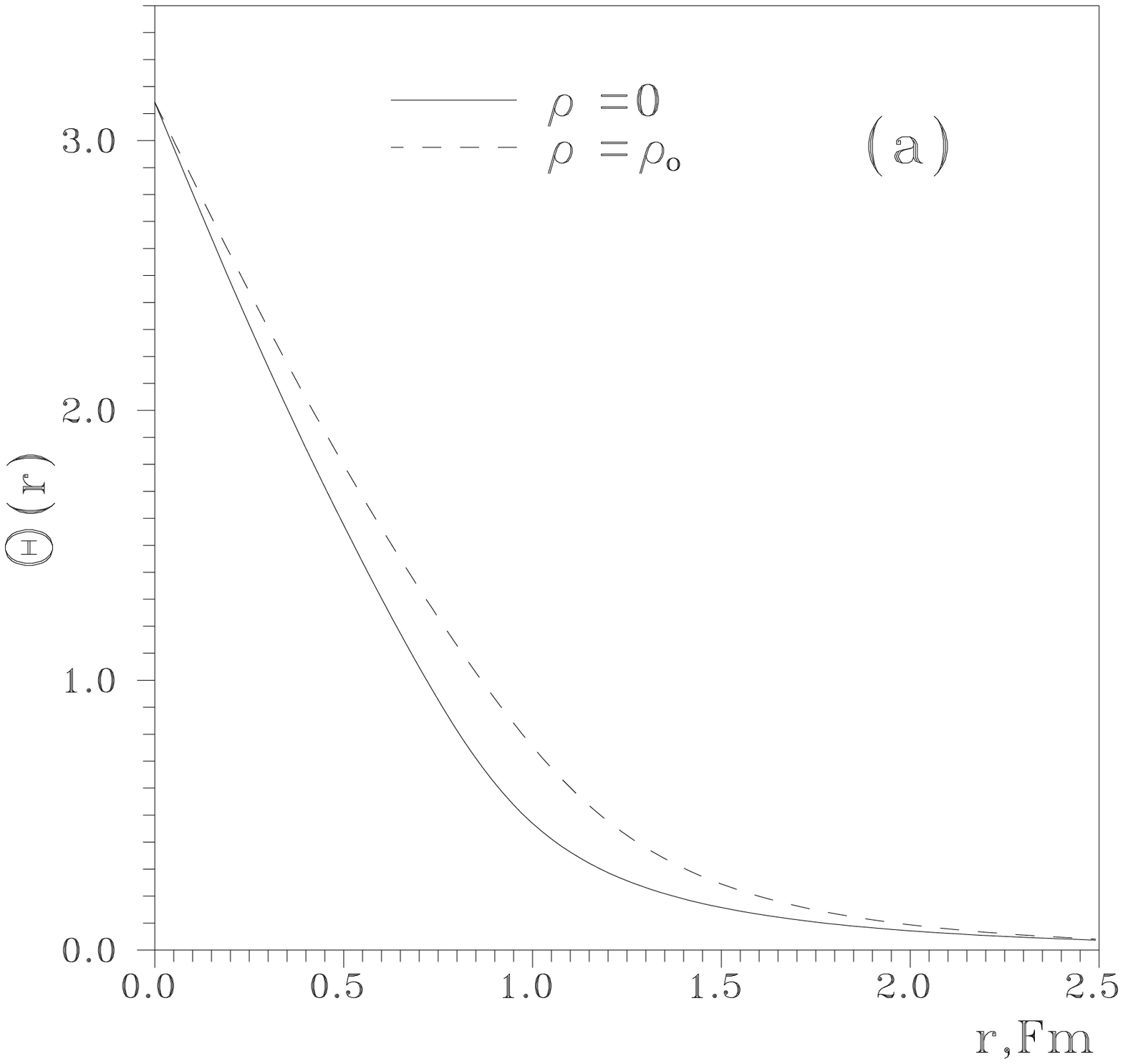} \ec
   \centerline{\parbox{15cm}
   {\caption{\label{npafig4a}
 The profile function $\theta(r) $
 for the $\rho=0$ (solid curve) and
$\rho=\rhozero$ (dashed curve). They are the solutions
 of equations of motion  \re{appeneq}  in the sector with $B=1$.
}}}
\end{figure}

\begin{figure}[htb]
\bc \epsfysize=14cm \epsfbox{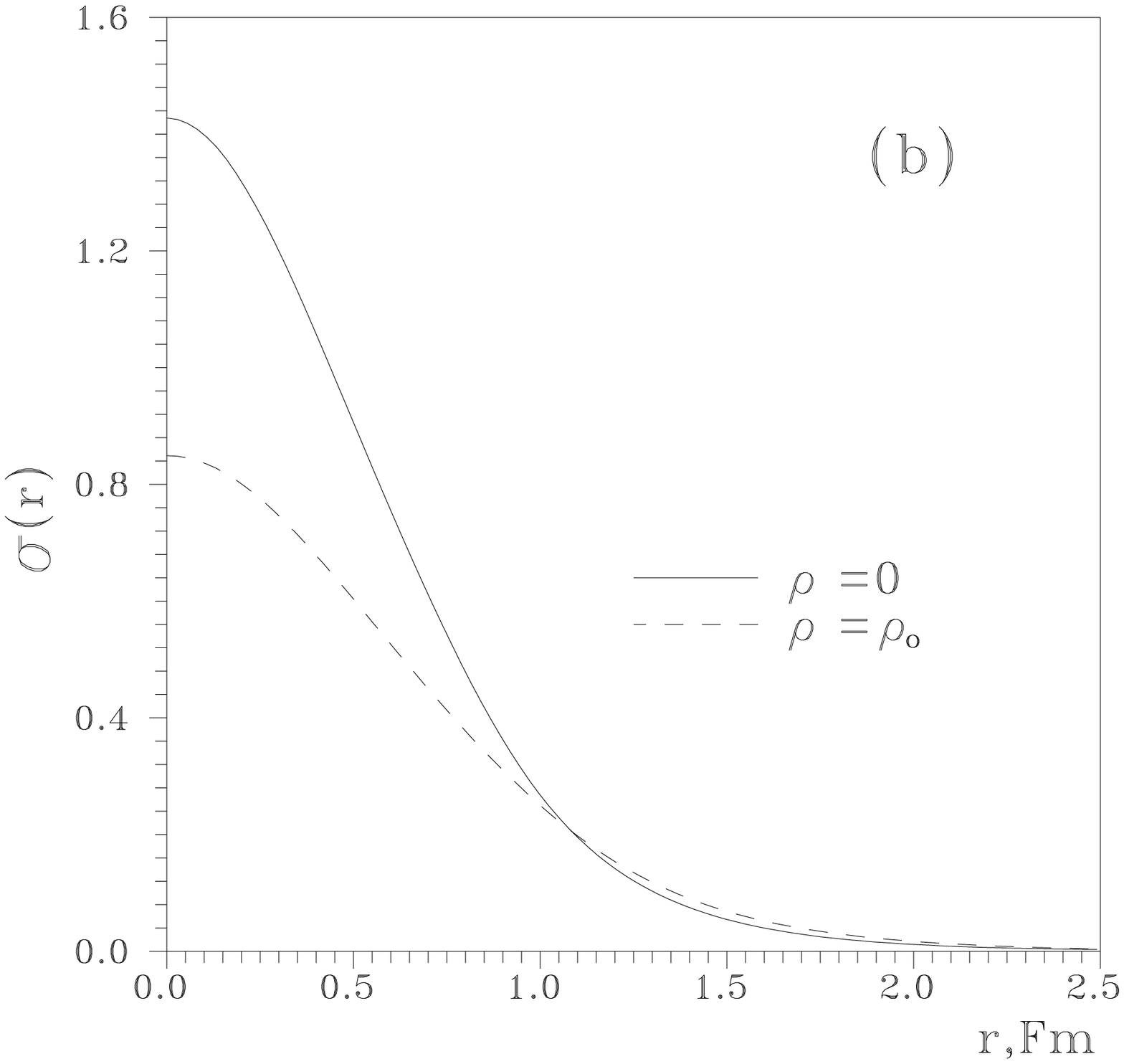} \ec
   \centerline{\parbox{15cm}
   {\caption{\label{npafig4b}
 The profile function
$\sigma (r)$  for the $\rho=0$ (solid curve) and $\rho=\rhozero$
(dashed curve). They are the solutions
 of equations of motion  \re{appeneq}  in the sector with $B=1$.
}}}
\end{figure}

A further interesting quantity is the pion-nucleon sigma term
$\Sigma_{{\pi}N}$. It is both the chiral symmetry breaking piece
of the nucleon mass and a measure of the scalar density of quarks
inside the nucleon. Due to the Hellman- Feynman theorem,
 it may be easily calculated \ci{hftheorem} in the Skyrme model that
\be \Sigma_{\pi{N}}=m_{q}\frac{\partial{M_N}}{\partial{m_q}}=
\frac{\partial{M_N}}{\partial{m_{\pi}^{2}}}m_{\pi}^{2}
\lab{sigman} \ee For free space ($\rho=0$) the Lagrangian
including scalar mesons \re{npa32}
 gives $\Sigma_{\pi{N}}\approx 20.1 MeV$ that is much smaller than that
obtained in the original Skyrme model \ci{skpi1,skpi2}. It may be
argued that $\Sigma_{\pi{N}}$ may also undergo changes in the
nuclear medium i.e.  $\Sigma_{\pi{N}}^*\ne\Sigma_{\pi{N}}$.
Actually, PCAC allows us to relate $\Sigma_{\pi{N}}$ to the
soft-pion limit of $\pi{N}$ scattering \ci{reua} whose parameters
may  not be the same in free space and the medium.
 So, defining an effective value of the in-medium
pion-nucleon sigma commutator as \be \Sigma_{\pi{N}}^*=m \
{}^*\!\bra{N}{\ds\int}d\vecr\bar\psi\psi\ket{N}^*=
\frac{\partial{M_N}^*}{
\partial{m_{\pi}^{*2}}}m_{\pi}^{*2}
\lab{sigmanstar} \ee where in the used notations $\ket{N}^*$ - is
the state of the nucleon bound in nuclear matter, we obtain
$\Sigma_{\pi{N}}^*\approx 40.1 MeV$ at normal nuclear density.
This means a large increase of the nucleon sigma term:
$\Sigma_{\pi{N}}^*/\Sigma_{\pi{N}}\approx 2$. Using appropriate
solutions of Eqs. \re{appeneq}, it is estimated that \be
\dsf{\Sigma_{\pi{N}}^*}{\Sigma_{\pi{N}}}{\approx}
\dsf{m_{\pi}^{*2}
\ds\int\limits_0^{\infty}e^{-3\sigma^*}[1-\cos(\theta^*)]x^2dx}
{m_{\pi}^{2}
\ds\int\limits_0^{\infty}e^{-3\sigma}[1-\cos(\theta)]x^2dx}=
1.88\dsf {m_{\pi}^{*2} }{m_{\pi}^{2} } \ , \ee where $\sigma^*(x)
$,  $ \theta^*(x)$ are profile functions at $ \rho=\rhozero$ and
$\sigma(x) $, $ \theta(x)$ are those at $\rho=0$. Thus,  the
medium renormalization of $\sum_{\pi{N}}$ is caused mainly due to
a large modification of the profile functions (see Figs.
\ref{npafig4a} and \ref{npafig4b}). On the other hand
$\Sigma_{\pi{N}}$ gives a good estimate for the quark condensate
at finite density: \be
\frac{\langle\bar{q}q\rangle_{\rho}}{\langle\bar{q}q\rangle_{vac}
}=1- \frac{\Sigma_{\pi{N}}^*\rho}{\mpi^{2}f_{\pi}^{2}} \ee Hence,
assuming the last equation  holds it is concluded that the in
medium enhancement of $\Sigma_{\pi{N}}$
 leads to further quenching
of the scalar density of quarks $\langle\bar{q}q\rangle_{\rho}$
in nuclear matter.

%%%%%%%%%%%%%%%%%%%%%%%%%%%%%%%%%%%%%%%%%%%%%%%%%%%%%%%%%%%%%%%%%%%
%%%%%%%%%%%%%%%%%%%%%%%%%%%%%%%%%%%%%%%%%%%%%%%%%%%%%%%%%%%%%%%%%%%
%%%%%%%%%%%%%%%%%%%%%%%%%%%%%%%%%%%%%%%%%%%%%%%%%%%%%%%%%%%%%%%%%%%

\section{Discussions and summary}
\indent

We have proposed a medium modified Skyrme like Lagrangian which
takes into account the distortion of the basic nonlinear meson
fields by the nuclear medium. It has been extended by including
the scalar - isoscalar sigma meson. The influence of the medium
on pion fields is introduced  by the self energy operators
$\hat\Pi_{p},(=\hat\Pi_{\Delta})$ and $\hat\Pi_{s}$  while the
effect of the  medium on the
 dilaton field
is limited to the renormalization of the gluon condensate. The
Lagrangian is applied to study changes in the hadron masses and
meson - nucleon vertex form - factors.

In particular,  the mass of the $\Delta$ - resonance decreases
more than that of the nucleon in the nuclear medium. Consequently
the pion requires lesser energy to excite the nucleon to the
$\Delta$ in the nuclear medium than it does for a free nucleon.
The mass difference between $\Delta$ and N decreases to $42\%$ of
that for free particles at the nuclear matter density. This is
quite consistent with earlier estimates \ci{meissmed,akira} but
contradicts the recent  theoretical results of Mukhopadhyay and
Vento \ci{vento}, who found $\delta M_{N\Delta}^{*}/\delta
M_{N\Delta}\approx 1.25$ for $ \rho=0.8\rhozero$. So, it would be
quite interesting to study N$\Delta$ mass splitting
experimentally by an analyses of the N$\Delta$ transitions in
heavy nuclei.

We have investigated the characteristic changes of the decay width
$\Gamma_{\sigma\rightarrow \pi\pi}$ at zero temperature. One may
expect that the temperature-dependence of the physical quantities
is qualitatively similar to the $\rho$ dependence.
 In this sense, our results are in good agreement
with predictions of the in - medium   NJL model
\ci{hatsudakunihiro}, that at sufficiently high temperatures the
$ \sigma$ meson becomes a sharp resonance and its width may even
vanish. Clearly, more precise predictions on
$\Gamma_{\sigma\rightarrow\pi\pi} (T,\rho)$ in the framework of
the present Lagrangian should be made by studying thermal Green's
function of $\sigma$ - meson e.g. within  Thermo Field Dynamics.

Furthermore the in - medium version of GT relation \re{npa49}
holds in the present approach. The renormalized pion decay
constant $f_{\pi}^{*}$ and nucleon mass $M_N^{*}$ do not satisfy
the BR scaling \ci{br}: the change in the nucleon mass is larger
than that in $f_{\pi}$.

The medium effects lead to a quenching of the meson - nucleon
form factors as well as the coupling constants   $g_{\pi NN} $
and $ g_{\sigma NN}$.
  The latter  should be
compared with the results of Banerjii and Tjon \ci{banerjiitjon}
obtained in the framework of CCM in the meanfield approximation.
There $\gsnn$ increased at low densities making a looping
behavior of the binding energy at saturation as a function of the
density. However we believe that the natural reduction of the
meson  masses and coupling strengths found in the present model
are expected to give a good description of the saturation
properties of nuclear matter.
 So, it would be quite interesting
to study the saturation properties of nuclear matter in
relativistic Brueckner approach by using density dependent meson
masses and coupling constants obtained in the present model.

By introducing a formal definition of the in - medium  pion -
nucleon sigma term $\Sigma_{\pi{N}}^*=m_{\pi}^{*2}
[{\partial{M_N}^*}/{\partial{m_{\pi}^{*2}}}]$, it is  found that
in contrast to the meson - nucleon coupling strengths the
$\Sigma_{\pi{N}} $ increased in the medium: $\Sigma_{\pi{N}}^* /
\Sigma_{\pi{N}}\approx 2 $. This enhancement could lead to a
decrease of the quark condensate $\langle{\bar
q}q\rangle_{\rho}$  in nuclear matter. However as it was recently
pointed out by Birse \ci{birse96} this change should not lead to a
drastic and rapid restoration of the chiral symmetry in nuclear
matter.

It is anticipated that ultrarelativistic heavy - ion collision
experiments (e.g. at RHIC) will provide  significant new
information on the strong interactions through the detection of
changes in hadronic properties \ci{last,quarkmatter}. This would
provide an impetus to consider refined models for strong
interactions in nuclear matter.

In conclusion, the modified in - medium Skyrme Lagrangian with
scale invariance provides a useful insight into the role of
medium in changing various properties of the mesons and nucleons.
Even though scale invariance is badly broken in strong
interactions its inclusion gives important information on the
role of this symmetry property in a many-particle system.

%\include{ch6jaf}
%%%%%%  ReMOVE  these 2 lines if you want your fig.s (tabs) to be numbered
%%as chapter.number like V.3
\setcounter{fignum}{\value{figure}}
\setcounter{tabnum}{\value{table}}
\chapter[
 Nucleon electromagnetic form -  factors in a nuclear medium at zero
temperature. ]{} \setcounter{figure}{\value{fignum}}
\setcounter{table}{\value{tabnum}}
\bc {\Large\bf Nucleon electromagnetic form - factors in a
nuclear medium at zero temperature.} \footnote {The  present
chapter is based on following articles by the author and his
collaborators: \ci{ourjafmed,ourjafmedff,ourprcmed, ournpamed}}
\ec

\section{Introduction}
\indent

The number of studies devoted to in - medium renormalization of
the properties of nucleons has ever been increasing in recent
years. Despite the absence of precise data from experiments from
heavy - ion collisions and deep inelastic nucleon scattering,
there are reasons
 to believe
that nucleons immersed in a nuclear medium undergo swelling.
Theoretically it is clear that this internuclear effect must be
due to the effect of strong meson fields on the quarks
constituting a given nucleon and on its mesonic cloud.

It is natural to assume that the change in the nucleon dimensions
due to the intra nuclear meson field is of a more general
character. Specifically nucleon elastic and inelastic form factors
in a nuclear medium are expected to be different from those in a
vacuum.

An effective Skyrme Lagrangian describing the properties of a
nucleon immersed in a nuclear medium has been given in the
previous  Chapter. Here we  analyze the nucleon electromagnetic
form factors in a nuclear medium.

\section{Electromagnetic properties of nucleons}
\indent

The problem of calculating of electromagnetic form factors for
the nucleons reduces to computing isoscalar and isovector currents
in a model. In topological models, in particular, in the Skyrme
model,
 the isoscalar current is proportional to the baryonic current

\beq B^{\mu}=\dsf{1}{24\pi^2}\epsilon^{\mu\nu\alpha\beta} \Tr
L_{\nu}L_{\alpha}L_{\beta}\,\,, \label{barcur} \eeq which is
seen to be model independent. This expression remains in force in
the theory based on the Lagrangian in Eq. \re{npa32}. Since the
isoscalar mean square radius ${\langle}r^2{\rangle}_{I=0}$ is
expressed in terms of the zero component of the baryon current as
\be {\langle}r^2{\rangle}_{I=0}=\ds\int\limits_0^{\infty}
B^0d^3r\,\,, \ee variations in it are not very important.
Nontrivial
 modifications are due exclusively to the distortion of the skyrmion
profile function in the nuclear matter. Similarly, the isovector
mean - square radius ${\langle}r^2{\rangle}_{I=1}$ which is
determined by the zero component of the vector current \beq
\vec{V}_{\mu}=-i\dsf{F_{\pi}^2}{16} C_{\mu}\Tr\,
\vectau(L_{\mu}+R_{\mu})+
\dsf{i}{16{e}^{2}}\Tr\,\vectau\{[L_{\nu}[L_{\mu},L_{\nu}]]+
[R_{\nu}[R_{\mu},R_{\nu}]]\}\,\,, \label{veccur} \eeq where
\[ L_{\mu}=U^+\partial_{\mu}U;\quad R_{\mu}=U\partial_{\mu}U^+;\quad
C_{\mu}=\left\{\begin{array}{ccc}
1 &, &\mbox{$\mu=0$}\\
(1-\chi_\Delta) &, & \mbox{$\mu=1,2,3$}\\
\end{array}\right. \]
[the vector current is treated as a Noether current that
corresponds to the transformation $U(x)\rightarrow
\exp(iQ_L)U(x)\exp(-iQ_R)$ ($Q_{L,R}$
 - $2\times 2$ matrices],
does not depend explicitly on the medium parameters either.
Relevant expressions for ${\langle}r^2{\rangle}$ are presented in
the Appendix B.

Proceeding to analyze the isoscalar and isovector magnetic
moments, we note that the former, \beq
\mu_{I=0}=\frac{1}{2}{\ds\int}d\vecr\;\;\; \vecr{\times}\vec{B}
\eeq does not depend on the medium parameters either. In
contrast, the latter, $\mu_{I=1}$, which is expressed in terms of
the spatial component of the vector current as \beq
\mu_{I=1}=\frac{1}{2}{\ds\int}d\vecr\;\;\;
\vecr{\times}\vec{V}_{3} \eeq involves the features of the
nuclear medium explicitly. In our approach
 these stem from the contribution of the
kinetic term in equation \re{npa32}.

The expressions for the electric and magnetic form factors, $G_E$
and  $G_M$ respectively, can easily be obtained from equations
\re{npa32} and  \re{veccur} by using the technique developed in
\cite{zb,skpi2}. The results are \beq
\begin{array}{l}
G_E^S(q)=-\dsf{2}{\pi}\int\limits_0^{\infty} s^2\theta^{\prime}j_0(qr)dr\,\,,\\
\quad \\
G_E^V(q)=\dsf{\ds\int\limits_0^{\infty} r^2s^2[\chi^2+
4(\theta^{\prime 2}+d)]j_0(qr)dr}{ \ds\int\limits_0^{\infty}
r^2s^2[\chi^2+4(\theta^{\prime 2}+d)]dr}\,\,, \label{emforfac}
\end{array}
\eeq \beq
\begin{array}{l}
G_{M}^S(q)=-\dsf{2M_N}{\pi
eF_{\pi}\lambda_{\mu}}\int\limits_0^{\infty}
x^2s^2\theta' \dsf{j_1(qx)}{qx}dx\,\,,\\
\quad \\
G_{M}^V(q)=-\dsf{2M_N\lambda_{M}
\ds\int\limits_0^{\infty}x^2s^2[\chi^2\alpha_p+4(\theta^{\prime
2}+d)] \dsf{j_1(qx)}{qx}dx }{
\ds\int\limits_0^{\infty}x^2s^2[\chi^2\alpha_p+4(\theta^{\prime
2}+d)]}
\end{array}
\eeq where \be
\lambda_{M}=\dsf{8\pi}{3e^2F_{\pi}}\int\limits_0^{\infty}dx\left\{x^2s^2
\left(\frac{\chi^2}{4}+(\theta'+d)\right)\right\}\,\,, \ee is the
skyrmion moment of inertia,
 $x\equiv eF_{\pi}r$ and $j_0,j_1$ are spherical Bessel functions.
Note that, in nuclear matter, the form factors do not obey the
well known relations  \cite{zb,skpi2} \beq
\begin{array}{l}
G_{M}^S(q)=-\dsf{2M_N}{\lambda_{M}}\frac{\partial}{
\partial q^2}G_E^S(q)\,\,,\\
\quad \\
G_E^V(q)=-\dsf{1}{M_N\lambda_{M}}\left(\dsf{3}{2}+q^2\frac{\partial}{
\partial q^2}\right)G_{M}^V(q)
\end{array}
\label{formfac} \eeq because the Lorentz invariance does not hold
here.

\section{Input Parameters}
\indent

In our calculations we set the input parameters as
$F_{\pi}=186MeV$ and $C_g=(260MeV)^4$, whereby the correct value
is reproduced for the sigma meson mass  $m_{\sigma}=550MeV$. The
Skyrme parameter $e$ was chosen in such a way as to fit the
nucleon and delta - isobar masses in vacuum ( $M_N=938MeV$ and
$M_{\Delta}=1232MeV$ respectively). The effect of nuclear medium
was specified by fixing the parameters of the polarization
operator
 $\Pi_\Delta$
at $g_0^{\prime}=0.6$ and  $c_0=0.13 m_{\pi}^{-3}$ \cite{polar}
as it was done in Chapter 5.

Let us discuss the density dependence of the input parameters. As
it was indicated in the previous chapter the renormalization of
the Skyrme parameter $e$ cannot be considered here because the
$\rho$ mesons have not been included on the  effective Lagrangian
\re{npa32} explicitly. For this reason we set $e^*=e$. The
renormalized pion mass  $m_{\pi}^*$ has the form \cite{ericsonmpi}
\beq m_{\pi}^*=\dsf{\sqrt{1-0.22\rho/\rho_0}}
{\sqrt{1-0.29\rho/\rho_0}}\,m_{\pi}\,\,, \label{mpi} \eeq where
$\rho_0=0.5 m_{\pi}^{3}$ - is the normal density of nuclear
matter. This parameterization also takes into account  the
correction to the contribution of the   $S$ - wave component of
the pion self energy $\Pi_S$.

The only input parameter in the dilaton sector is the gluon
condensate $C_g^*$, whose in - medium renormalization has to be
clarified conclusively. In the proposed model, this quantity is
given by Eq. \re{npa52} and can be expressed in terms of sigma
meson mass as \be C_g^*=\dsf {3F_{\pi}^{2}(m_{\sigma}^{*})^2N_f}
{4(4-\varepsilon)} \ee
 Unfortunately, no information about the renormalization of $m_\sigma$
has been deduced so far within the present approach. In view of
this,
 the parametrization
\be m_{\sigma}^{*}=[1-0.12\frac{\rho}{\rho_0}]m_\sigma \ee which
was obtained in \ci{saitothomas}
 within the quark - meson coupling model, is taken for the sigma meson mass.
\section{Results and Discussions}
\indent

Let us now proceed to discuss our results. \normalsize
\begin{table}[htb]
\caption{\it Ratio of the nucleon mass, nucleon root -mean -
square radii, and nucleon magnetic moments in a nuclear medium
(labelled with an asterisk) to the corresponding vacuum values. }
\begin{center}
\begin{tabular}{cccccccccc}
\hline
\quad \\
$\rho/\rho_0$& $\dsf{M_N^*}{M_N}$&
$\sqrt{\dsf{{\langle}r^2{\rangle}_{I=0}^*}{{\langle}r^2{\rangle}_{I=0}}}$&
$\sqrt{\dsf{{\langle}r^2{\rangle}_{I=1}^*}{{\langle}r^2{\rangle}_{I=1}}}$&
$\sqrt{\dsf{{\langle}r^2{\rangle}_p^*}{{\langle}r^2{\rangle}_p}}$&
$\sqrt{\dsf{{\langle}r^2{\rangle}_n^*}{{\langle}r^2{\rangle}_n}}$&
$\dsf{\mu_{I=0}^*}{\mu_{I=0}}$ &$\dsf{\mu_{I=1}^*}{\mu_{I=1}}$ &
$\dsf{\mu_p^*}{\mu_p}$ &$\dsf{\mu_n^*}{\mu_n}$ \\
\quad \\
\hline
0.0 &1.000 & 1.000 & 1.000 & 1.000 & 1.000 & 1.000 & 1.000& 1.000 & 1.000 \\
0.5 &0.900 & 1.101 & 1.069 & 1.079 & 1.044 & 0.856 & 1.081& 1.052 & 1.119 \\
1.0 &0.819 & 1.196 & 1.122 & 1.145 & 1.061 & 0.817 & 1.145& 1.103 & 1.200 \\
\hline
\end{tabular}
\end{center}
\label{tab1jafmed}
\end{table}
%\large

Table \ref{tab1jafmed} presents the calculated   root -mean -
square radii and magnetic moments of the nucleons in the nuclear
medium. It can be seen that, in - medium modifications to the
root -mean - square radii of the proton are greater than those to
the root -mean - square radii of the neutron. The effect is
inverse for the magnetic moments - that is , the neutron magnetic
moment is more sensitive to the impact of the medium than the
proton magnetic moment.

In nuclear matter of normal density, the proton charge radius
$\langle r_{p}^{2}\rangle ^{1/2}$ increases by some $14\%$, which
is in agreement with the result presented in
\ci{meisprl,bunat88}.We note that, only $5\%$ increase in this
quantity was found by Lu et al. \ci{lu}, who took into account
solely the renormalization of the nucleon mass, disregarding the
renormalization of the  pion - nucleon coupling constant
$\gpinn$.  However, in previous Chapter it has been shown that,
the in - medium modifications of meson - nucleon coupling
constants are  less sizeable than the corresponding modifications
of the nucleon mass.

From Table \ref{tab1jafmed}, it can be seen that the effect of a
nuclear medium leads to an increase in the nucleon magnetic
moment.
  As it was first revealed in
\ci{meisprl,meispl}, the isoscalar magnetic moment changes more
pronouncedly than the isovector one.
 In our calculations, the former and the latter change by
about $14\%$ and $8\%$ respectively, at $\rho=0.5\rho_0$ and by
$18\%$ and $15\%$ respectively, at $\rho=\rho_0$.

\begin{figure}[htb]
\bc \epsfysize=14cm \epsfbox{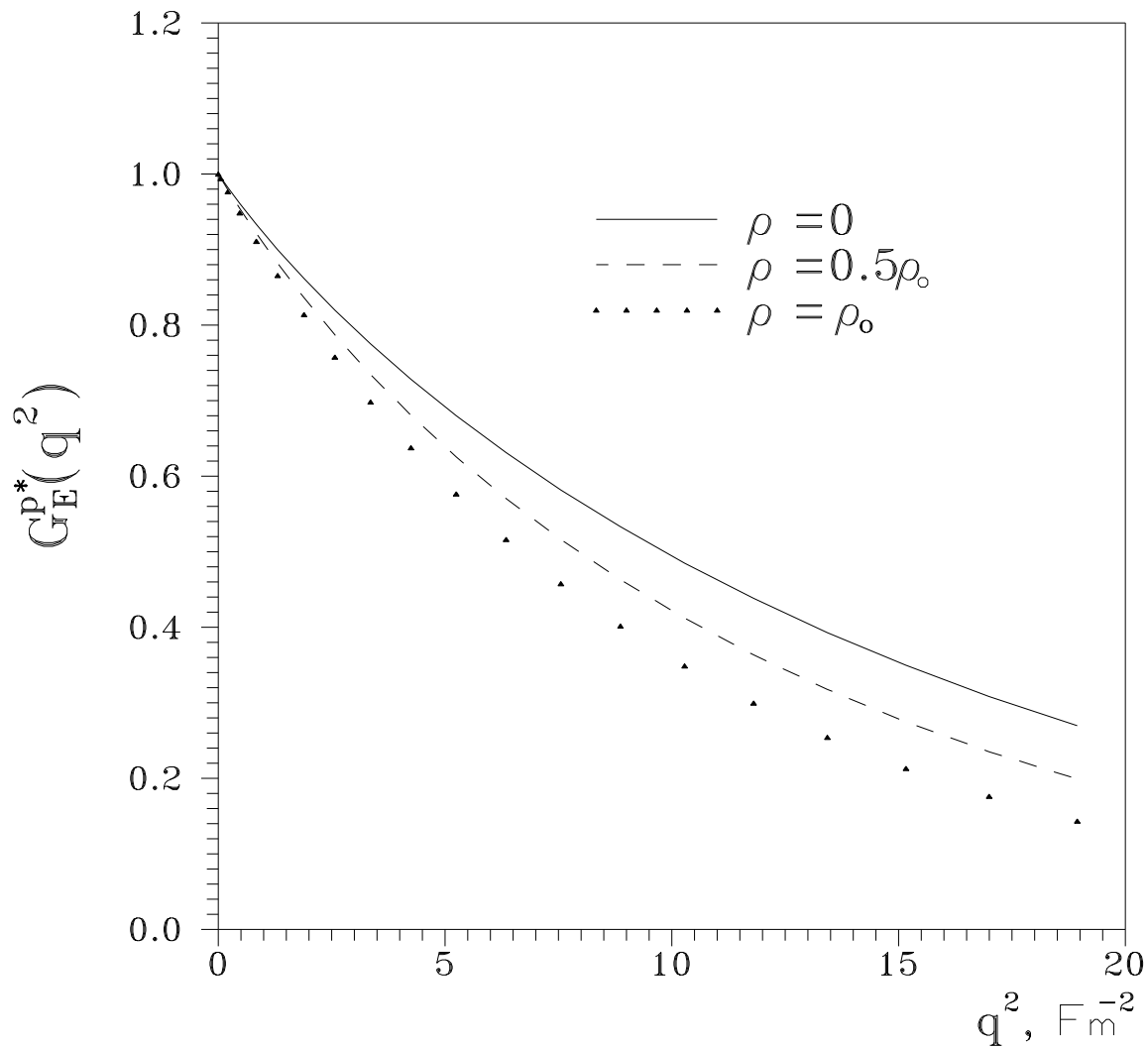} \ec
   \centerline{\parbox{15cm}
   {\caption{\label{f1ajafmed}
 Proton electric form factor
  $G^{p*}_E(q^2)$.
 The solid, dashed and dotted curves correspond to the
nuclear - density value of $\rho=0$, $\rho=0.5\rho_0$ and
 $\rho=\rho_0$ respectively.
}}}
\end{figure}

\begin{figure}[htb]
\bc \epsfysize=14cm \epsfbox{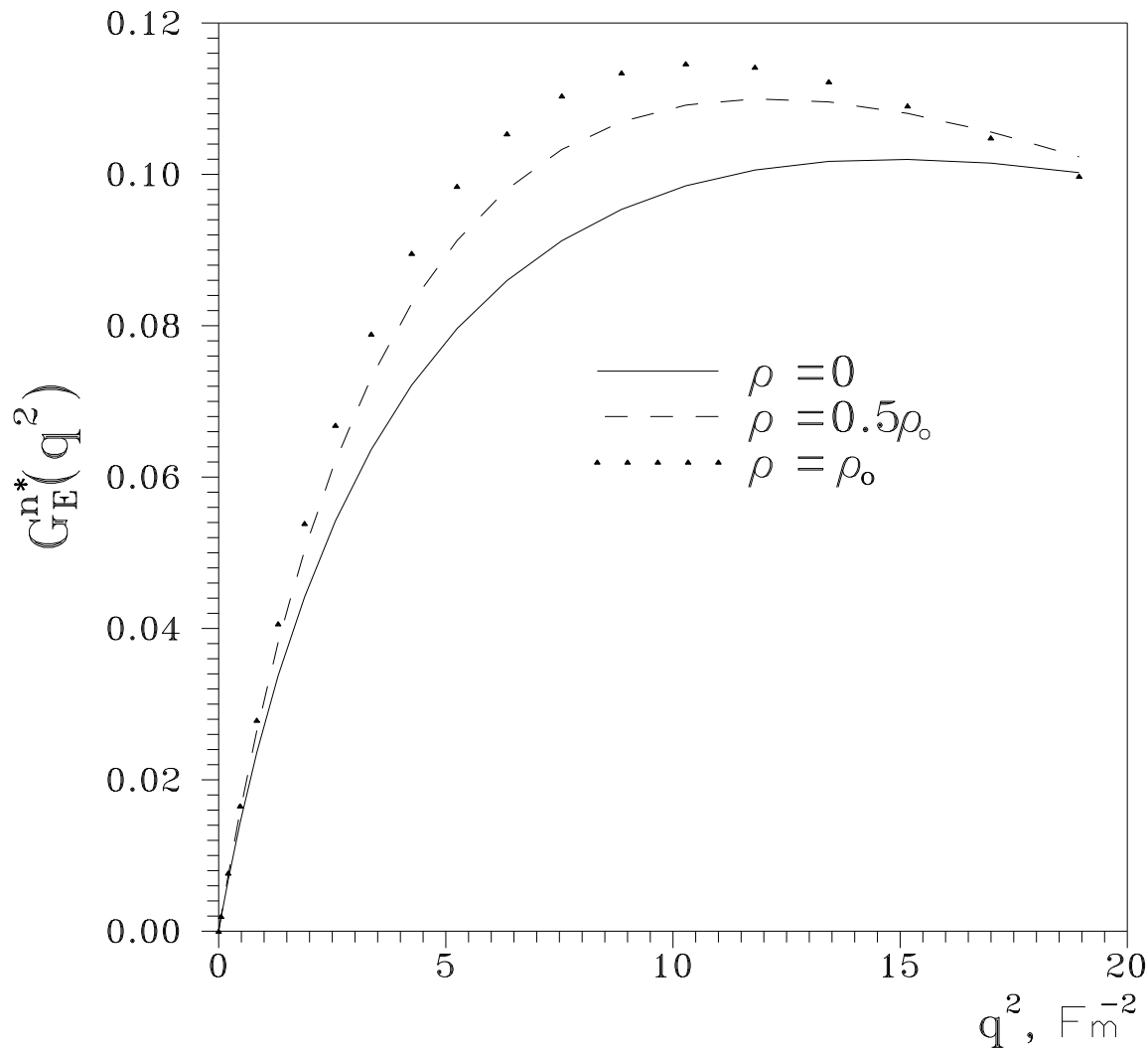} \ec
   \centerline{\parbox{15cm}
   {\caption{\label{f1bjafmed}
 The same as in Fig. \ref{f1ajafmed}
but for neutron electric  form factor. }}}
\end{figure}

Figures \ref{f1ajafmed}  and \ref{f1bjafmed} display the proton
and neutron electric  form factors respectively, at the nuclear
density values of  $\rho=0$, $\rho=0.5\rho_0$ and $\rho=\rho_0$
(the solid, dashed and dotted curves respectively).

%%%%%%%%%%%%%%%%%%%%%  FIG 1a of JAfmedFF

%%%%%%%%%%%%%%%%%%%%%  FIG 1b of JAfmedFF

\begin{figure}[!htb]
\bc \epsfysize=14cm \epsfbox{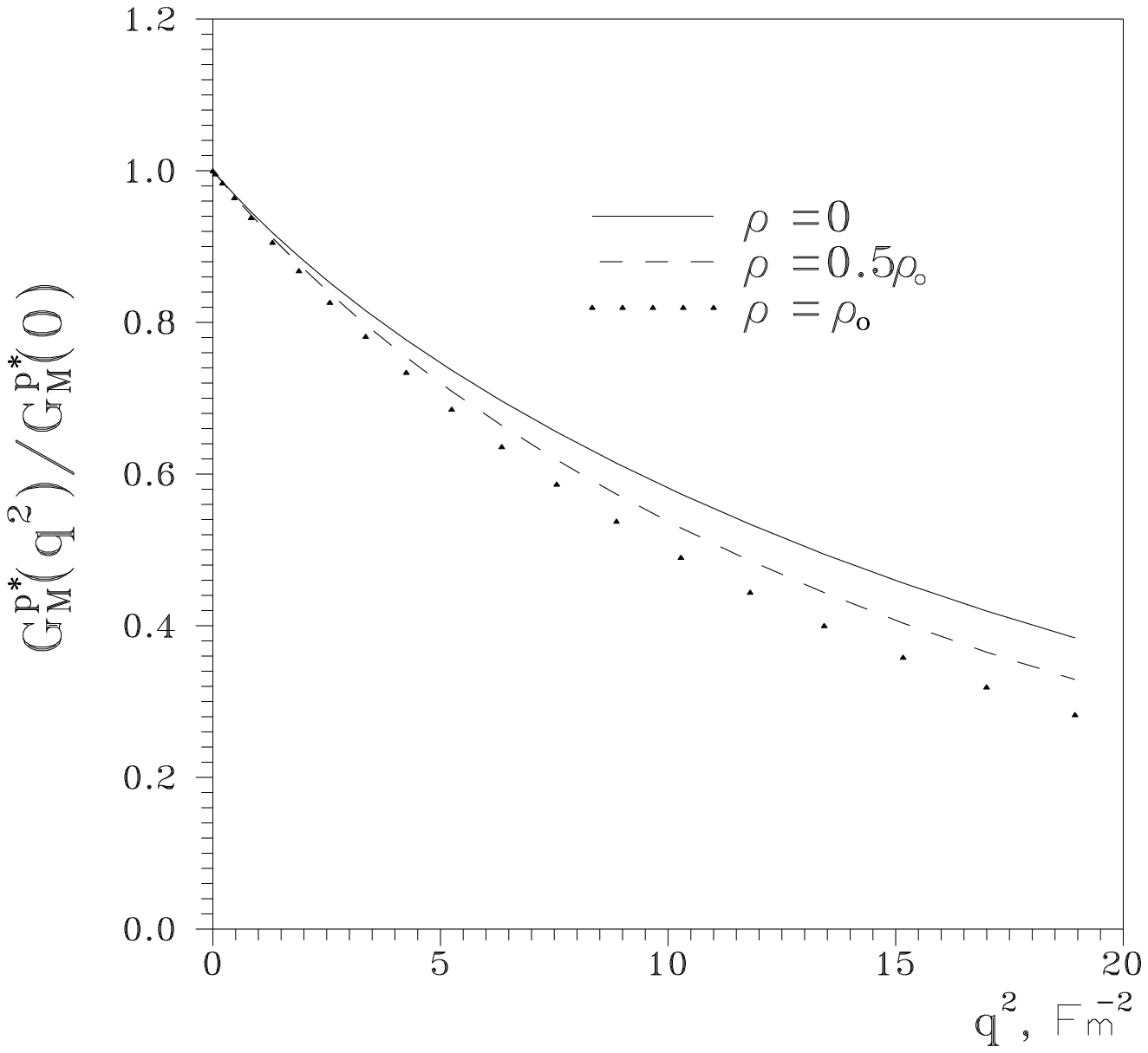} \ec
   \centerline{\parbox{15cm}
   {\caption{\label{f2ajafmed}
 Normalized proton magnetic form factor
$G^{p*}_M(q^2)/G^{p*}_M(0)$.
 The solid, dashed and dotted curves correspond to the
nuclear - density value of $\rho=0$, $\rho=0.5\rho_0$ and
 $\rho=\rho_0$ respectively.
}}}
\end{figure}

\begin{figure}[!htb]
\bc \epsfysize=14cm \epsfbox{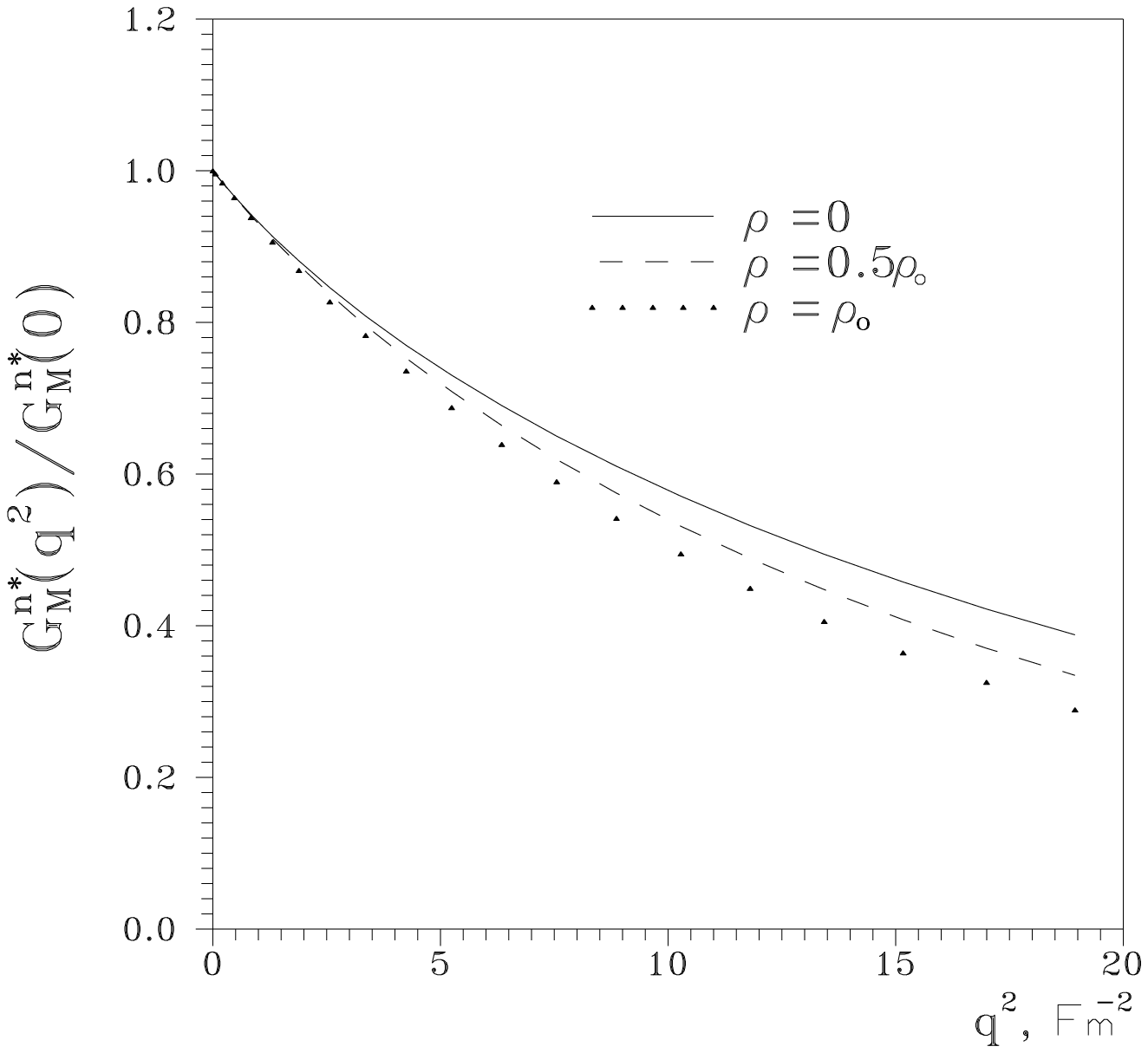} \ec
   \centerline{\parbox{15cm}
   {\caption{\label{f2bjafmed}
 Normalized neutron magnetic form factor
$G^{n*}_M(q^2)/G^{n*}_M(0)$.
 The solid, dashed and dotted curves correspond to the
nuclear - density value of $\rho=0$, $\rho=0.5\rho_0$ and
 $\rho=\rho_0$ respectively.
}}}
\end{figure}

\begin{table}[!htb]
\caption{\it Form factors versus the nuclear density
($\rho=\lambda\rho_0$) at some values of the momentum transfer
Here the magnetic form factors are normalized to their values at
 $q=0$.}
\begin{center}
\begin{tabular}{cccccc}
\hline
\quad \\
$q^2(Fm^{-2})$& $\lambda$ & $G_E^{p*}(q^2)$ & $G_E^{n*}(q^2)$ &
$\dsf{G_M^{p*}(q^2)}{
G_M^{p*}(0)}$ & $\dsf{G_M^{n*}(q^2)}{G_M^{n*}(0)}$\\
\quad \\
\hline
      &  0.0  & 0.680 & 0.080 & 0.737 & 0.730\\
5.25  &  0.5  & 0.635 & 0.078 & 0.709 & 0.709\\
      &  1.0  & 0.599 & 0.076 & 0.686 & 0.688\\
\hline
      &  0.0  & 0.582 & 0.091 & 0.655 & 0.650\\
7.55  &  0.5  & 0.529 & 0.089 & 0.618 & 0.620\\
      &  1.0  & 0.481 & 0.087 & 0.586 & 0.590\\
\hline
\end{tabular}
\end{center}
\label{tab2jafmed}
\end{table}
%\large
It can be seen that with increasing $q^2$ the form factors
decrease much more slowly in a nuclear medium than in a vacuum.
Figures \ref{f2ajafmed} and \ref{f2bjafmed} show the  normalized
magnetic form factors $G^{*}_M(q^2)/G^{*}_M(0)$ for the proton
and neutron respectively. At     $q^2\sim 0.3GeV^2$  ($7.5
Fm^{-2}$) the proton and neutron charge form factors decrease by
about
 $9\%$ and $2\%$ respectively
 at $\rho=0.5\rhozero$ and by about
$17\%$ and $4\%$ at normal nuclear density. Similarly, the proton
and neutron magnetic form factors
       decrease by about
 $6\%$ and $5\%$, respectively,
 at $\rho=0.5\rhozero$ and by about
$11\%$ and $9\%$ at normal nuclear density.
%%%%%%%%%%%%%%%%%%%%%  FIG 2a of JAfmedFF

%%%%%%%%%%%%%%%%%%%%%  FIG 2b of JAfmedFF
From a comparison of the data in Figs.
\ref{f1ajafmed}-\ref{f2bjafmed} it can be seen that the charge
form factors are more sensitive to medium effects than the
magnetic form factors. Qualitatively  this conclusion complies
with the results obtained in \ci{lu} on the basis of the  quark -
meson coupling model.

We note , however, that at the above momentum transfer value
 that is
at  $q^2\sim 0.3GeV^2$
 the proton and neutron charge form factors
       decrease by about
 $5\%$ and $6\%$, respectively,
 at $\rho=0.5\rhozero$ and by about
$8\%$  at normal nuclear density. A similar situation is observed
for the magnetic form factors as well. That Lui et al.\ci{lu}
found so weak
 suppression of the form factors in nuclear matter stems from their
  disregarding
of the renormalization of meson -  nucleon coupling constants and
masses in their study.

In ref. \cite{cheon} the form factors for bound nucleons were
calculated on the basis of the bag model. By preliminary fitting
the input parameters of the model, the experimental values of the
root - mean - square radii and magnetic moments were reproduced
there rather well for the free nucleons. Further, medium effects
were taken into account in that study through    the
renormalization of meson and   nucleon masses. The analysis
performed in \ci{cheon}
 revealed that the nucleon electromagnetic
form factors are suppressed in the $^{12}C$, $^{40}Ca$ and
$^{56}Fe$ nuclei. The nucleon magnetic moments in medium were
enhanced  by $2\%$ to $20\%$ for various nuclear species. These
results agree with ours and the ones from refs.
\ci{meisprl,meispl} However, the proton charge radius in the
$^{40}Ca$ nucleus is enhanced in ref. \ci{cheon} by nearly $40\%$,
 which seems too large
  in comparison with our results (about
$\sim 15\%$  at the normal nuclear density).
 It should also be noted that the results reported in
\ci{cheon} do not show the suppression of the axial coupling
constant in nuclei. The failure to reproduce this well known
effect casts serious doubts on the concept behind the
calculations from \ci{cheon}.

It is well known that, in a vacuum, the neutron and proton form
factors calculated in topological models at low momentum transfers
obey following  relations \ci{anw,zb} \beq
\dsf{G_{M}^p(q^2)}{G_{M}^p(0)}\approx
\dsf{G_{M}^n(q^2)}{G_{M}^n(0)} \approx  G_E^p(q^2)\,\,.
\label{skeyling} \eeq Our calculations have demonstrated that,
because of violation of Lorentz invariance, this relation is
fulfilled only partly: \beq
\dsf{G_{M}^{p*}(q^2)}{G_{M}^{p*}(0)}\approx \dsf{G_{M}^{n*}(q^2)}{
G_{M}^{n*}(0)} \not\approx G_E^{p*}(q^2)\,\,. \label{skeyour} \eeq
We have also established that, at low momentum transfers
($q^2<7.5 Fm^{-2}$), the form factors show a nearly linear
decrease with increasing a nuclear density. This results are in
line with the results obtained in refs. \ci{meisprl,meispl}. The
nuclear density dependence of the form factors is illustrated in
Tabl. \ref{tab2jafmed} for various values of the momentum
transfer.
%%%%%%%%%%%%%%% the second table
%\normalsize

\section{Conclusions and  Summary}
\indent

On the basis of chiral and scale invariant Lagrangian, we have
investigated the possible modifications to the nucleon
electromagnetic form factors in a nuclear medium. Our results
demonstrate that this model  faithfully reproduces the features
of hadrons as functions of the medium density.
 The calculations have revealed that the  isoscalar nucleon
form factors are more sensitive to the medium effects than the
isovector ones. On the whole, our predictions are consistent
 with the results of other authors.
This proves that, qualitatively, the resulting behavior of the
features of nucleons in a nuclear medium is model independent.
Specifically the coupling constants, masses and form factors are
suppressed by the effective fields of the nuclear medium,
wehereas the dimensions and magnetic moments are enhanced. By way
of example we indicate that, at the normal nuclear density the
root mean square radii for protons and neutrons increase by some
$\sim$ 15\%
 $\sim$ 6\%  respectively.
With increasing density the nucleon form factors exhibit a nearly
linear decrease in the region
 $q^2\le 7 Fm^{-2}$.

Here we have taken no account of recoil effects.
 Both prior to and after collision
with a photon, the nucleon involved is assumed here to be frozen
at the center of nucleus where the density is constant. As it was
shown by Bunatian \ci{bunat90} the contribution to
 $\langle{r^2}\rangle_N$
from recoil effects is  small, but it can be expected that
modifications to the nucleon structure will greatly depend on the
nucleon momentum within Fermi sphere. The inclusion of recoil
effects in the calculation of the nucleon electromagnetic form
factors for finite nuclei would be of great interest in the
problems of electron nucleon scattering. Calculations of this
type my also be helpful in processing data from future MIT Bates
and TJNAF experiments
 \cite{dak} that will measure the neutron electromagnetic form factor.
Since there is no neutron free target  relevant information will
be extracted from data on electron nucleus (say $e\,{}^3He$)
scattering where it is necessary to take onto account nuclear
medium effects.

Note that, the modification of nucleon electromagnetic
 form-factors in finite nuclei
has been considered recently in ref. \cite{ulugepj}.
%%%%%%%%%%%\ END JAFMEDFF%%%%%%%%%%%%%%%%%%%%%%%%%%%%%%%%%%%

%\include{ch7tfd}
%%%%%%  ReMOVE  these 2 lines if you want your fig.s (tabs) to be numbered
%%as chapter.number like V.3
\setcounter{fignum}{\value{figure}}
\setcounter{tabnum}{\value{table}}
%%%%%%%%%%%%%%%%%%%
\chapter[
 Meson - nucleon vertex form -  factors at finite temperature.
]{} \setcounter{figure}{\value{fignum}}
\setcounter{table}{\value{tabnum}}
\bc { \Large\bf
 Meson - Nucleon Vertex Form  Factors at Finite Temperature.
}\footnote {The  present chapter is based on following articles
by the author and his collaborators: \ci{ourprctfd,ourreact}} \ec

\section{Introduction}
\indent

 In-medium properties of hadrons and their
interactions is a field of high current interest. For studying
phase transitions and thermodynamical properties of the nuclear
system under extreme conditions, it is essential to determine the
temperature ($T$) and density ($\rho$) dependence of the nuclear
force. Below a critical temperature $T_c$, meson exchange picture
of nuclear physics provides the natural description for the
nucleon -  nucleon ($NN$) interaction. Quarks and gluons are
certainly present in all $NN$ interactions, but it is not always
necessary to take them explicitly into account especially below a
critical temperature
 $T_c$, when the transition into a quark gluon phase takes
place. At distances smaller than a fermi,  the inner structure of
hadrons is probed since it involves the short distance or the
high momentum components of the wave function. This
 structure is typically taken into
account by including  vertex form  factors. In hot and dense
matter, the  structure of hadrons undergoes changes which should
lead to a modification of the meson - nucleon coupling constants
or, in general, the form factors.
 The temperature  \ci{dom}
and density dependence of some coupling constants have been
investigated recently
 (see Chapters 5, 6),
 while that of form factors is still fairly unknown and will be
studied in the present Chapter for the first time. For this
purpose we shall use the formalism of Thermo Field Dynamics (TFD)
in order to obtain the required temperature and density
dependence.

The Thermo Field Dynamics (TFD)  which  was  first suggested by
Takahashi and Umezawa  \ci{tfd1} - \ci{tfd5}  is  a
 real  time  operator formalism  of  finite  temperature  quantum field
 theory.  In this framework,  all the operator  formalism
 of  quantum  field theory at zero temperature can be extended
 directly to finite temperature and density. Therefore in TFD
 it is possible to  continue  using  Wick's theorem and the
 Feynman diagrammatic approach as in the case of zero temperature
 theory.  An  auxiliary  field  is introduced which leads to  a  field
doublet;  the  propagators  become $ 2\times 2$
 matrices,  the Feynman  rules are now algebraic operational rules
 in the space of $2\times 2$  matrices.  Recently,  some  attempts  to
 study  in  medium properties of hadrons at finite temperature
using TFD \ci{zhang,oset} have been made.

A study of relativistic heavy-ion collisions   \ci{tfdrev}
 is expected to lead
us to a proper equation of state of hot and dense nuclear matter.
The behaviour of nuclear matter as a function of temperature and
density is relevant for investigation in  nuclear physics,
astrophysics, cosmology and particle physics. The phase
transition from hadronic system to a quark-gluon plasma phase and
the subsequent hadronisation of the quark-gluon plasma phase can
provide information on the asymptotic freedom as well as
confinement behaviour in Quantum Chromodynamics (QCD). Therefore
it is important to study theoretically the behaviour of the
hadronic system at high temperatures. In this Chapter we shall
make an
 attempt to study such
a behaviour.
 In Sec.2, a brief review of the
TFD is given. In Sec.3 we calculate the renormalization of meson
nucleon vertices using the Feynman propagators from TFD. The
results of calculations and their discussions will be presented
in Sec.4.

\section{Finite temperature formalism for meson and nucleon propagators.}
\indent

The Thermo Field Dynamics is a real time operator formalism of
quantum field theory at finite temperature. The main feature of
the TFD is that thermal average of operator $\hat{A}$ is defined
as the expectation value with respect to a temperature dependent
vacuum, $\ket{0(\beta)}$, which is obtained from the regular
vacuum by a  Bogoliubov transformation. Therefore, we have \be
<\hat{A}>\equiv \Tr
(\hat{A}e^{-\beta(H-\mu)})/\Tr(e^{-\beta(H-\mu)})
=\matel{0(\beta)}{\hat A}{0(\beta)}\vergul \label{tfd1} \ee where
$\beta \equiv 1/k_{B}T$ with $k_{B}$ being the Boltzmann
constant, H is the total Hamiltonian of the system, and $\mu$ is
the chemical potential. Due to the doubling of all degrees of
freedom, every field is represented in  the thermal doublet form
in TFD. For example, the thermal doublet for the nucleon field is
\begin{equation}
\psi^{(a)} (x) \equiv \left\{
\begin{array}{c}
\psi (x)\\
i \  ^t\tilde{\psi}^\dagger
\end{array}
\right\}\vergul \lab{tfd2}
\end{equation}

\noindent where $\psi(x) $ is the ordinary nucleon field and
$\tilde{\psi}$ is  the doublet partner of $\psi(x)$. Here the
superscript represents the transpose operation on the vector
index. As a consequence, the thermal propagator of a fermion
field is a $2\times 2$ matrix  defined as
\begin{equation}
i S_{F}^{(a,b)} (x_{1}, x_{2}) = <0(\beta) | T [\psi^{(a)} (x_{1})
\bar{\psi}^{(b)} (x_{2})] | 0 (\beta) > \label{3}
\end{equation}

The free thermal propagator  $S_{F}$
  to be used in perturbation theory
can be separated into two parts, i.e. the usual Feynman part
$S_{0}$, propagator at zero temperature, and the temperature
dependent part $S_T$ such that
$S_{F}^{(ab)}=S_{0}^{(ab)}+S_{T}^{(ab)}$. For a fermion with mass
$M$, the propagators have the form
\begin{eqnarray}
S_{0}^{(ab)}=(\not p+M)_{(ab)}\left(
\begin{array}{cc}
G_{0}(p^2) & 0\\
0& G_{0}^{*}(p^2)
\end{array}\right)\\
\label{tfd4} S_{T}^{(ab)}=2\pi i\delta(p^2-M^2)(\not
p+M)_{ab}\left(
\begin{array}{cc}
\sin^2\theta_{p_0} & \frac{1}{2}\sin 2\theta_{p_0} \\
 \frac{1}{2}\sin 2\theta_{p_0} &-\sin^2\theta_{p_0}
\end{array}\right)
\lab{tfd5}
\end{eqnarray}
with \beq
\begin{array}{l}
G_{0}(p^2)=\dsf{1}{p^2-M^2+i\epsilon} \vergul \\
\\
\cos\theta_{p_0}=\dsf{\theta{(p_0)}}{(1+e^{-x})^{1/2}}+
\dsf{\theta({-p_0})}{(1+e^{x})^{1/2}},\\
\quad \\
\sin\theta({p_0})=\dsf{e^{-x/2}\theta({p_0})}{(1+e^{-x})^{1/2}}-
\dsf{e^{x/2}\theta{(-p_0)}}{(1+e^{x})^{1/2}},
\end{array}
\lab{tfd6} \eeq where $x=\beta(p_0-\mu)$, $a$ and $b$ are Dirac
indices, and $\theta(x)$ is the step function. The propagator of
the scalar particle with mass $m_B$ is given by \beq
\delb(p)=\delb_0(p)+\delb_T(p), \lab{tfd7} \eeq
\begin{eqnarray}
\delb_{0}=\left(
\begin{array}{cc}
D_0(p^2) & 0\\
0&-D_{0}^{*}(p^2)
\end{array}\right)\\
\\
\label{tfd8} D_{T}=-2\pi i\delta(p^2-m_B^2)\left(
\begin{array}{cc}
\mbox{sinh}^2\phi_{p_0} & \frac{1}{2}\mbox{sinh}2\phi_{p_0} \\
 \frac{1}{2}\mbox{sinh}2\phi_{p_0} &\mbox{sinh}^2\phi_{p_0}
\end{array}\right)
\label{tfd9}
\end{eqnarray}
with \beq
\begin{array}{l}
D_0(p^2)=\dsf{1}{p^2-m_B^2+i\epsilon}\vergul\\
\\
\mbox{cosh}\phi_{p_0}=\dsf{1}{(1-e^{-|y|})^{1/2}},\\
\quad\\
\mbox{sinh}\phi_{p_0}=\dsf{e^{-|y|/2}}{(1-e^{-|y|})^{1/2}},
\end{array}
\lab{tfd10} \eeq where $y=\beta p_0$.
 Note that
the propagator of vector mesons would be similar to that of the
scalar mesons except for an additional factor of
$(-g_{\mu\nu}+k_{\mu}k_{\nu}/m^2)$.

\section{Vertex form factors}
\indent

 The OBE model of the N-N interaction
 \ci{machrep}
 includes exchange of several
 mesons. Here we consider the meson-nucleon interactions of $\pi$ ,
 $\sigma$ , $\omega$- and $\rho$-mesons:
\be \ba {\cal L}_{\pi NN}=-ig_{\pi NN}\bar\psi
\gamma_5\vec\tau\psi\vec\varphi_{\pi}\,; \qquad
{\cal L}_{\sigma NN}=g_{\sigma NN}\bar\psi\psi \varphi_{\sigma}\,;\\
\quad \\
{\cal L}_{\omega NN}=-g_{\omega NN}\bar\psi
\gamma_\mu\psi\omega^\mu- \ds\frac{f_{\omega
NN}}{4M}\bar\psi\sigma_{\mu\nu}\psi
[\partial^{\mu}\omega_\nu-\partial^{\nu}\omega_\mu]\,;\\
\quad \\
{\cal L}_{\rho NN}= [ -g_{\rho NN}\bar\psi
\gamma_\mu\vec\tau\psi\vec\rho^\mu- \ds\frac{f_{\rho
NN}}{4M}\bar\psi\vec\tau\sigma_{\mu\nu}\psi
(\partial^{\mu}\vec\rho_\nu-\partial^{\nu}\vec\rho_\mu)]\,.
\label{tfd11} \ea \ee Here $g_{BNN}$ are the meson-nucleon
coupling constants and $f_{\omega NN}$ and $f_{\rho NN}$ are the
tensor coupling constants for $\omega$- and $\rho$- mesons
respectively.

The corresponding meson nucleon form factors are usually defined
as: \be \ba \matel{N(p')}{\Gamma_{\pi}^\alpha}{N(p)}=i
\gpinn (t)\bar{u}(p')\gamma_{5}\tau_\alpha u(p)\\
\quad\\
\matel{N(p')}{\Gamma_{\sigma}}{N(p)}=
-\gsnn(t)\bar{u}(p') u(p)\\
\quad\\
\matel{N(p')}{\Gamma_{\omega}^{\mu}   }{N(p)}=
\bar{u}(p')\left[\gamma^\mu\gomegann(t)+\ds\frac{i}{2M}\fomegann(t)
\sigma^{\mu\nu}q_\nu\right]u(p)\\
\quad\\
\matel{N(p')}{\Gamma_{\rho}^{\alpha,\mu}   }{N(p)}=
\bar{u}(p')\tau_{a}\left[\gamma^\mu\grhonn(t)+\ds\frac{i}{2M}\frhonn(t)
\sigma^{\mu\nu}q_\nu\right] u(p)\vergul\\
\label{tfd12} \ea \ee where $\gpinn $, $\gsnn$, $\gomegann$,
$(\fomegann)$ , $\grhonn$ and  $(\frhonn)$ are form factors of
 meson - nucleon  vertices;
  $t=q^2=q_{0}^{2}-\ds{(\vec{q})^2}=(p-p')^2$ is
the 4-momentum transfer and $M$ is the  nucleon mass.

%%%%%%%%%%%%%%%%%%PICTURE%%%%%%%%%%%%%%%%%%%%%%%%%%%%%%%%%%%%%%%%
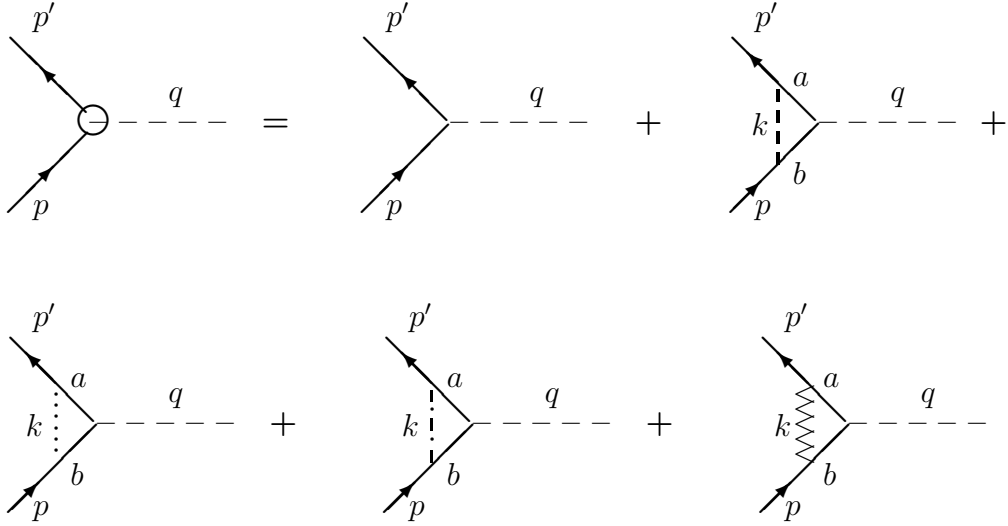
\begin{figure}[htb]
\setlength{\unitlength}{0.1mm}
\begin{picture}(1500,800)
\thicklines \normalsize \put(250,600){$q$}
\put(100,625){\line(-1,1){60}} \put(140,585){\vector(-1,1){60}}
\put(82,498){\line(1,1){60}} \put(37,453){\vector(1,1){60}}
\put(150,575){\circle{40}} \put(70,700){$p^{\prime}$}
\put(70,450){$p$}
%%%%2
\put(730,600){$q$} \put(145,560){\bf -- -- -- -- --\quad  =}
\put(570,625){\line(-1,1){60}} \put(620,575){\vector(-1,1){60}}
\put(562,508){\line(1,1){60}} \put(507,453){\vector(1,1){60}}
\put(625,560){\bf -- -- -- -- -- \quad +}
\put(540,700){$p^{\prime}$} \put(540,450){$p$}
%%%%3
\put(1210,600){$q$} \put(1025,555){$k$}
\put(1060,625){\line(-1,1){60}} \put(1110,575){\vector(-1,1){90}}
\put(1034,490){\line(1,1){80}} \put(997,453){\vector(1,1){40}}
\put(1115,560){\bf -- -- -- -- --\,\, +}
\put(1030,700){$p^{\prime}$} \put(1030,450){$p$}
\put(1080,620){$a$} \put(1080,490){$b$}
\put(1060,600){\line(0,1){15}} \put(1060,572){\line(0,1){15}}
\put(1060,544){\line(0,1){15}} \put(1060,516){\line(0,1){15}}
%%%%4
\put(60,155){$k$} \put(120,220){$a$} \put(120,90){$b$}
\put(250,200){$q$} \put(100,225){\line(-1,1){60}}
\put(150,175){\vector(-1,1){90}} \put(74,90){\line(1,1){80}}
\put(37,53){\vector(1,1){40}} \put(155,160){\bf -- -- -- --
--\quad +} \put(70,300){$p^{\prime}$} \put(70,50){$p$}
\put(95,200){$\cdot$} \put(95,185){$\cdot$} \put(95,170){$\cdot$}
\put(95,155){$\cdot$} \put(95,140){$\cdot$} \put(95,125){$\cdot$}
%%%%5
\put(560,155){$k$} \put(620,220){$a$} \put(620,90){$b$}
\put(750,200){$q$} \put(600,225){\line(-1,1){60}}
\put(650,175){\vector(-1,1){90}} \put(574,90){\line(1,1){80}}
\put(537,53){\vector(1,1){40}} \put(655,160){\bf -- -- -- --
--\quad +} \put(570,300){$p^{\prime}$} \put(570,50){$p$}
\put(600,200){\line(0,1){15}} \put(595,180){$\cdot$}
\put(600,162){\line(0,1){15}} \put(595,140){$\cdot$}
\put(600,120){\line(0,1){15}}
%%%%6
\put(1055,155){$k$} \put(1120,220){$a$} \put(1120,90){$b$}
\put(1250,200){$q$} \put(1100,225){\line(-1,1){60}}
\put(1150,175){\vector(-1,1){90}} \put(1074,90){\line(1,1){80}}
\put(1037,53){\vector(1,1){40}} \put(1155,160){\bf -- -- -- -- --
} \put(1070,300){$p^{\prime}$} \put(1070,50){$p$}
\put(1080,200){$<$} \put(1080,180){$<$} \put(1080,160){$<$}
\put(1080,140){$<$} \put(1080,120){$<$}
\end{picture}
   \vspace{1.0cm}
\medskip
   \centerline{\parbox{15cm}
   {\caption{\label{figtfd1}
Feynman diagrams for pion - nucleon vertex. The solid line is for
nucleon. Dashed, dotted, dot - dashed and wavy lines  are for
$\pi$, $\sigma$, $\omega$ and  $\rho$ mesons respectively. }}}
\end{figure}
%%%%%%%%%%%%%%%%%%%%%%%%%%%%%%%%%%%%%%%%%%%%%%%%%%%%%%%%%%%%%%%%%%%
%\large

In order to investigate the $T$ dependence we calculate three
line vertex correction, as is illustrated in Fig. \ref{figtfd1}
e.g. for
 $\gpinn (t,T)$.
In accordance with Feynman rules this may be written as: \be \ba
\Gamma_A(t,T)=\Gamma_A(t)+i\ds\sum_B\Lambda_{AB}(t,T),\\
\quad \\
\Lambda_{AB}(t,T)=\ds\int\frac{d^4k}{(2\pi)^4}
\Gamma_B(k^2)S_F(a)\Gamma_A(t)S_F(b)\Gamma_B(k^2)\delb(k^2)
\label{tfd13} \ea \ee where $A, B$ - denote
$\pi,\sigma,\omega,\rho$ mesons, e.g. $\Gamma_{\pi}=\gpinn(t)$,
$a=p'-k, b=p-k$. Here $S_F(a)$
 and $\Delta^{B}(k^2)$
are the in - medium nucleon and meson propagators respectively.
In the framework of TFD the thermal propagator has a $2\times 2$
matrix  structure, but only the $11$-component refers to the
physical field. So, we may use the following representation: \be
\ba S_{F}^{11}(a)\equiv S_F(a)=(\hat a+M)[G_0(a)+G_T(a)]\,, \nwl
\Delta^{11}\equiv \Delta(k^{2})=D_0(k^2)+D_T(k^2)\,, \nwl
D_0(k^2)=\ds\frac{1}{k^2-m^2+i\varepsilon}\,,\quad D_T(k^2)=-2\pi
i n_B(k)\delta(k^2-m^2)\,, \nwl
G_0(a)=\ds\frac{1}{a^2-M^2+i\varepsilon}\,,\quad G_T(a)=2\pi i
\delta(a^2-M^2)N_F(a)\,, \label{tfd14} \ea \ee where   $\hat a
=a_\mu \gamma^\mu$,
       $N_F(a)=\theta(a_0)n_F(a)+\theta(-a_0)\bar n_F(a)$,
 $m$ is the mass
of corresponding meson and \be \ba
n_F(a)=\dsf{1}{e^{\beta(|a_0|-\mu)}+1}\,,\qquad \bar
n_F(a)=\dsf{1}{e^{\beta(|a_0|+\mu)}+1}
\\
\\
n_{B}(k)=\dsf{1}{e^{\beta|k_0|}-1} \label{tfd15} \ea \ee are the
fermion, antifermion and boson distribution functions
respectively. It is clear that when one substitutes Eq. \re{tfd14}
into Eq. \re{tfd13},  $\Lambda_{AB}(t,T)$ will
 be separated into two parts:
one refers to the naive zero temperature  contribution, while
the other one depends on the density and temperature \ci{zhang}.
 We shall concentrate on the latter part, which may be rewritten
as follows: \be \ba
\Lambda_{AB}(t,T)=\ds\int\dsf{d^4k}{(2\pi)^4}\Gamma_B(k^2)
(\hat a+M)\Gamma_A(t)(\hat b+M)\Gamma_B(k^2)\times\\
\quad \\
\times[G_0(a)G_0(b)D_T^B(k^2)+ 2G_0(a)G_T(b)D_0^B(k^2)]
\lab{tfd16} \ea \ee Note that we have neglected terms that are
quadratic in the temperature dependent distribution function.
Now, using \re{tfd16}  and \re{tfd14} in \re{tfd13} we get the
following expressions for the form factors: \be \ba
G_{ANN}(t,T)/G_{ANN}(t,T=0)=1+
i\ds\sum_B\int\dsf{d^4k}{(2\pi)^4}W_{AB}(t,k^2,T)\times \nwl
\\
\times[G_0(a)G_0(b)D_T^B(k^2)+ 2G_0(a)G_T(b)D_0^B(k^2)]
\lab{tfd17} \ea \ee where the explicit formulas for
$W_{AB}(t,k^2,T)$  may
 be found in the Appendix C.
Here  we present the expressions for $W_{\omega\pi}(t,k^2,T)  $
and $W_{\rho\pi}(t,k^2,T)  $ for illustration: \be \ba
W_{\omega\pi}(t,k^2,T) =\dsf{ 3G_{\pi NN}^2(k^2) } { 4 }
\left\{2[2M^2-(ab)]- \dsf{3F_{\omega NN}(t)t} {2G_{\omega
NN}(t)}\right\}\vergul
\\
\\
W_{\rho\pi}(t,k^2,T) =\dsf{-G_{\pi NN}^2(k^2)}{4}
\left\{2[2M^2-(ab)]- \dsf{3F_{\rho NN}(t)t}{2G_{\rho
NN}(t)}\right\}\vergul \label{tfd18} \ea \ee where
$W_{\omega\pi}(t,k^2,T)$ and $W_{\rho \pi} (t, k^{2}, T)$ denote
the contribution from the $\pi$- exchange diagram to the
$\gomegann(t,T)$ and $G_{\rho NN}(t, T)$ respectively.

\section{Results and discussions}
\indent

\begin{figure}[htb]
\bc \epsfysize=14cm \epsfbox{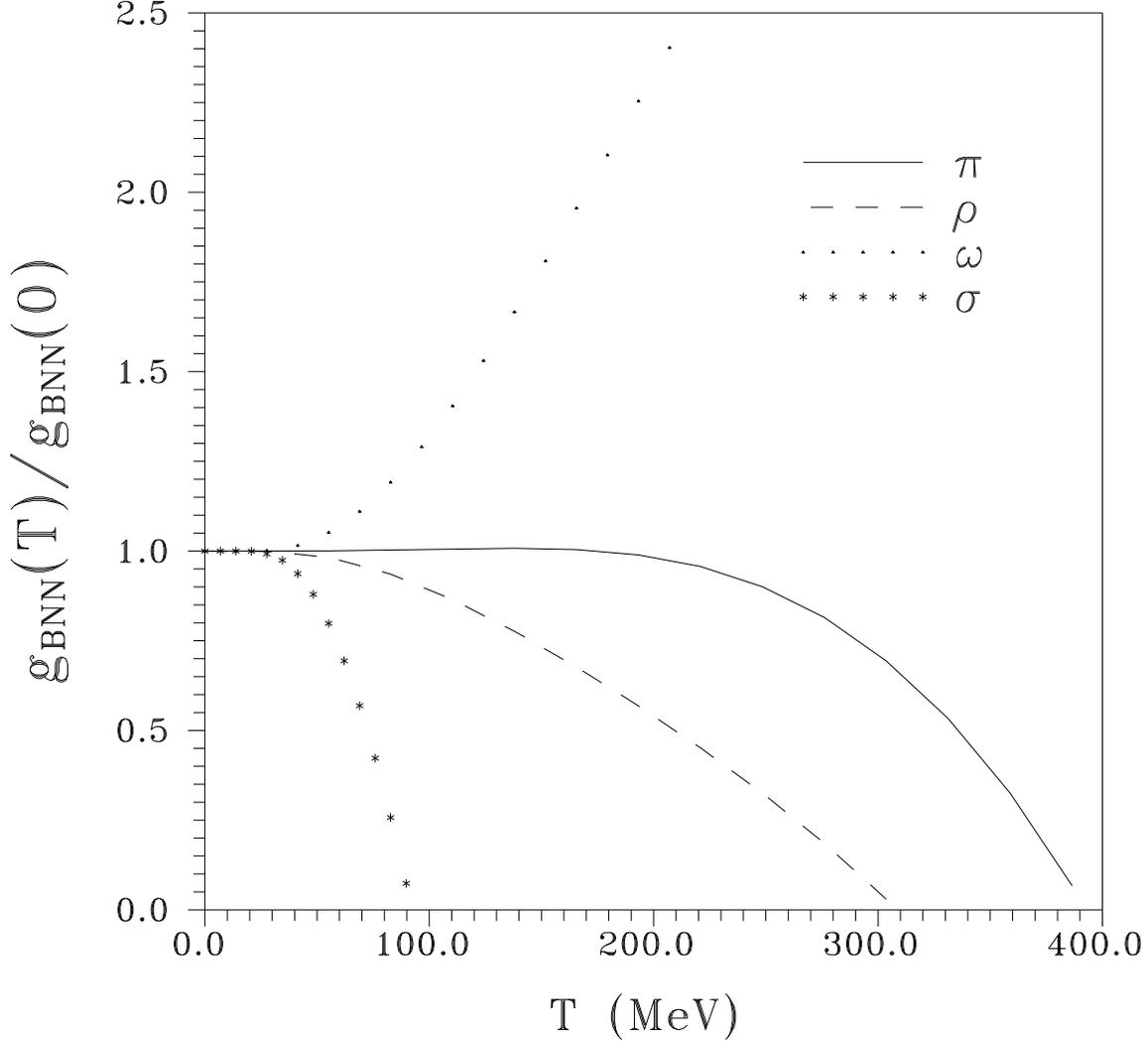} \ec
   \centerline{\parbox{15cm}
   {\caption{\label{figtfd2}
The ratio of meson-nucleon coupling constants at finite
temperature T at T=0 as a function of temperature; $\rho=0$.
 }}}
\end{figure}

In order to start the  calculations, a set of free space meson -
nucleon form factors are chosen. We choose the  OBE monopole form
factors \ci{machrep}
 (Bonn A):
 $G_{BNN}(t)=g_{BNN}(\Lambda_{B}^{2}-m_{B}^{2})/(\Lambda_{B}^{2}-t) $
,where, for example, $g_{\pi NN}^{2}/4\pi=14.09$,
 $\Lambda_{\pi NN}=1005MeV$, $m_\sigma= 550 $MeV.
Let us first discuss  the $T$ dependence of meson - nucleon
coupling constants - $g_{BNN}(T)\equiv G_{BNN}(t=m_{B}^{2},T)$.
The variation of the coupling constants $g_{\pi NN}(T)/g_{\pi NN}(T=0),\\
 g_{\sigma NN}(T)/g_{\sigma NN}(T=0),
g_{\omega NN}(T)/g_{\omega NN}(T=0)$ and $g_{\rho NN}(T)/g_{\rho
NN}(T=0)$ with temperature are displayed in Fig. \ref{figtfd2}.
%%%%%%%%%%%%%%FIG 2%%%%%%%%%%%%%%%%%%%%%%%%%%%%%%%%%%%%%%%%%%%%%%%%%%%%%
%%%%%%%%%%%%%%FIG 3%%%%%%%%%%%%%%%%%%%%%%%%%%%%%%%%%%%%%%%%%%%%%%%%%%%%%
\begin{figure}[htb]
\bc \epsfysize=14cm \epsfbox{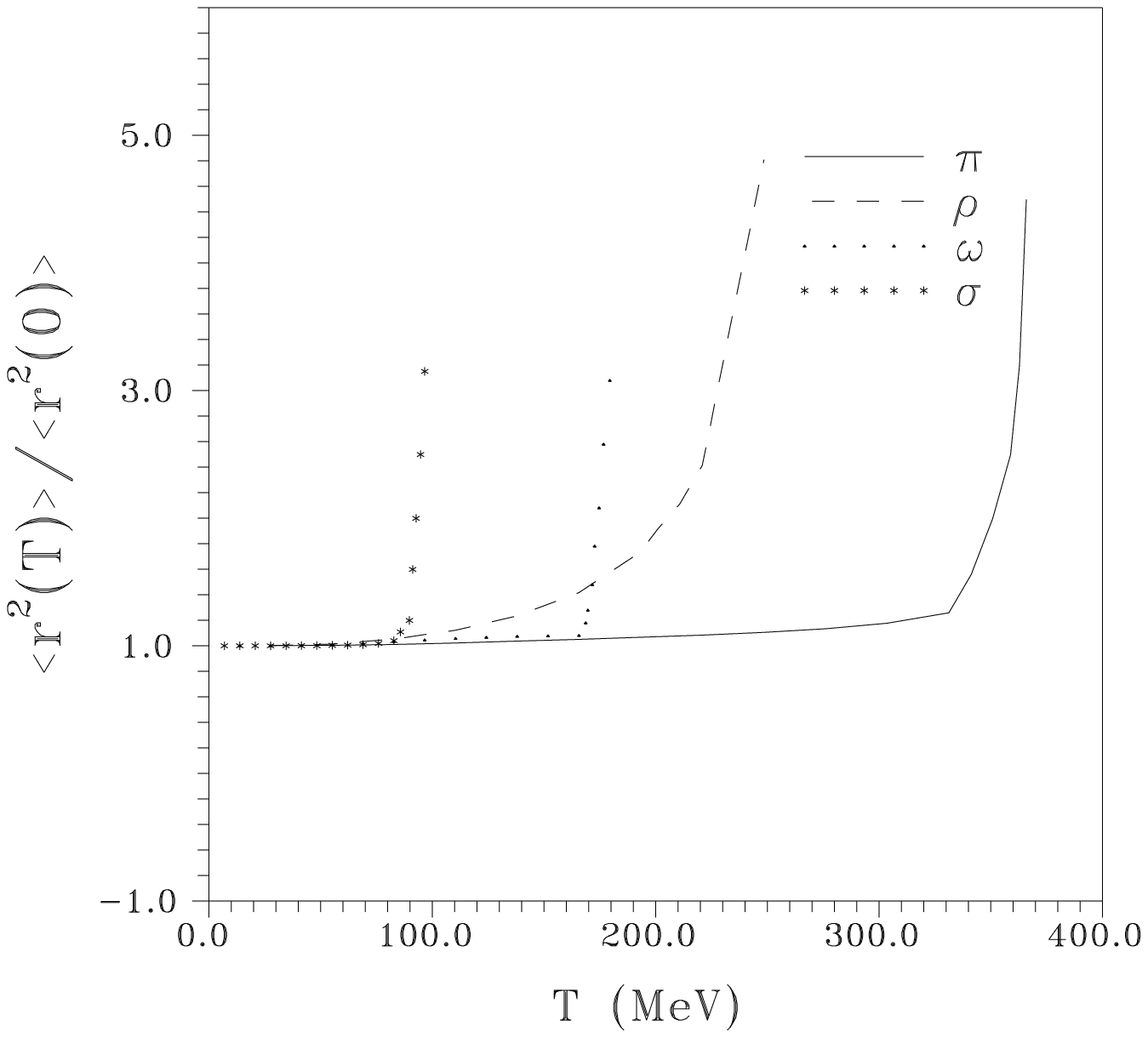} \ec
   \centerline{\parbox{15cm}
   {\caption{\label{figtfd3}
The ratio of mean square radius at finite temperature T and at T
= 0 as a function of temperature; $\rho=0$.
 }}}
\end{figure}
%%%%%%%%%%%%%%%%%%%%%%%%%%%%%%%%%%%%%%%%%%%%%%%%%
It is clear that they are nearly independent of  the temperature
below  $T_{c}^{B}$, then change rapidly for $T > T_{c}^{B}$. A
similar behavior was also  predicted by Zhang et al.   \ci{zhang}
and by Dominguez et al. \ci{dom}
 for pion - nucleon coupling constant
 $g_{\pi NN}(T)/g_{\pi NN}(T=0)$.
But here in Fig. \ref{figtfd2}, an unexpected result is that, the
$\omega -N$ coupling constant, $g_{\omega NN}(T)$ increases while
the coupling constants of all other mesons decrease! Let's
consider a possible origin of this controversy, by comparing the
medium modification of $\omega NN$ and $\rho NN$ coupling
constants. In the present model the modifications arise from the
triangle diagrams (Fig.  \ref{figtfd1}). Actual calculations show
that, for any $G_{B NN}(t,T)$ the triangle diagram with pion
exchange gives a dominant contribution especially at small
density. It is clear from Eq. \re{tfd18} that at small $t$ the
explicit expressions for $g_{\omega NN}(T)$ and  $g_{\rho NN}(T)$
 formally coincide. However, $W_{\rho\pi}=-\frac{ W_{\omega\pi} }{3}$.
and hence the two have an opposite sign. The factor $(-1/3)$
arises from the isotopic spin structure. In fact, using  Eq.
\re{tfd13} and Fig.\ref{figtfd1}, $g_{\omega NN}(T)$ and
$g_{\rho NN}(T)$ may be written in a schematic way: \be \ba
g_{\omega NN}(T)\approx g_{\omega NN}(T=0)+
 g_{\pi NN}^{2} \ds\sum_{\beta}  \tau_{\beta}
                   g_{\omega NN}\tau_{\beta} +\dots
\nwl g_{\rho NN}(T)\tau_\alpha\approx g_{\rho NN}(T=0)\tau_\alpha+
 g_{\pi NN}^{2}\ds\sum_{\beta}  \tau_{\beta}\tau_\alpha
         g_{\rho NN}\tau_{\beta} +\dots
\ea \lab{tfd19} \ee where $\alpha$ and $\beta$ are isospin
indices.
%%%%%%%%%%%  TABLES   1,2,3
%\normalsize
\begin{center}
\begin{table}[htb]
\caption{\it Parameters  of vertex form factors in Eq.s
\re{tfd20}-\re{tfd21} at half of normal nuclear
 matter density  $\rho=0.5\rho_0.$}
\vskip 0.2cm
\begin{center}
\begin{tabular}{|c|c|c|c|c|c|}\hline
Meson&$\alpha_g$&$\beta_g$&$\alpha_\lambda$&$\beta_\lambda$&$T_c$\\
\hline
$\pi$   & 0.9925&-1.3538& 0.6789&-1.0397 &$\approx 180$\\
$\sigma$& 0.2572&-1.2507& 1.0413&-1.6472&$\approx 90$\\
$\rho$  & 0.3678&-0.8579 &-0.4821&-0.4655&$\approx 180$\\
$\omega$& 2.1001&-1.8110& 1.1595&- 4.44165&$\approx 175$\\
\hline
\end{tabular}
\end{center}
\label{tftb1}
\end{table}
\end{center}
\begin{center}
\begin{table}[htb]
\caption{\it The same as in Table \ref{tftb1} , but for
$\rho=\rho_{0}$} \vskip 0.2cm
\begin{center}
\begin{tabular}{|c|c|c|c|c|c|}\hline
Meson&$\alpha_g$&$\beta_g$&$\alpha_\lambda$&$\beta_\lambda$&$T_c$\\
\hline
$\pi$   &1.5047&- 1.7415&1.5816& -  2.2862  &$\approx 185$\\
$\sigma$&0.8238&- 1.8623&1.5911& -  2.0814&$\approx 90$\\
$\rho$  &0.6556&- 0.9325 &- 0.1778&- 0.5858&$\approx 180$\\
$\omega$&3.0774&- 2.6030&3.0351&- 9.8837&$\approx 175$\\
\hline
\end{tabular}
\end{center}
\label{tftb2}
\end{table}
\end{center}
\begin{center}
\begin{table}[htb]
\caption{\it The same as in Table \ref{tftb1} , but for
$\rho=3\rho_{0}$} \vskip 0.2cm
\begin{center}
\begin{tabular}{|c|c|c|c|c|c|}\hline
Meson&$\alpha_g$&$\beta_g$&$\alpha_\lambda$&$\beta_\lambda$&$T_c$\\
\hline
$\pi$   &3.5869&- 3.4716&5.5941&- 8.5846 &$\approx 220$\\
$\sigma$&3.0715&-4.2687 &3.8743&- 4.0881&$\approx 90$\\
$\rho$  &2.3344&- 2.5660 &1.3642&- 2.3675&$\approx 210$\\
$\omega$&6.4629&- 5.0795&9.2984&-26.3338&$\approx 175$\\
\hline
\end{tabular}
\end{center}
\label{tftb3}
\end{table}
\end{center}
We omit the spin variables, since $\omega$ and $\rho$ have the
same spin structure. Now, using $\ds\sum_{\beta}\tau_{\beta}^2=3$
and
$\ds\sum_{\beta}\tau_{\beta}\tau_\alpha\tau_{\beta}=-\tau_\alpha$,
it is clear that the contribution of the leading triangle diagram
with pion exchange to   $g_{\omega NN}(T)$ and  $g_{\rho NN}(T)$
has different signs. In other words, the term $\Lambda_{\rho\pi}$
in Eq. \re{tfd13} is negative, while $\Lambda_{\omega\pi}$   is
positive.

In general, we conclude  from the results (Fig. \ref{figtfd2})
that at a certain critical temperature $T_c^{B}$, the couplings
$g_{BNN}$ change drastically. The thermal behavior of $g_{\pi NN}
(T) = \gpinn(t,T)|_{t\to 0}$ has been investigated earlier
  \ci{dom} and we get here a similar behaviour of
$g_{\pi NN} (T)$ as a function of temperature.
 The temperature $T_c$, where $ g_{\pi NN} (T) $
changes dramatically,  was interpreted in  \ci{dom} as a signal
for the quark - gluon deconfinement phase transition. Indeed,
near $T_c$ the associated mean square radius
 $<r^{2}_{BNN}> = 6[\frac{ \partial}{\partial t} \ln G_{BNN}(t,T)]|_
{t \to 0}  $ is a monotonically increasing function of $T$ and in
fact diverges at the critical temperature. A similar behavior of
$<r^{2}_{BNN}>$ is also found in our calculations and is
illustrated in Fig. \ref{figtfd3}.
 As the critical temperature, $T_{c}^{B}$, is approached the
 strength of coupling of $\pi, \sigma$ and $\rho$ mesons
to nucleons is quenched, at the same time, the size of nucleons
as probed by the appropriate meson gets bigger. However,
 it is to be stressed that
$T_{c}^{B}$ is not the same for all mesons: $T_{c}^{\pi} \approx
360 MeV$, $T_{c}^{\sigma} \approx 95 MeV$, $T_{c}^{\omega}
\approx 175 MeV$ and $T_{c}^{\rho} \approx 200 MeV$. In OBE
picture this means that , in the temperature region e.g. $200
MeV< T <300 MeV $, the $\rho$  meson exchange is no longer
important while the pion exchange is still important. On the
other hand, the $\sigma$ and $\pi$ mesons are mainly responsible
for the attraction between nucleons. So, the quenching of the
$\sigma NN$ coupling constant at $T>T_{c}^{\sigma}$ leads to the
vanishing of the bound state, as it was predicted earlier
\ci{zhang}.

The density dependence of vertex form factors has been studied in
greater detail \ci{ournpamed,meissmed}. It was found that most of
the vertex form factors are quenched at high densities and it was
anticipated that the temperature dependence is likely to yield
results that are qualitatively similar to those of density
dependence.
%%%%%%%%%%%%%%FIG 4a%%%%%%%%%%%%%%%%%%%%%%%%%%%%%%%%%%%%%%%%%%%%%%%%%%%%%
\begin{figure}[!htb]
\bc \epsfysize=14cm \epsfbox{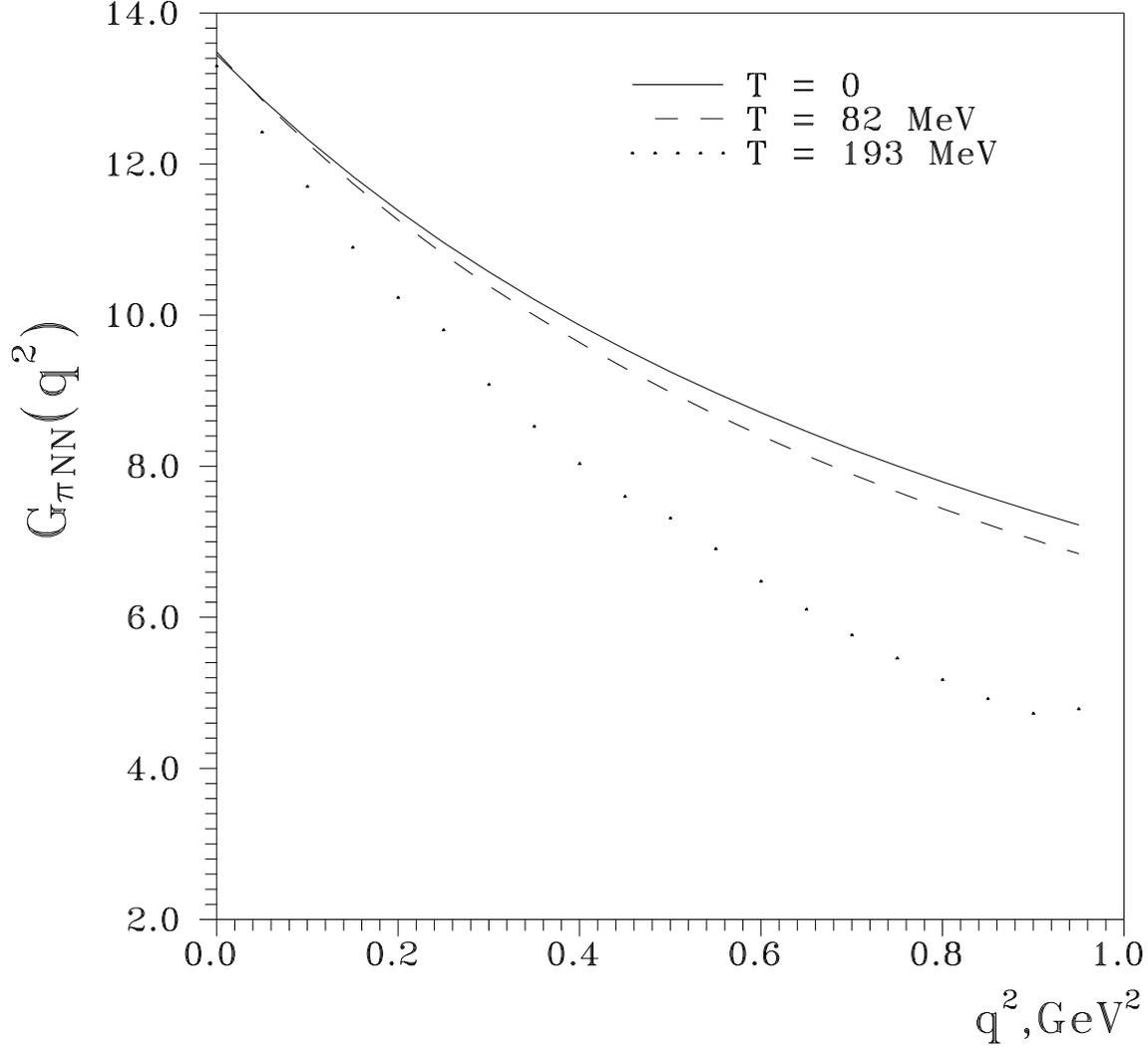} \ec
   \centerline{\parbox{15cm}
   {\caption{\label{figtfd4a}
Meson nucleon form factors at several temperatures as a function
of $q^{2}$ using parametrisation of BONN group \ci{machrep} for
  $\pi NN$ vertex
at $\rho=0$. }}}
\end{figure}
%%%%%%%%%%%%%%%%%%%%%%%%%%%%%%%%%%%%%%%%%%%%%%%%%

%%%%%%%%%%%%%%FIG 4b%%%%%%%%%%%%%%%%%%%%%%%%%%%%%%%%%%%%%%%%%%%%%%%%%%%%%
\begin{figure}[htb]
\bc \epsfysize=14cm \epsfbox{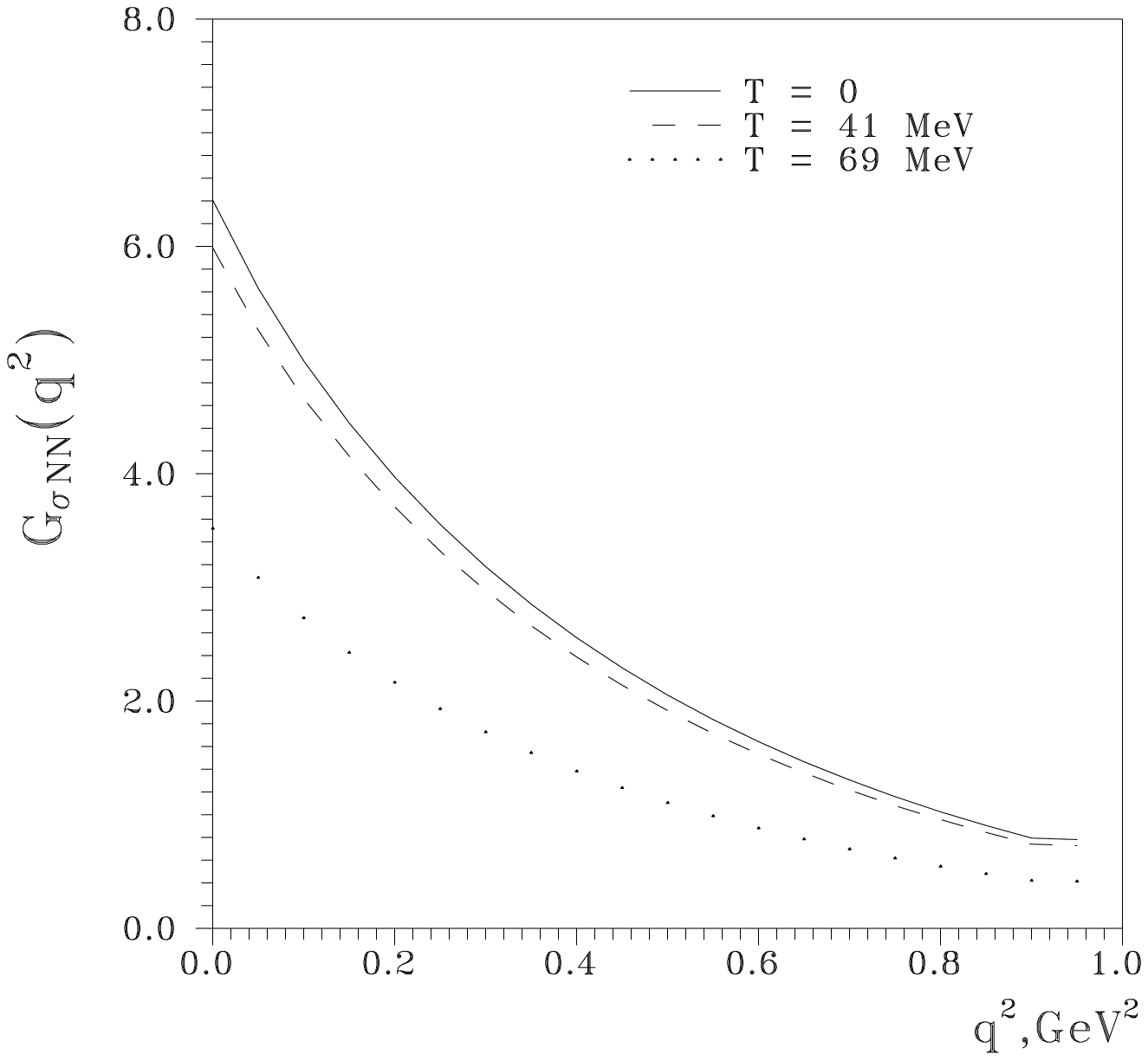} \ec
   \centerline{\parbox{15cm}
   {\caption{\label{figtfd4b}
The same as in Fig \ref{figtfd4a} but for  $\sigma NN$ vertex. }}}
\end{figure}
%%%%%%%%%%%%%%%%%%%%%%%%%%%%%%%%%%%%%%%%%%%%%%%%%

%%%%%%%%%%%%%%FIG 4c%%%%%%%%%%%%%%%%%%%%%%%%%%%%%%%%%%%%%%%%%%%%%%%%%%%%%
\begin{figure}[!htb]
\bc \epsfysize=14cm \epsfbox{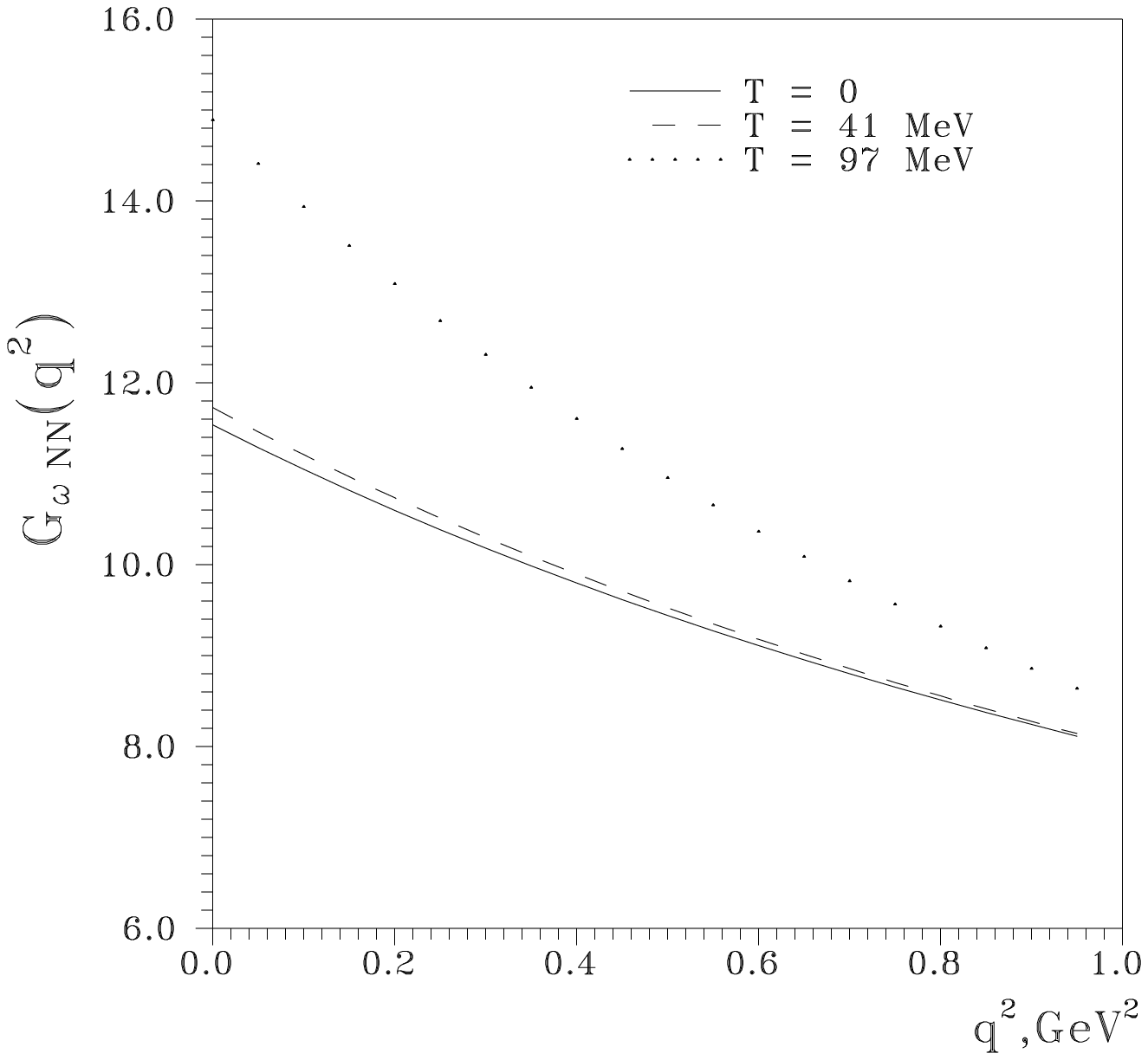} \ec
   \centerline{\parbox{15cm}
   {\caption{\label{figtfd4c}
The same as in Fig \ref{figtfd4a} but for  $\omega NN$ vertex. }}}
\end{figure}
%%%%%%%%%%%%%%%%%%%%%%%%%%%%%%%%%%%%%%%%%%%%%%%%%
%%%%%%%%%%%%%%FIG 4d%%%%%%%%%%%%%%%%%%%%%%%%%%%%%%%%%%%%%%%%%%%%%%%%%%%%%

  Now, let's consider their temperature dependence in some detail.
The form factors as a function of momentum transfer at several
temperatures are displayed in Figs. \ref{figtfd4a} -
\ref{figtfd4c}.
 It is seen that, the in-medium effects
lead to the suppression of $\gpinn(t,T)$,  $\gsnn(t,T)$ and
$\grhonn(t,T)$.
\begin{figure}[!htb]
\bc \epsfysize=14cm \epsfbox{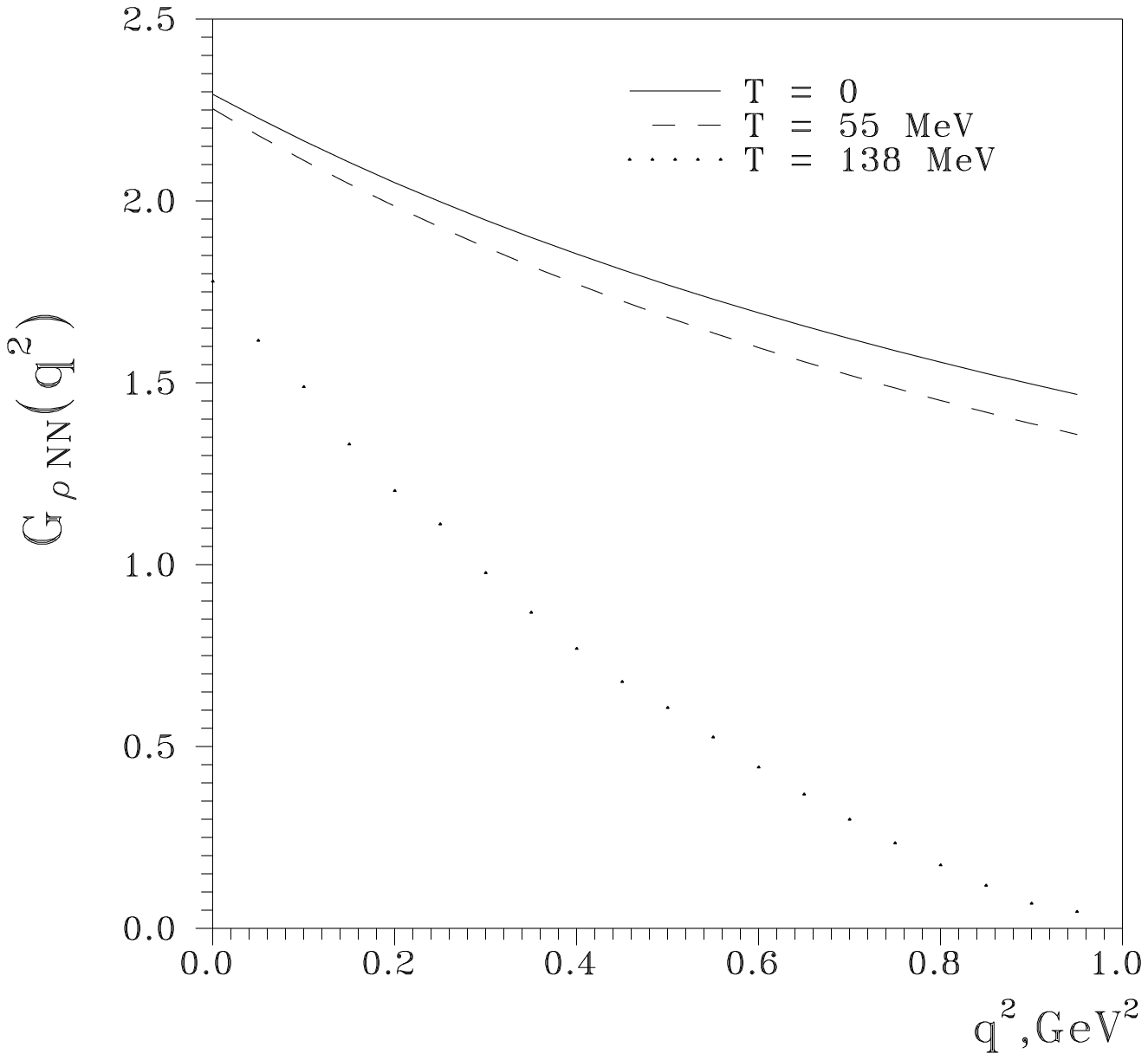} \ec
   \centerline{\parbox{15cm}
   {\caption{\label{figtfd4d}
The same as in Fig \ref{figtfd4a} but for $\rho NN$ vertex. }}}
\end{figure}
%%%%%%%%%%%%%%%%%%%%%%%%%%%%%%%%%%%%%%%%%%%%%%%%%
The temperature dependence of $\gomegann (t,T)$, is opposite to
that of $\grhonn$ due to it's isotopic structure as outlined
above.
%           INSERTED TEXT%%%%%%%%%%%%%%%%%
The temperature dependence of tensor couplings of vector mesons
$F_{VNN}(t,T) $ are quite similar to that of vector couplings
$G_{VNN}(t,T)$. Particularly, the ratio
$\kappa_v=F_{VNN}/G_{VNN}$, where $\kappa_\rho=6.1$ and
$\kappa_\omega=0$ in free space, remains constant in a wide range
of temperature.
%%%%%%   %%%%%%%%%%%%%%%%%%%%%%%%%%%%%%%%%%%%%%%%%%
For practical calculations a parametrization of these
$G_{BNN}(t,T,\rho)$ form factors is needed. At small momentum
transfer   we can
 parametrize them by a monopole form :
\be
G_{BNN}(t,T,\rho)=g_{B}(T,\rho)(\Lambda_{B}^{2}(T,\rho)-m_{B}^{2})
/(\Lambda_{B}^{2}(T,\rho)+t) \label{tfd20} \ee
 where in general the
 effective mass of a meson,
$m_{B}$, is also temperature dependent. Consideration of this
dependence is beyond the scope of the present Chapter.
 But here, for simplicity, one may consider this as
 just a parameterization  and
choose the parameters $g_{B}(T,\rho)$ and
$\Lambda_{B}^{2}(T,\rho)$. Their  temperature dependence is still
unknown. Here  the T dependence  may be represented, for $T<<
T_{c}^{B}$, in a polynomial form as:
%%%%%%%%  this formula has been rewritten%%%%%%%%%%
\be \ba
\dsf{g_{B}(T,\rho)}{g_{B}(T=0,\rho=0)}=\Phi(\rho)[1+\ds\alpha_B^g(\rho)(T/T_c^B)
^2+
\ds\beta_B^g(\rho)(T/T_c^B)^4],\\
\quad\\
\dsf{\Lambda_{B}(T,\rho)}{\Lambda_{B}(T=0,\rho=0)}=\Phi(\rho)[
1+\alpha_B^\lambda(\rho)(T/T_c^B)^2+
\beta_B^\lambda(\rho)(T/T_c^B)^4],\\
\ea \label{tfd21} \ee where $\Phi(\rho)=1/(1+C_0 \rho/\rho_0)$.
The calculated form factors are fitted to this form and the
paramters   $C_0$,  $\alpha$ and $\beta$ are determined. The
results are presented in Tables \ref{tftb1}, \ref{tftb2} and
\ref{tftb3} for densities $\rho=0.5\rho_0$,  $\rho=\rho_0$ and
$\rho=3\rho_0$, respectively (with $\rho_0=0.17fm^{-3}$ - the
density of normal nuclear matter). The results show that the
density dependence of parameters, and hence, the form factors
 is not so drastic (sharp) as their
temperature dependence in the range of moderate densities. So, at
$T\approx 0$ the Eq. \re{tfd21} with $C_0=0.26$ is in good
agreement with Brown Rho scaling law \ci{frrhosong} (where
$C_0=0.28$).

Summarizing this Chapter, we have considered the temperature
dependence of meson - nucleon form factors and coupling constants
in TFD formalism. It is shown that at a critical temperature,
where the coupling constant changes drastically and the
associated mean square radius diverges is different for different
mesons. The temperature dependent vertex form factors are
parametrized in a simple monopole form and the T-dependence of
these parameters is clarified. These form factors may be used in
the calculation of the in - medium NN cross sections
  \ci{limach}
and in investigations of the properties of hot dense matter.

%\large

\begin{figure}[!htb]
\bc \epsfysize=14cm \epsfbox{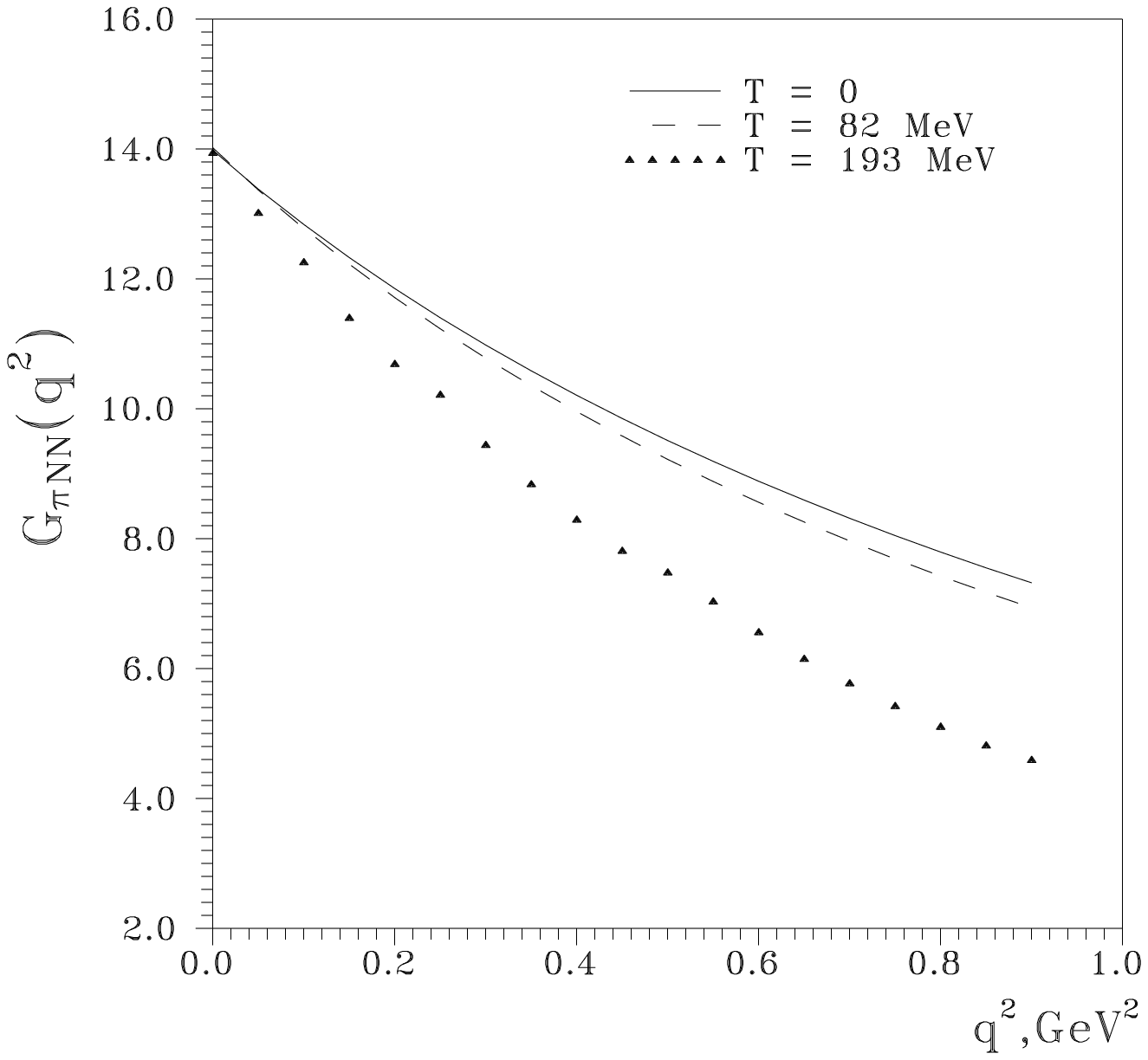} \ec
   \centerline{\parbox{15cm}
   {\caption{\label{figtfd5a}
Meson-nucleon form factors at several temperatures as a function
of $q^{2}$ using parametrisation of Meissnar et al. \ci{ourpirho}
for  $\pi NN$ vertex at $\rho=0$. }}}
\end{figure}
In the present calculations, the monopole form factors with the
parameters (at T = 0) given by the Bonn group \ci{machrep} are
used. But the next question which may arise is: are the results
sensitive to the shape of the input form factors? To answer this
question a different set of form factors
 obtained in Chapter IV in the framework
of  a topological soliton model have been used.
\begin{figure}[!htb]
\bc \epsfysize=14cm \epsfbox{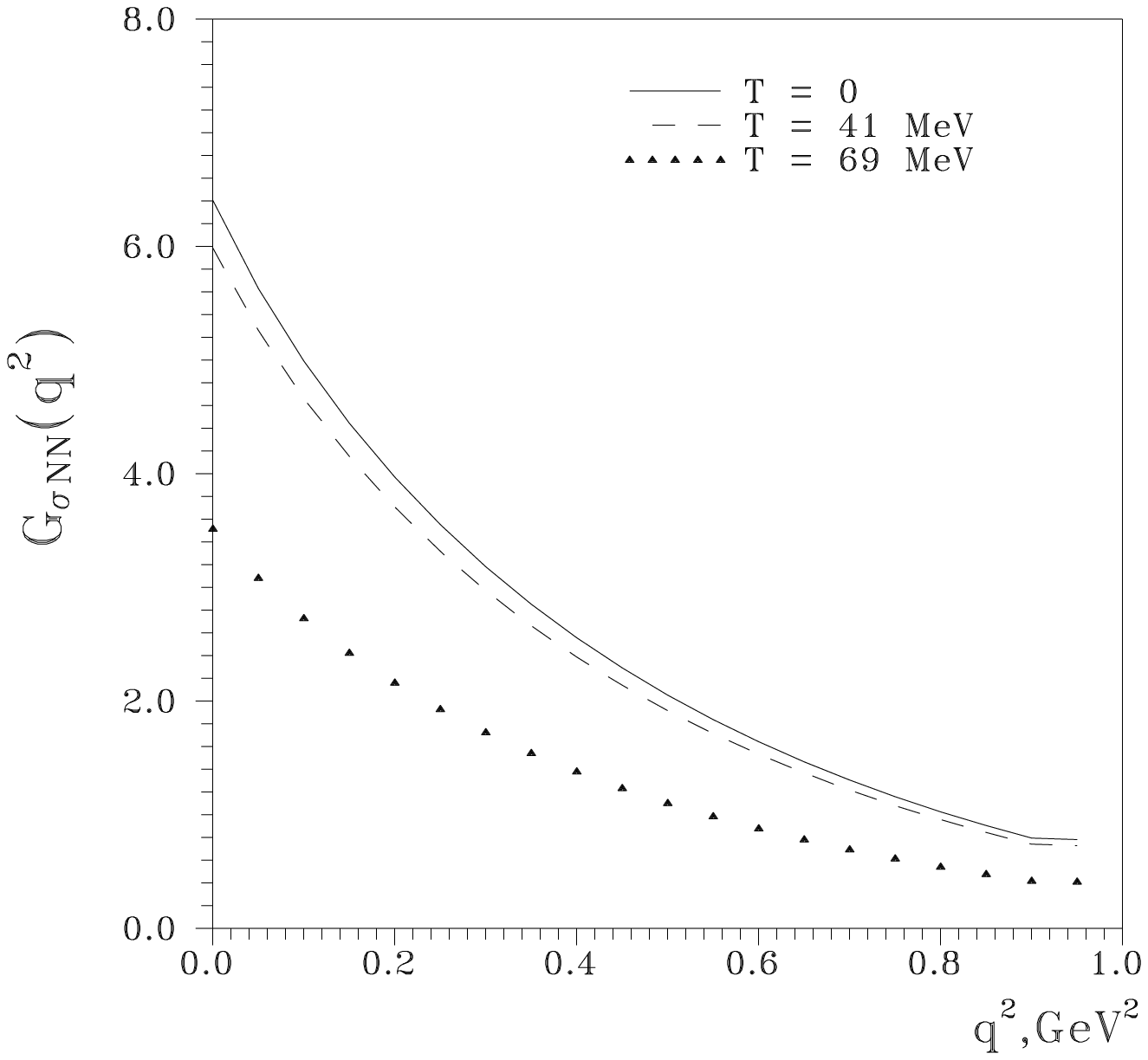} \ec
   \centerline{\parbox{15cm}
   {\caption{\label{figtfd5b}
The same as in Fig. \ref{figtfd5a} but for  $\sigma NN$ vertex.
}}}
\end{figure}
The thermal behaviour of such form factors  is shown in Figs.
\ref{figtfd5a} - \ref{figtfd5d}.
 Now, by comparing Figs. \ref{figtfd4a} - \ref{figtfd4d} and
 Figs.\ref{figtfd5a} - \ref{figtfd5d} it is clear
that only the $\sigma NN$ form factor is affected the most. The
reason is that even in free space and at $T=0$ the $\sigma NN$
form factor in the topological soliton model   \ci{ourpirho} is
much  harder ($\Lambda_{\sigma}\approx 600 MeV$)
 than those of the Bonn potential ($\Lambda_{\sigma}\approx 2000 MeV$).
Nevertheless, our main conclusions about thermal behavior of
different meson nucleon vertices remain valid.
%%%%%%%%%%%%%%FIG 5a%%%%%%%%%%%%%%%%%%%%%%%%%%%%%%%%%%%%%%%%%%%%%%%%%%%%%

%%%%%%%%%%%%FIG 5b%%%%%%%%%%%%%%%%%%%%%%%%%%%%%%%%%%%%%

%%%%%%%%%%%%%%%%%%%%%%%%%%%%%%%%%%%%%%%%%%%%%%%%%
%%%%%%%%%%%%FIG 5c%%%%%%%%%%%%%%%%%%%%%%%%%%%%%%%%%%%%%
\begin{figure}[!htb]
\bc \epsfysize=14cm \epsfbox{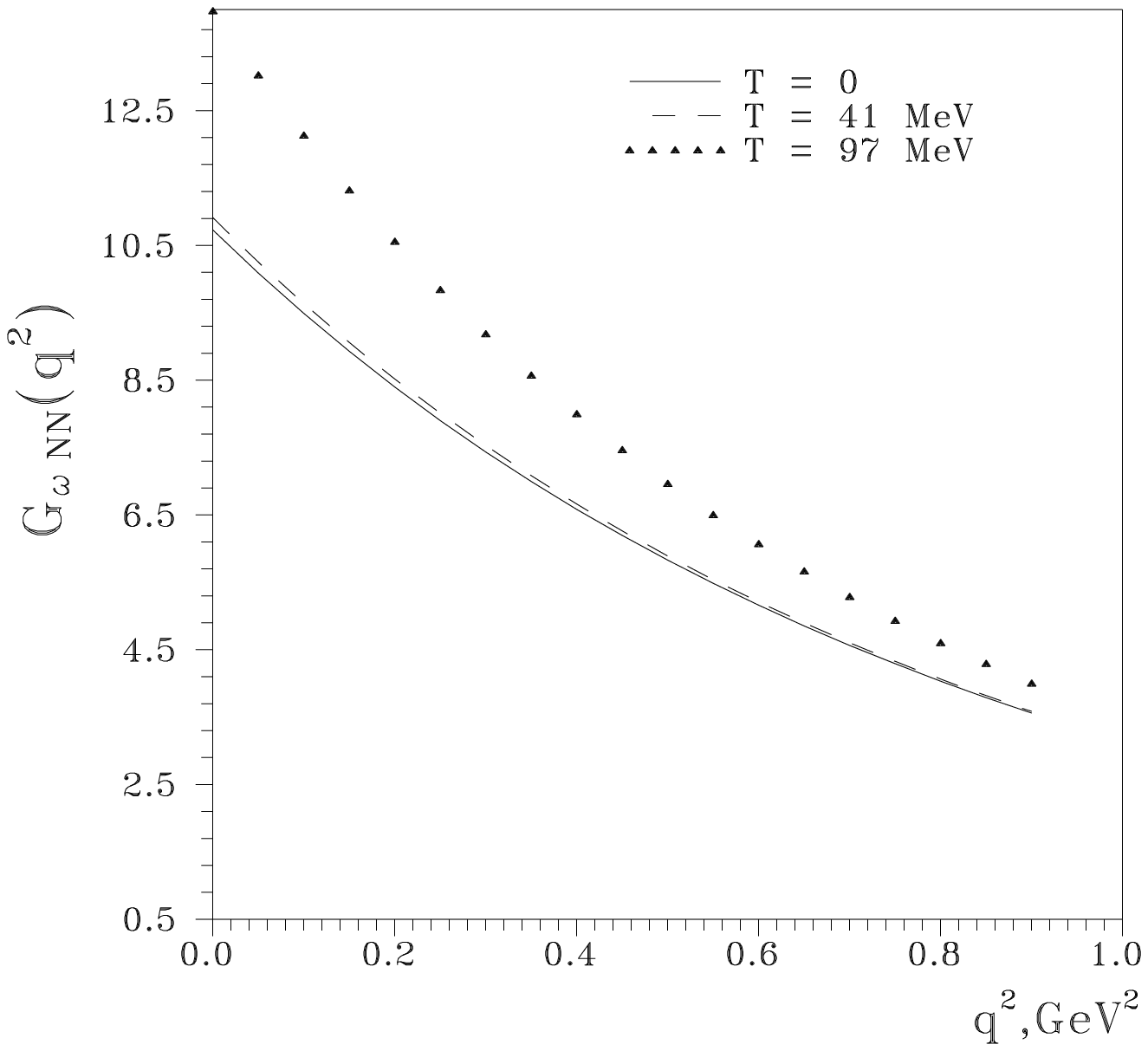} \ec
   \centerline{\parbox{15cm}
   {\caption{\label{figtfd5c}
The same as in Fig. \ref{figtfd5a} but for   $\omega NN$ vertex.
}}}
\end{figure}
%%%%%%%%%%%%%%%%%%%%%%%%%%%%%%%%%%%%%%%%%%%%%%%%%
The variation of the meson-nucleon coupling constant with
temperature (Fig. \ref{figtfd2}) clearly indicates that the
attractive NN force due to $\sigma$-exchange decreases quite
rapidly while the repulsive N-N force due to $\omega$-exchanges
increases. This would lead to the fact that the nuclear matter is
quite likely un-bound at high temperature. In fact it will look
more like a hard sphere gas since the repulsive interaction at
short distances will dominate. The attractive interactions in
such a case play only a small perturbative role. The hot and
dense nuclear matter could be approximated as a free gas with an
excluded volume around each nucleon. Perhaps the extensive studies
\ci{haged1}-\ci{haged4} along this line can be justified on the
basis of the results obtained in this Chapter. The nuclear matter
would be hard to compress.
%%%%%%%%%%%%FIG 5d%%%%%%%%%%%%%%%%%%%%%%%%%%%%%%%%%%%%%
\begin{figure}[!htb]
\bc \epsfysize=14cm \epsfbox{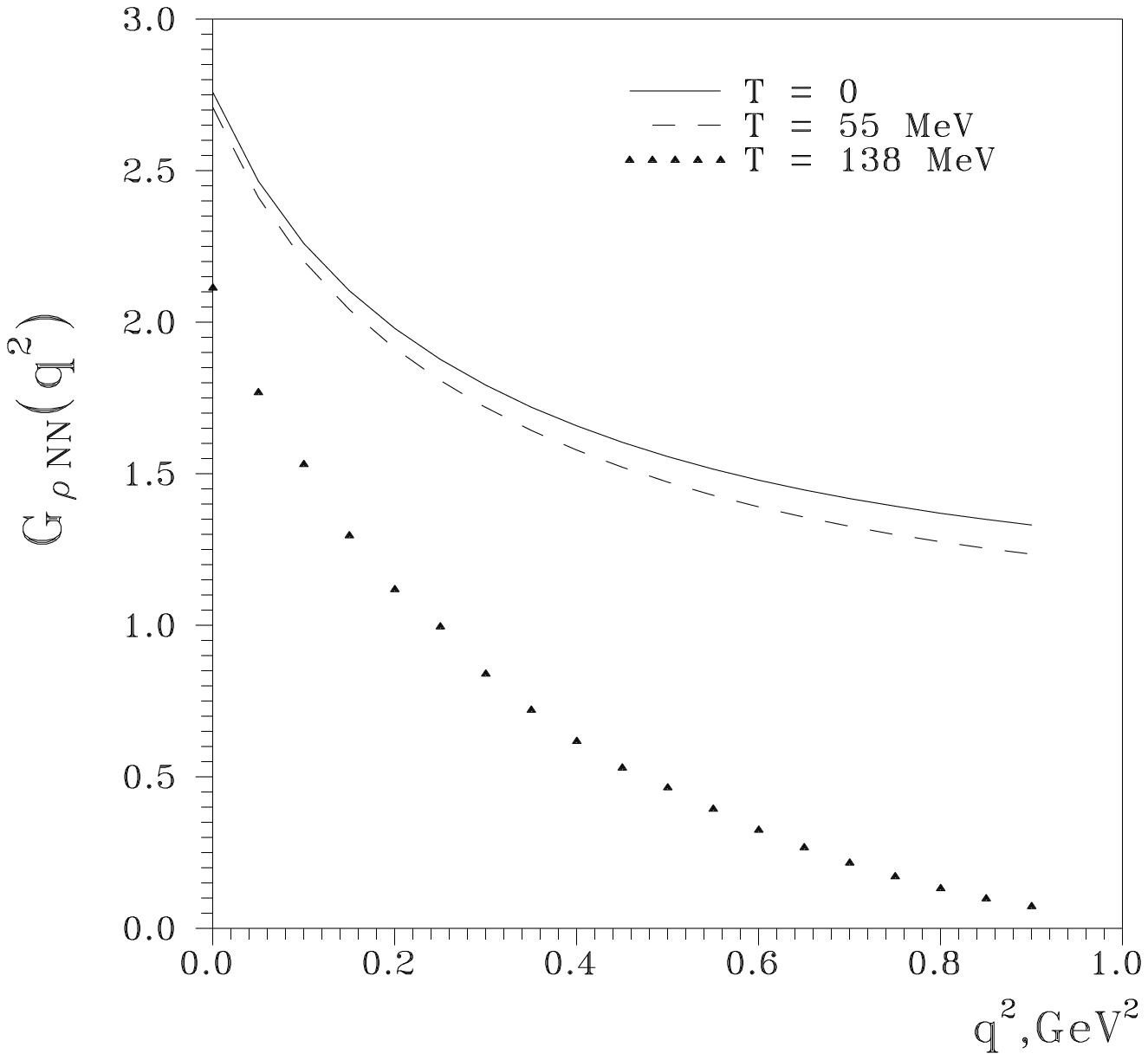} \ec
   \centerline{\parbox{15cm}
   {\caption{\label{figtfd5d}
The same as in Fig. \ref{figtfd5a} but for   $\rho NN$ vertex. }}}
\end{figure}
%%%%%%%%%%%%%%%%%%%%%%%%%%%%%%%%%%%%%%%%%%%%%%%%%

It would be quite interesting to know, if the critical
temperature, $T_{c}^{B}$, maybe uniquely considered as a point of
transition from the hadron state to a quark gluon state, as it
was suggested   \ci{dom}. The usual way of estimating $T_c$  is
based on QCD lattice calculations \ci{satztfd}  or temperature
 dependent quark - gluon potentials  \ci{alber}.
In such calculations it is usual to start from high temperatures
and then decreases the  temperature, looking for a phase
transition from the quark gluon state to a hadron state i.e.
hadronization of the quarks. Clearly this method gives a unique
$T_c$, since at any stage there is either a hadron or a quark
gluon plasma phase. Contrary to such studies the present
calculations explore the transition from a hadron to a
quark-gluon phase and find that the critical temperature for
hadron $\rightarrow$ quark-gluon transition depends on the kind
of hadrons
 under consideration. A similar result, where  hadrons and
quark - gluon plasma coexist have been obtained in lattice
calculations \ci{satztfd}. In conclusion it is important to
emphasize that the dynamics of hadrons and of quark-gluon plama
at finite temperature is poorly understood. The question of
critical phenomenon, in particular critical temperature, in the
hadronic systems is not well-known. Studies in nuclear matter are
needed urgently to clarify the critical transition from a
hadronic system to the quark gluon phase which will depend on
both density and temperature. The experiments with relativistic
heavy ion colliders would need such an understanding to clarify
the production of quark-gluon plasma from hadrons and the
eventual hadronisation of the quark-gluon system. It is possible
that there will be clear and unambiguous signals for the
formation and subsequent hadronisation of the quark-gluon plasma
\ci{singh1,singh2}.
%%%%%%%%%%%   END   TFD %%%%%%%%%%%%%%%%%%%%%%%%%%%%%

%\include{conclus}
%\include{thanks} They are in the Conclusion
%\newpage
%\chapter*[Appendices]{}
\addcontentsline{toc}{chapter}{{\bf Appendices}}
\addcontentsline{toc}{section}{{\bf Appendix 1}}
\def\theequation{A1.\arabic{equation}}
\setcounter{equation}{0}
%\flushright{
%{\Large\bf Appendix 1}
%}
\bc {\Large\bf Appendix 1} \ec \indent

In the present Appendix, following T. Cohen \ci{meisff},
 we derive explicit expressions
 for $G_{\pi NN}^{*}(q^{2})$ and $G_{\pi NN}^{*}(q^{2})$
used in Chapter 5. The small fluctuations around the vacuum value
are related to the pion field by \be \ba U=\exp(2i\vec\tau\vec\pi
/F_{\pi})\approx 1+2i \vec \tau\vec
\pi/{F_{\pi}}-2{{\vec{\pi}}^2}/\fpi^2+\dots \ea \ee which gives
the following approximation for the Lagrangian in Eq. \re{npa32}:
\be \ba {\cal L}\approx -\dsf{1}{2}(\vec \nabla \vec
\pi)^2\alpha_p- \dsf{1}{2}m_{\pi}^{*2}\vec \pi^2 \ea \ee
 and for the equation of motion:
\be \ba
 -(\vec \nabla^2 \pi)\alpha_p+m_{\pi}^{*2}\vec \pi=0
\ea \ee The factor $\alpha_p=1-\chi_{\Delta}$ in the last
equations is obtained by renormalization of the pion propagator
in the medium i.e. by the $\Delta$ - hole self energy term
$\hat\Pi_{\Delta}$. We define the in-medium $\pi NN$ coupling
constant and the vertex form factor by introducing the source term
\be \vec j=iG_{\pi NN}^*\bar \psi\vec \tau\gamma_5\psi \ee
 into the
equation of motion: \be -\nabla^2\vec
\pi\alpha_p+m_{\pi}^{*2}\vec \pi=iG_{\pi NN}^* \bar
\psi\gamma_5\vec \tau\psi , \lab{appkg} \ee where
\begin{eqnarray}
\psi\ket{N(\vec P)}=\dsf{e^{i\vec P\vec x}}{(2\pi)^{3/2}}
\left(\dsf{E^*+M_N^*}{2E^*}\right)^{1/2} \left(\begin{array}{c}
1\\
(\vec \sigma\vec P)/(E^*+M_N^*) \lab{psi}
\end{array}\right)\chi_S,
\end{eqnarray}
where $E^{*2}=\vec P^2+M^{*2}$. In the Breit frame $(\vec P
'=-\vec P$,
 $\vec q=\vec P -\vec P')$ the matrix element of the source
evaluated between nucleon states is given by: \be \bra{N(\vec
P')}\vec j(r)\ket{N(\vec P)}= \dsf{e^{i\vec q\,\vec
r}\bra{N}(\vec \sigma\vec q)\ {\vec \tau}\ket{N}} {(2\pi)^{3} \ \
2M_N^*}. \lab{appmelj} \ee Using the quantization rules \be \ba
\pi_{\alpha}(r)=-\dsf{iF_{\pi}}{4}\Tr[\tau_{\alpha}AU_0(r-R)A^+],\\
\quad \\
\bra{N}\Tr\tau_{\alpha}A\tau_{\beta}A^+\ket{N}=
-\dsf{2}{3}\bra{N}\sigma_{\alpha}\tau_{\beta}\ket{N} \ea \ee we
have \bea\ba \bra{N'(\vec P')}\pi_{\alpha}(\vec r,\vec
R)\ket{N(\vec P)}= \dsf{e^{i\vec q\,\vec r}}{(2\pi)^{3}}\ds\int
\bra{N'}\pi_{\alpha}(x)\ket{N}e^{-i\vec q\,\vec x}d\vec x=\\
\quad \\
=\dsf{F_{\pi}e^{i\vec q\,\vec r}}{6(2\pi)^{3}}\ds\int
\bra{N'}\sigma_{\alpha}(\vec \tau\vec x)\ket{N} e^{-i\vec q\,\vec
x}\sin(\theta)d\vec x \lab{appmelpi} \ea\eea where $\theta$ is
defined by the hedgehog ansatz $U_0(r)=e^{i(\vec \tau\vec
r)\theta(r)}$. Now the matrix element of Eq. \re{appkg} is
evaluated between collective wave functions for spin-up proton
$|p\uparrow\rangle$ with momentum $\vec P$ and
 $\vec P'$ in the Breit frame
and using equations \re{psi} - \re{appmelpi} to obtain \bea\ba
G_{\pi NN}^*(\vec{q}^{\ 2})=-\dsf{iF_{\pi}\alpha_pM_N^*}{3q}\int
(-\nabla^2+m_{\pi}^{*2}/\alpha_p)\hat x_3\sin(\theta)
e^{-i\vec q\,\vec x}d\vec x=\\
\quad \\
=\dsf{4\pi F_{\pi}\alpha_pM_N^*}{3}\ds\int\limits_o^{\infty}
\dsf{j_1(qx)}{qx}S_{\pi}(x)x^3dx \ea\eea where \be \ba
S_{\pi}(x)=-2\theta'c/x-\theta''c+\theta's+2s/x^2+
m_{\pi}^{*2}s/\alpha_p \ea \ee
 with $c=\cos(\theta)$, $s=\sin(\theta)$.

Similarly, the coupling constant at the $\sigma NN$
vertex is defined by the equation:\\
$(-\vec \nabla^2 +m_{\sigma}^{*2})\sigma=G_{\sigma
NN}^{*}\bar\psi\psi$. Evaluating matrix elements of both sides of
this equation it is easy to obtain the $\sigma NN$ form factor:
\bea\ba G_{\sigma NN}^*(\vec{q}^{\ 2})=2\pi F_{\pi}
\ds\int\limits_o^{\infty}  j_0(qx)S_{\sigma}(x)dx\    , \ea\eea
with
$S_{\sigma}(x)=-x^2\sigma''+2x\sigma'+x^2m_{\sigma}^{*2}\sigma$.

Note that the profile functions $\theta(r)$ and $\sigma(r)$ in
$S_{\pi}(x)$ and  $S_{\sigma}(x)$ are the solutions of the
equations of motion: \bea\ba
\theta''x^2\chi^2\alpha_p+4s_2\theta^{\prime
2}+8s^2\theta^{\prime\prime}+
2x\theta'\chi^2\alpha_p+2x^2\theta'\chi\chi'\alpha_p-\\
\qquad\qquad
-\chi^2\alpha_ps_2-4s_2d-x^2\beta^2\chi^3s=0\\
x^2\chi''+2\chi\chi'-2\chi\alpha_{p}x^2(\theta^{\prime 2}/2+d)-
16x^2{\cal D}_{eff}(\chi^3-\chi^{\varepsilon-1})-\\
\qquad\qquad -3x^2\beta^2(1-c)\chi^2=0. \lab{appeneq} \ea\eea
where $x\equiv eF_{\pi}r$, $s_2\equiv\sin (2\theta)$, $d\equiv
s^2/x^2$, $\beta=m_{\pi}^*/e F_{\pi}$, ${\cal
D}_{eff}=C_g^*/24e^2F_{\pi}^4$,\\ $\chi\equiv\exp(-\sigma(x))$.
The boundary conditions are:\\
 $\theta=\pi$, $\sigma'=0$ for $x\rightarrow 0$ and
$\theta \sim (1+\beta x)e^{-\beta x}/x^2$, $\sigma\sim \theta^2$
for large $x$.

For completeness we write also the explicit expressions for
$I=I_0^*+I_{rt}$: \be \ba
I_0^*=\dsf{2\pi}{3e^3F_{\pi}}\int\limits_0^{\infty}
s^2\{e^{-2\sigma}+ 4(\theta^{\prime 2}+d)\}x^2dx, \nwl
I_{rt}=\dsf{\pi\alpha_{rt}}{3\omega_{\Delta}^2eF_{\pi}}
\int\limits_0^{\infty} \{\theta^{\prime 2}c^2+2d\}x^2dx. \ea \ee

\addcontentsline{toc}{section}{{\bf Appendix 2}}
\def\theequation{A2.\arabic{equation}}
\setcounter{equation}{0} \bc {\Large\bf Appendix 2} \ec \indent

Here some auxiliary formulas for Chapter 6 are presented. The
isoscalar and isovector mean square radii are given by \beq
\begin{array}{l}
{\langle}r^2{\rangle}_{I=0}=-2\ds\int\limits_0^{\infty}r^2\theta's^2dr\,\,,\\
{\langle}r^2{\rangle}_{I=1}=\dsf{\ds\int\limits_0^{\infty}r^4
s^2[\chi^2+4(\theta^{\prime 2}+d)]dr}{
\ds\int\limits_0^{\infty}r^2 s^2[\chi^2+4(\theta^{\prime
2}+d)]dr}\vergul
\end{array}
\eeq where the notations $\theta$, $s$ etc are given in Appendix
A. The nucleon mean square radii  ${\langle}r^2{\rangle}_{p,n}$
are expressed in terms of ${\langle}r^2{\rangle}_{I=0}$ and
${\langle}r^2{\rangle}_{I=1}$ as:
$$
{\langle}r^2{\rangle}_{p,n}=\frac{1}{2}[{\langle}r^2{\rangle}_{I=0}\pm
{\langle}r^2{\rangle}_{I=1}].
$$
We note that the expression for mean square radius in a medium is
similar to that in a vacuum, a solution to the equation
\re{appeneq} with medium parameters must be taken in former case.

The proton and neutron magnetic moments are given by \beq
\begin{array}{l}
\mu_{p,n}=\dsf{1}{2}[\mu_{p,n}^{I=0}+\mu_{p,n}^{I=1}]\,\,,\\
\quad\\
\mu_{p,n}^{I=0}=\dsf{2}{9}M_N(M_{\Delta}^*-M_N^*)
{\langle}r^2{\rangle}_{I=0}\mu_B\,\,,\\
\quad\\
\mu_{p}^{I=1}=-\mu_{n}^{I=1}=\dsf{1}{3}\cdot
\dsf{8\pi}{3e^2F_{\pi}}\int\limits_0^{\infty}dx\left\{x^2s^2
\left(\frac{\alpha_p}{4}+(\theta'+d)\right)\right\}\,\,,
\end{array}
\eeq where $\mu_B$ is the Bohr magneton. For the nucleon form
factors we have \beq
\begin{array}{l}
G_{E}^p=\dsf{1}{2}(G_{E}^S+G_{E}^V)\,\,,\quad
G_{E}^n=\dsf{1}{2}(G_{E}^S-G_{E}^V)\,\,;\\
\quad \\
G_{M}^p=\dsf{1}{2}(G_{M}^S+G_{M}^V)\,\,,\quad
G_{M}^n=\dsf{1}{2}(G_{M}^S-G_{M}^V)\,\,.
\end{array}
\eeq

\addcontentsline{toc}{section}{{\bf Appendix 3}}
\def\theequation{A3.\arabic{equation}}
\setcounter{equation}{0} \bc {\Large\bf Appendix 3} \ec

\indent

Here the  explicit expressions for $W_{AB}(t,k^2,T)$ introduced
in Chapter 7 are given. \be \ba W_{\pi\pi}=G_{\pi}(k^2)[M^2-(ab)]
\quad , W_{\pi\sigma}=W_{\pi\pi}|_{G_{\pi}\rightarrow
G_{\sigma}}\, , \nwl
W_{\pi\omega}=A_{\pi}^\omega[M^2-(ab)]+3F_1^\omega(k^2)F_2^\omega(k^2)
z\, \quad , \nwl W_{\pi\rho}=-W_{\pi\omega}|_{
(F_i^\omega\rightarrow F_i^\rho \;,\; A_\pi^\omega\rightarrow
A_\pi^\rho)} \ea \ee
\medskip
\be \ba
A_{\pi}^\omega=[F_1^\omega(k^2)]^2[4-\ds\frac{k^2}{m_\omega^2}]+
\frac{3k^2[F_2^\omega(k^2)]^2}{4M^2}\,, \ea \ee
\medskip
\be \ba W_{\sigma\pi}=-3g_\pi^2(k^2)[M^2+(ab)]\,,\quad
W_{\sigma\sigma}=-\ds\frac{1}{3}W_{\sigma\pi}|_{G_\pi\rightarrow
G_\sigma}\,, \nwl
W_{\sigma\omega}=[M^2+(ab)]A_\sigma^\omega\,;\quad
W_{\sigma\rho}=3W_{\sigma\omega}|_{ A_\sigma^\omega\rightarrow
A_\sigma^\rho}\,; \nwl
A_{\sigma}^\omega=-[F_1^\omega(k^2)]^2[4-\ds\frac{k^2}{m_\omega^2}]+
\frac{3k^2[F_2^\omega(k^2)]^2}{4M^2}\,;\\
%%\quad \\
W_{\omega\pi} =\dsf{3g_{\pi}^2(k^2)}{4}\left\{2[2M^2-(ab)]-
\dsf{3F_2^\omega(t)t}{2F_1^\omega(t)}\right\}\,;\quad
W_{\omega\sigma}=\dsf{1}{3}W_{\omega\pi}|_{G_{\pi}\rightarrow G_\sigma}\,;\\
%%\quad \\
W_{\omega\omega}=\dsf{1}{4}\left\{2\Gamma_1^\omega[2M^2-(ab)]+
\dsf{\Gamma_2^\omega}{M^2}[k^2[M^2-(ab)]+2(ak)(bk)]+\right.\\
%%\quad \\
+3z \Gamma_3^\omega-\dsf{F_2^\omega(t)}{2M^2F_1^\omega(t)}
[3\Gamma_1^\omega t+\Gamma_2^\omega(k^2t-z^2)]+\\
%%\quad \\
+\left.\dsf{\Gamma_3^\omega}{M^2}[z[(ab)+3M^2]-
2(ak)(bq)-2(aq)(bk)]\right\}\,; \ea \ee
\medskip
\be \ba
\Gamma_1^\omega=[F_1^\omega(k^2)]^2[2-\ds\frac{k^2}{m_\omega^2}]+
\frac{k^2[F_2^\omega(k^2)]^2}{4M^2},\,\quad
\quad\\
\Gamma_2^\omega=\dsf{2M^2}{m_\omega^2}[F_1^\omega(k^2)]^2
-[F_2^\omega]^2\,;
\quad\\
\Gamma_3^\omega=F_1^\omega(k^2) F_2^\omega(k^2)\,,\quad \quad
W_{\omega\rho}=3 W_{\omega\omega}|_{\Gamma_i^\omega\rightarrow
\Gamma_i^\rho}\,,\quad
W_{\rho\pi}=-\dsf{1}{3}W_{\omega\pi}|_{\omega\rightarrow\pi}\,;\\
W_{\rho\sigma}=-W_{\rho\pi}|_{G_\pi\rightarrow G_\sigma}\,;\quad
W_{\rho\omega}=W_{\omega\omega}\left|_{\footnotesize\ba{l}
\Gamma_i^\omega\rightarrow\Gamma_i^\omega\\
F_i^\omega\rightarrow F_i^\rho \ea}\right.\,;\quad
W_{\rho\rho}=-W_{\omega\omega}|_{\omega\rightarrow\rho}\,; \ea \ee
where $G_{\pi}\equiv G_{\pi NN}(k^2)$, $G_{\sigma}\equiv
G_{\sigma NN}(k^2)$, $F_1^\omega\equiv G_{\omega NN}$,
$F_2^\omega\equiv F_{\omega NN}$ , $z=(kq)$ and $q^2=t$.

\edc